\documentclass[12pt]{article}
\usepackage{amssymb,amsmath}
\usepackage{cancel}
\usepackage{palatino}
\usepackage{longtable}
\usepackage{color}
\numberwithin{equation}{section}
\newtheorem{theorem}{Theorem}[section]     
\newtheorem{definition}[theorem]{Definition}
\newtheorem{defi}[theorem]{Definition}
\newtheorem{proposition}[theorem]{Proposition}
\newtheorem{lemma}[theorem]{Lemma}
\newtheorem{rmk}[theorem]{Remark}

\newtheorem{remark}[theorem]{Remark}

\def\d{\partial}
\def\n{\noindent}
\def\f{\frac}

\def\proof{\noindent\hspace{2em}{\itshape Proof: }}
\def\QEDclosed{\mbox{\rule[0pt]{1.3ex}{1.3ex}}} 

\def\QED{\QEDclosed} 
\def\endproof{\hspace*{\fill}~\QED\par\endtrivlist\unskip}
\newcommand{\eqa}{\begin{eqnarray}}
\newcommand{\eeqa}{\end{eqnarray}}
\newcommand{\beq}{\begin{equation}}
\newcommand{\eeq}{\end{equation}}

\begin{document}
\title{Complex reflection groups,  logarithmic connections and bi-flat F-manifolds}
\author{Alessandro Arsie* and Paolo Lorenzoni**\\
\\
{\small *Department of Mathematics and Statistics}\\
{\small The University of Toledo,}
{\small 2801 W. Bancroft St., 43606 Toledo, OH, USA}\\
{\small **Dipartimento di Matematica e Applicazioni}\\
{\small Universit\`a di Milano-Bicocca,}
{\small Via Roberto Cozzi 53, I-20125 Milano, Italy}\\
{\small *alessandro.arsie@utoledo.edu,  **paolo.lorenzoni@unimib.it}}

\date{}

\maketitle

\begin{abstract}
We show that bi-flat $F$-manifolds can be interpreted as natural geometrical structures encoding the almost duality for Frobenius manifolds without metric.
Using this framework, we extend Dubrovin's duality between orbit spaces of Coxeter groups and Veselov's $\vee$-systems, to the orbit spaces of exceptional well-generated complex reflection groups of rank $2$ and $3$. On the Veselov's $\vee$-systems side, we provide a generalization of the notion of $\vee$-systems that gives rise to a dual connection which coincides with a Dunkl-Kohno-type connection associated with such groups. In particular, this allows us to treat on the same ground several different examples including Coxeter and Shephard groups. Remarkably, as a byproduct of our results, we prove that in some examples  basic flat invariants are not uniquely defined. As far as we know, such a phenomenon has never been pointed out before.

\end{abstract}
\tableofcontents
\section{Introduction}
A complex (pseudo)-reflection is a unitary transformation of $\mathbb{C}^n$ of finite period that leaves invariant a hyperplane. It is characterized by the property that all the eingevalues of the associated matrix representation are equal to $1$, except for one. The remaining eigenvalue is a $k$-th primitive root of unity, where $k$ is the period of the transformation.

A finite subgroup of the group of unitary transformations is a finite complex reflection groups if it is generated by complex reflections. Irreducible finite complex reflection groups were classified by Shephard and Todd in \cite{ST}, and consist in an infinite family depending on $3$ positive integers and $34$ exceptional cases.
They proved that the ring of invariant polynomials of a complex reflection group is generated by $n$ algebraically independent invariant polynomials, where $n$ is the dimension of the complex vector space on which the group acts. Well-generated irreducible complex reflection groups are irreducible complex reflection groups of rank $n$, whose minimal generating set consists of $n$ reflections. In the sequel we will restrict our attention only to well-generated complex reflection groups, even if it is not explicitly stated.

In this paper, we consider two affine flat connections naturally associated with an irreducible well-generated complex reflection group $G$: 
\begin{enumerate}
\item The first one is defined by the Saito coordinates, which correspond to a special choice of basic invariant polynomials. They were originally introduced for Coxeter groups in \cite{SYS,Sa} and play a crucial role in the construction of the Frobenius manifolds associated with Coxeter groups \cite{DubrovinCoxeter}  (see also \cite{Sa2} for the case of singularity theory). The notion of  Saito flat coordinates (also called flat basic invariants) was extended to Shephard groups in \cite{OS2} (see also \cite{OT}). Recently, the existence of a system of flat basic invariants for well-generated complex reflection groups has been proved in \cite{KMS2}.  
\item The second one is a flat torsionless  logarithmic connection  associated with the arrangement of hyperplanes $H$ fixed by some elements of $G$, given by
\beq\label{kohno}\tilde \nabla=\nabla-\f{1}{N}\sum_{H\in \mathcal{H}}\frac{d\alpha_H}{\alpha_H}\otimes\kappa_H\pi_H,\eeq
where $\nabla$ is the standard flat connection on $\mathbb{C}^n$,  $\alpha_H$ is a linear form defining the mirror $H$, $\mathcal{H}$
 is  the collection of the mirrors $H$, $\pi_H$ denotes the unitary projection onto the unitary complement of the hyperplane $H$,  computed with respect to a suitable Hermitian metric, the weights $\kappa_H$ are suitable complex numbers and $N$ is a normalizing factor chosen in such 
a way that
\beq\label{normc}
\sum_{H\in \mathcal{H}}\kappa_H\pi_H={\rm Id}_{\mathbb{C}^n}.
\eeq
The above connection is a particular case of {\it Dunkl-Kohno-type} connections considered in \cite{CHL}  (see also similar structures considered in \cite{BMR} and \cite{DO}). 
If the collection $\{k_H\}_{H\in \mathcal{H}}$ of complex weights is $G$-invariant, then the connection defined in \eqref{kohno} is flat (see  \cite{Looijenga}). 
In a more general setting, different flatness conditions for this kind of connections were first studied by Kohno (see \cite{Ko,Ko2}). 
In the case of Coxeter groups, these Dunk-Kohno--type connections coincide with the flat connection appearing in the theory of $\vee$-system \cite{Ve,Ve2}. In this case Kohno's  flatness conditions are equivalent to Veselov's definition of $\vee$-system \cite{ALjmp,FV}. Important examples of these  logarithmic connections appear also in the Physics literature and are called KZ connections \cite{KZ}.
\end{enumerate}
The aim of this paper is to show how the two flat structures above are related. 
\newline
\newline
The main results of the paper are the following.
\begin{enumerate}
\item  We show that the two affine flat connections described above endow the space of orbits of well-generated irreducible complex reflection groups with the structure of a bi-flat $F$-manifold. 
The notion of bi-flat $F$ manifold has been introduced in \cite{AL-Bi-flat} as a natural generalization of Frobenius manifolds in the framework of integrable dispersionless PDEs.  It was later showed that tridimensional semisimple bi-flat $F$ are parameterized by  solutions of generic Painlev\'e VI equation \cite{Limrn} (see also \cite{AL-Bi-flat}) and, in the non-semisimple (regular) case, by solutions of generic  Painlev\'e IV  and  Painlev\'e V equations, depending on the Jordan canonical form of the linear endomorphism given by multiplication by the Euler vector field \cite{ALmulti}. We believe that these results combined with the results  of the present paper unveil the relevance of the notion of bi-flat $F$-manifold as a meaningful generalization of the notion of Frobenius manifold.  

Due to computational difficulties we focus on the rank $2$ and rank $3$ cases. However we believe that our results hold true also for higher rank groups. 
\item We prove that for the Shephard groups $G_5$, $G_6$, $G_9$, $G_{10}$, $G_{14}$, $G_{17}$, $G_{18}$, $G_{21}$, $G_{26}$ the bi-flat $F$-structure is not uniquely defined but depends on a parameter. For each of these groups, there is a single value of the parameter that corresponds to the associated Frobenius manifold structure. In other words, the Frobenius structure of each of these orbit spaces is naturally embedded in a one-parameter family of bi-flat $F$-manifold structures, which can be viewed as a kind of deformation of the Frobenius structure itself. 

\item It is a well-known result that the Frobenius structure of each Shephard group coincides with the Frobenius structure of the underlying Coxeter group (\cite{Dad}) as a consequence of Hertling's theorem \cite{CH}. In complete analogy, we prove that on the orbit spaces of all Shephard groups mentioned above, the bi-flat $F$-manifold structure coincide with the bi-flat $F$-structure   of the orbit space of the underlying Coxeter group. More precisely:
\begin{itemize}
\item the one parameter families of bi-flat $F$-structures on the orbit space of the groups $G_5$, $G_{10}$, $G_{18}$  coincide with the  one parameter family of bi-flat  $F$-structures on the orbit space of $B_2$.
\item  the one parameter families of bi-flat  $F$-structures on the orbit space of the groups $G_6$, $G_{9}$, $G_{17}$ coincide with the  one parameter family of bi-flat $F$-structures on the orbit space of $I_2(6)$.
\item the one parameter family of bi-flat $F$-structures on the orbit space of $G_{14}$ coincides with the  one parameter family of bi-flat  $F$-structures on the orbit space of $I_2(8)$.
\item the one parameter family of bi-flat  $F$-structures on the orbit space of $G_{21}$ coincides with the  one parameter family of bi-flat  $F$-structures on the orbit space of $I_2(10)$.
\item the one parameter family of bi-flat  $F$-structures on the orbit space of $G_{26}$  coincides with the  one parameter family of bi-flat  $F$-structures on the orbit space of $B_3$.
\end{itemize}
In all these cases, for specific values of the parameter we recover the standard Frobenius manifold structure on the space of orbits of Coxeter groups. This fact leads us to introduce the notion of {\em generalized} Saito flat coordinates. Remarkably, 
 there is a standard choice of the bi-flat structure on the orbit space of well-generated complex reflection groups such that
 the structure constants of the dual product and the Christoffel symbols of the dual connection have the following form:
\beq\label{main-intro}
c^{*i}_{jk}(p)=-\Gamma^{*i}_{jk}(p)=\frac{1}{d_l-1}\f{\d^2 u^l}{\d p^j\d p^k}(J^{-1})^i_l,
\eeq
where $\{p^1 \dots, p^n\}$ are standard coordinates on $\mathbb{C}^n$, $\{u^1, \dots, u^n\}$ are the set of basic invariant polynomials defining the generalized Saito flat coordinates, $J^i_j=\f{\d u^i}{\d p^j}$ and $d_i$ are the degrees of the invariant
 polynomials.

\item Inspired by Veselov's $\vee$-system idea, we prove that in all rank $2$ and $3$ exceptional well-generated complex reflection groups, the dual connection can be expressed as a connection of Dunkl-Kohno-type built from the reflecting hyperplanes. As a byproduct, our construction provides also a generalization of $\vee$-systems to the case of complex roots in $\mathbb{C}^n$ and to the case of \emph{Hermitian metrics}. Indeed, while in the case of $\vee$-systems one uses a non-degenerate symmetric bilinear form (either in $\mathbb{R}^n$ or in $\mathbb{C}^n$, see \cite{FV}), in our case, to identify the relevant connection as a connection of Dunkl-Kohno-type, one is forced to use a non-degenerate sesquilinear form (in all cases we considered it is just the standard Hermitian form on $\mathbb{C}^n$, except in the case of $G_{27}$). Remarkably, in all the examples the dual product has the form
\beq\label{main-intro2}
X*Y_{p}=\f{1}{N}\left(\sum_{s=1}^{n}\frac{\kappa_s}{||\alpha_s||^2}\frac{\alpha_s(X)\,\alpha_s(Y)\,\check{\bar\alpha}_s}{\alpha_s(p)}\right)
\eeq
where $\kappa_s$ is the order of the corresponding (pseudo)-reflection, $X$ and $Y$ are arbitrary vector fields, $\alpha_s$ are the covectors defining the hyperplanes and $\check{\bar\alpha}_s$ are vectors obtained from $\alpha_s$ using a suitable Hermitian metric. Combining the above formula  with the formula \eqref{main-intro} we obtain the following system of PDEs for the standard generalized Saito flat coordinates:  
\beq\label{main-intro3}
\f{\d^2 u^i}{\d p^j\d p^k}=(d_i-1)c^{*s}_{jk}\f{\d u^i}{\d p^s},
\eeq
where the dual product is given by the formula \eqref{main-intro2}. In other words \emph{each Saito flat coordinate $u^i$ is also a flat coordinate of the flat connection} $\nabla^{(0)}+(d_i-1) *$. Using conditions \eqref{main-intro3} one can easily find
 standard flat basic invariants starting from general basic invariants. A similar formula holds true also in the non-standard case. 
  As far as we know, many of the explicit formulas obtained 
 in this paper for flat basic invariants have never appeared in literature before.
\end{enumerate}

As it is often the case in the mathematical literature concerning well-generated complex reflection groups, our results have been obtained via a case by case analysis for each group. This is consistent with the fact that very often there does not seem to exist an overarching strategy to tackle properties of well-generated complex reflection groups (see for instance the comments in \cite{VR}).  

In our case this is also due to the fact that while we present general statements, there are always some ``misbehaved" groups. For instance, when we prove that the dual connection can be expressed as a connection of Dunkl-Kohno type using a sort of generalization of Veselov's $\vee$-systems, the Hermitian metric appearing in the construction is always the standard one, except for $G_{27}$ (see Table \ref{tab3}). Similarly, the (generalized) flat Saito coordinates are in some case uniquely determined, while in other cases they appear in families depending on a parameter. These facts suggest that it might be unlikely to find  general a priori proofs of most of the results appearing in this work.

The paper is organized as follows. In Section 2, we introduce the notion of Frobenius structures and their almost dual and we recall the Dubrovin-Saito construction on the space of orbits of Coxeter groups, and  Dubrovin's construction on the space of orbits of Shephard groups. In Section 3, we introduce Veselov's  $\vee$-systems focusing our attention on the case of Coxeter groups. In Section 4, we briefly recall the definition of flat and bi-flat $F$-manifolds and we show how the bi-flat structure arises
 naturally when one tries to extend Dubrovin's almost duality to Frobenius manifolds  ``without metric".  In Section 5 we show how to endow with a bi-flat $F$-structure the space of orbits of well-generated complex reflection groups. In Section 6, we show that for all rank 2 and 3 exceptional well-generated irreducible complex reflection groups, the standard dual connection coincides with the Dunkl-Kohno connection obtained choosing as weights $k_{H}$ the orders of the corresponding (pseudo)-reflections.
 
 In general, in the case of Shephard groups the standard bi-flat $F$-structure does not coincide with the structure of Frobenius manifolds introduced by Dubrovin. In Section 7 we explain how to recover the Dubrovin's structure considering a family of deformed
 Saito flat coordinates (generalized Saito Coordinates). It turns out that, also in this case,  the dual connection is of Dunkl-Kohno type but with a different choice of the weights. In the final Section 8, we show that semisimple bi-flat $F$-manifolds are parametrized in flat coordinates by solutions of generalized WDVV equations. These equations are the oriented
 associativity equations \cite{LM} with an additional homogeneity condition. They have been recently introduced in \cite{KMS1} in the study of Saito structures without metrics \cite{S}. These structures are completely characterized by a flat meromorphic
 connection on the bundle $\pi^*TM$ on $\mathbb{P}\times M$. The result of Section $8$ implies that this connection plays the same role in the theory of semisimple bi-flat $F$-manifolds as the role played by Dubrovin's extended connection in the theory of Frobenius manifolds.  It also implies that one can construct a bi-flat structure on the orbit space of a well-generated complex reflection group,  starting from any solution of generalized WDVV equations.  
The paper ends with three Appendices. In the first one we list all the reflecting hyperplanes for all exceptional well-generated complex reflection groups up to $G_{27}$, in the second one,  using our procedure, we provide (conjectural) Saito flat coordinates for $G_{29}$, $G_{32}$ and $G_{33}$. In the last one we briefly discuss what happens in the non-well-generated cases.
\section*{Ackowledgements} 
We would like to thank Vic Reiner for having suggested reference \cite{LT} and Alexander Veselov for reference \cite{FS} and for his interest in this work.
 This research was partially supported by GNFM through the 2015 Visitors Program.

\section{Frobenius manifolds and their almost dual}
Originally introduced by Dubrovin as a geometric framework to study the WDVV equations of topological field theory, Frobenius manifolds play nowadays
 an important role in several areas of mathematics including singularity theory, quantum cohomology and integrable systems.
 Let us recall their definition.
\begin{defi} 
\label{defi:fmancc}
A  \emph{Frobenius manifold} $(M,\circ,\eta,e,E)$ is a manifold equipped with an associative commutative product $\circ$ on sections of its tangent bundle, two distinguished vector fields $e$ (unit) and $E$ (Euler vector field) and 
 a flat pseudo-metric $\eta$ satisfying the following requirements:
\begin{itemize}
\item  $\eta$ is invariant with respect to the product: $\eta_{il}c^l_{jk}=\eta_{jl}c^l_{ik}$, where $c^i_{jk}$ are the structure constants for $\circ$. 
\item the Levi-Civita connection $\nabla$ associated to $\eta$ is  compatible with the product:
\begin{equation*}
\nabla_k c^i_{jl}=\nabla_j c^i_{kl}.
\end{equation*}
\item $e$ is the unit of the product and it is flat: $\nabla e=0$.
\item Furthermore, the following conditions must hold: $$\nabla\nabla E=0,\quad [e,E]=e, \quad {\rm Lie}_Ec^i_{jk}=c^i_{jk},\quad {\rm Lie}_E\eta=D\eta,$$	
\end{itemize}
where $D$ is a constant. 
\end{defi}

The product is called \emph{semisimple} if there exist a distinguished coordinates system such that $c^i_{jk}=\delta^i_j\delta^i_k$.

The existence of the Euler vector field allows one to define a second commutative associative product on sections of the tangent bundle, called {\it the dual product} and defined as
\begin{equation}
X*Y:=E^{-1}\circ X\circ Y,
\end{equation}
and a second contravariant flat metric $g$, called the intersection form, and defined as $g^{ij}=\eta^{il}c^j_{lk}E^k$. 

Furthermore, the very definition of Frobenius manifold is constructed in such a way that in flat coordinates $\{u_1, \dots, u_n\}$ for $\eta$, there exists a smooth function $F(u_1, \dots, u_n)$ such that 
$c^i_{jk}=\eta^{il}\frac{\d^3 F}{\d u_l\d u_j \d u_k}.$ This function is called \emph{Frobenius potential}. 

In the next Section we will study in details the case of Frobenius manifolds associated with Coxeter groups.

\subsection{Frobenius manifolds and space of orbits of Coxeter groups}
One of the main examples of Frobenius manifold is provided by the space of orbits of Coxeter groups \cite{DubrovinCoxeter}. The realization of Frobenius manifold structure on these spaces  is based on the notion of \emph{flat pencil of metrics} and relies on the existence of a distinguished set of basic 
 invariants of the group, called \emph{Saito flat coordinates} \cite{Sa}. 

A flat pencil of contravariant metrics is a one-parameter family of contravariant metrics $g_{\lambda}=g-\lambda\eta$ satisfying the following two properties \cite{du97}: 
\begin{itemize}
\item $g_{\lambda}$ is flat for any $\lambda$,
\item the pencil of Christoffel symbols coincides with the Christoffel symbols of the pencil.
\end{itemize}
In the case of Frobenius manifolds, it is well known that the inverse $\eta^{ij}$ of the invariant (pseudo)-metric $\eta_{ij}$ and the intersection form $g^{ij}$ satisfy the above conditions (see \cite{du97} for details). They satisfy also two additional properties called \emph{quasi-homogeneity} and \emph{exactness}:
\begin{itemize}
\item  Quasi-homogeneity implies the existence of a constant $c$ such that ${\rm Lie}_E \eta=c\,\eta$ (here and below ${\rm Lie}_{X}$ denotes the Lie derivative with respect to a vector field $X$).
\item Exactness means that there exists a vector field (the unit vector field $e$ in the Frobenius case) such that
\beq 
{\rm Lie }_e g=\eta,\qquad {\rm  Lie}_e \eta=0.
\eeq
Since the unit vector field is flat (i.e. $\nabla e=0$), one can choose flat coordinates $\{u^1, \dots u^n\}$ such that $e=\f{\d}{\d u^1}$. In such coordinates one obtains $\eta$ just shifting $g$ along the variable $u^1$.
\end{itemize}
Conversely, given an exact quasi-homogeneous flat pencil of metrics, one can reconstruct a Frobenius manifold. In particular the Frobenius potential $F(u_1, \dots, u_n)$ is obtained
 in flat coordinates $(u^1,...,u^n)$ for $\eta$ (the Saito flat coordinates) by means of the following formula \cite{DubrovinCoxeter}
\beq\label{Ffromg}
\eta^{il}\eta^{jm}\f{\d^2 F}{\d u^l\d u^m}=\frac{g^{ij}}{{\rm deg}(g^{ij})}
\eeq
where ${\rm deg}(g^{ij})=\f{E(g^{ij})}{g^{ij}}$. 

Given a finite Coxeter group,  the crucial observation that allowed Dubrovin to define an exact flat pencil of metrics is that, due to the properties of the degrees of basic invariant polynomials, the Euclidean metric $g$,
 written with respect to a set of basic invariants, depends linearly on the invariant polynomial of highest degree. This implies that the metrics $g$ and $\eta:={\rm Lie}_e g$ (where $e$ is the generator of the shift along the highest invariant polynomial) form an exact flat pencil of metrics. Moreover, using the homogeneity properties of basic invariant polynomials, one can easily prove 
 that this pencil of metrics  is quasi-homogeneous. Starting from this pencil one can easily reconstruct the full Forbenius structure using the formula \eqref{Ffromg}. In order to illustrate Dubrovin's construction we consider two simple examples.
 
\subsubsection{The case of $A_3$}
$A_3$ is a finite Coxeter group that can be realized as a real reflection group. Any real reflection group gives rise to a well-generated complex reflection group. $A_3$ can be viewed as a well-generated complex reflection group of rank three. Since it also coincides with the symmetric group on $4$ letters $\rm{Sym}(4)$ (with a suitable representation) its invariants are constructed starting from $\mathbb{C}^4$, considering polynomials in $p_0, p_1, p_2, p_3$ that are invariants under $\rm{Sym}(4)$ and then restricting them to the hyperplane given by the equation $p_0+p_1+p_2+p_3=0$. 
A general basis for the ring of invariants, restricted already to the hyperplane $p_0+p_1+p_2+p_3=0$ is provided by the following polynomials of degree $2,3,4$, where an arbitrary constant $c$ has been introduced:
\begin{eqnarray*}
u_1&=&-p_1^2-p_2^2-p_3^2-p_1p_2-p_1p_3-p_2p_3,\\
u_2&=&p_1^2(p_2+p_3)+p_2^2(p_1+p_3)+p_3^2(p_1+p_2)+2p_1p_2p_3,\\
u_3&=&-p_1^2p_2p_3-p_1p_2^2p_3-p_1p_2p_3^2-cu_1^2.\\
\end{eqnarray*}
The Euclidean covariant metric $dp_0^2+dp_1^2+dp_2^2+dp_3^2$ restricted to the hyperplane $p_0+p_1+p_2+p_3=0$ is expressed via the following matrix (in the coordinates $p_1, p_2, p_3$):
$$g=\left[\begin{array}{ccc}
2 & 1 & 1\\
1 & 2 & 1\\
1& 1 & 2
\end{array} \right],$$
and inverting it, one gets the contravariant metric 
$$g^{-1}=\left[\begin{array}{ccc}
\f{3}{4} & -\f{1}{4} & -\f{1}{4}\\
-\f{1}{4} & \f{3}{4} & -\f{1}{4}\\
-\f{1}{4} & -\f{1}{4} & \f{3}{4}
\end{array} \right].$$
 Using the tensorial properties of the metric, we can express the metric $g$ in the coordinates $u_1, u_2, u_3$ as follows:
$$g=\left[\begin{array}{ccc}
-2u_1 &-3u_2& -4u_3\\
-3u_2 & -4u_3+(1-4c)u_1^2 &  \left(6c+\frac{1}{2}\right)u_1u_2\\
-4u_3 &  \left(6c+\frac{1}{2}\right)u_1u_2 & \frac{3}{4}u_2^2+(16c-2)u_1u_3+(8c^2-2c)u_1^3
\end{array} \right].$$
In particular in the Saito flat coordinates, which correspond to the value $c=\f{1}{8}$, we obtain
$$
g=\begin{pmatrix}
-2u_1 & -3u_2 & -4u_3\cr
-3u_2 & -4u_3+\f{1}{2}u_1^2 & \f{5}{4}u_1u_2\\
-4u_3 & \f{5}{4}u_1u_2 & \f{3}{4}u_2^2-\f{1}{8}u_1^3
\end{pmatrix}.
$$
Up to an inessential constant factor the metric $\eta$ is given by
$$
\eta={\rm Lie}_eg=\begin{pmatrix}
0 & 0 & 1\cr
0 & 1 & 0\\
1 & 0 & 0
\end{pmatrix},
$$
and, solving the equations \eqref{Ffromg} one obtains the following Frobenius potential:
\beq\label{F-A3}
F= \f{1}{8}u_3^2u_1+\f{1}{8}u_3u_2^2-\f{1}{64}u_1^2u_2^2+\f{1}{3840}u_1^5.
\eeq

\subsubsection{The case of $H_3$ ($G_{23}$)}
A second example is the Coxeter group $H_3$. It corresponds to the group $G_{23}$ of the Shephard-Todd list. 
The basic invariants are given by (\cite{Me,IKM})
\begin{eqnarray*}
U_1&=&p_1^2+p_2^2+p_3^2\\
U_2&=&2(p_1^6+p_2^6+p_3^6)-15(p_1^4p_2^2+p_1^4p_3^2
+p_2^4p_1^2+p_2^4p_3^2+p_3^4p_1^2+p_3^4p_2^2)+180p_1^2p_2^2p_3^2\\
&&+21\sqrt{5}\Delta\\
U_3&=&5\left(2\,(p_1^{10}+p_2^{10}+p_3^{10})-45(p_1^8p_2^2+p_1^8p_3^2+p_2^8p_1^2+p_2^8p_3^2
+p_3^8p_1^2+p_3^8p_2^2)\right.\\
&&\left.+42\,(p_1^6p_2^4+p_1^6p_3^4+p_2^6p_1^4+p_2^6p_3^4+p_3^6p_1^4
+p_3^6p_2^4)\right.\\
&&\left.+1008\,(p_1^6p_2^2p_3^2+p_2^6p_1^2p_3^2+p_3^6p_1^2p_2^2)
-1260\,(p_1^4p_2^4p_3^2+p_1^4p_3^4p_2^2
+p_2^4p_3^4p_1^2)\right)+\\
&&-33\sqrt{5}\Delta\left(3(p_1^4+p_2^4+p_3^4)-11(p_1^2p_2^2+p_1^2p_3^2
+p_2^2p_3^2)\right).
\end{eqnarray*}
where $\Delta=(p_1^2-p_2^2)(p_1^2-p_3^2)(p_2^2-p_3^2)$.
It turns out that Saito flat coordinates in this case are provided  by the following polynomials:
$$u_1=U_1,\quad u_2=U_2+c_1U_1^3,\quad u_3=U_3+c_2U_1^2U_2+c_3U_1^5,$$
where $c_1=-2$, $c_2=\f{33}{7}$ and $c_3=-\f{2091}{175}$.

The Euclidean contravariant metric  written in the Saito flat coordinates is given by
$$g=
\begin{pmatrix}
4u_1 & 12u_2 & 20u_3\\
 12u_2 &  -\f{3528}{187}u_3-168u_2u_1^2+\f{3528}{25}u_1^5 &  \f{748}{3}u_2^2u_1-\f{5236}{5}u_2u_1^4\\
 20u_3 & \f{748}{3}u_2^2u_1-\f{5236}{5}u_2u_1^4 & -\f{349690}{3087}u_2^3+\f{139876}{49}u_2^2u_1^3+\f{139876}{25}u_1^9
\end{pmatrix},
$$
and the metric $\eta={\rm Lie}_eg$ is given by
$$
\eta=\begin{pmatrix}
0 & 0 & 20\cr
0 & -\f{3528}{187} & 0\\
20 & 0 & 0
\end{pmatrix}.
$$
Solving the equations \eqref{Ffromg} one obtains the Frobenius potential for this case, namely
\beq\label{F-H3}
F= \f{1}{400}u_1u_3^2-\f{187}{70560}u_3u_2^2
+\f{34969}{1764000}u_2^2u_1^5-\f{34969}{4445280}u_2^3u_1^2.
\eeq

\subsection{Frobenius manifolds and space of orbits of Shephard groups}
A Shephard group is the symmetry group of a regular complex polytope \cite{Sh1}. The space of orbits of a Shephard group is endowed with the structure of a Frobenius manifold. The definition of this structure was given in \cite{Dad}. The crucial observation in this case is that, due to the results of \cite{OS}, the inverse of the Hessian of the basic invariant polynomial of lowest degree defines a flat (pseudo)-metric which depends linearly on the highest degree polynomial (in a coordinate system given by basic invariant polynomials). Using this fact, one can use the same arguments that work in the case of Coxeter groups.  In particular, in the case of Coxeter groups, the lowest degree basic invariant polynomial is the sum of the squares of the coordinates and its Hessian coincides with the standard Euclidean metric. In this way one recovers the standard construction presented in the examples above. We present as an example the case of $G_{26}$, which is a Shephard group.

\subsubsection{The case of $G_{26}$}
Starting from the invariant polynomial of lowest degree
$$u_1 = p_1^6-10p_1^3p_2^3-10p_1^3p_3^3+p_2^6-10p_2^3p_3^3+p_3^6,$$
one can define the covariant (pseudo)-metric given by the Hessian:
$$
\f{\d^2 u_1}{\d p^i\d p^j}=
\begin{pmatrix}
30p_1^4-60p_1p_2^3-60p_1p_3^3 & -90p_1^2p_2^2 & -90p_1^2p_3^2\\
 -90p_1^2p_2^2 & -60p_1^3p_2+30p_2^4-60p_2p_3^3 & -90p_2^2p_3^2\\
 -90p_1^2p_3^2 & -90p_2^2p_3^2 & -60p_1^3p^3-60p_2^3p_3+30p_3^4
\end{pmatrix}.
$$
The inverse of this metric  in the Saito coordinates $(u_1,u_2,u_3)$ with $u_2$ and $u_3$ defined as
\begin{eqnarray*}
u_2 &=& (p_1^3+p_2^3+p_3^3)(216p_1^3p_2^3p_3^3+(p_1^3+p_2^3+p_3^3)^3)-\f{1}{4}u_1^2,\\
u_3 &=&(p_1^3+p_2^3+p_3^3)^6-540p_1^3p_2^3p_3^3(p_1^3+p_2^3+p_3^3)^3-5832p_1^6p_2^6p_3^6-\f{3}{4}u_1u_2+\f{3}{16}u_1^3.
\end{eqnarray*}
 reads
$$
g=
\begin{pmatrix}
\f{6}{5}u_1 & \f{9}{5}u_2 & \f{12}{5}u_3 \\
\f{9}{5}u_2 & -\f{1}{160}u_3+\f{3}{1280}u_1^2 & -\f{9}{4}u_1u_2\\
\f{12}{5}u_3 & -\f{9}{4}u_1u_2 & \f{27}{40}u_1^3+\f{2592}{5}u_2^2
\end{pmatrix}.
$$
Notice that $g$ does indeed depend linearly on $u_3$, the highest degree invariant. Then defining $\eta:={\rm Lie}_e g$ (where $e$ in this case is the vector field $\frac{\d}{\d u_3}$), it is easy to check that the pencil
$$
g-\lambda\eta=
\begin{pmatrix}
\f{6}{5}u_1 & \f{9}{5}u_2 & \f{12}{5}u_3 \\
\f{9}{5}u_2 & -\f{1}{160}u_3+\f{3}{1280}u_1^2 & -\f{9}{4}u_1u_2\\
\f{12}{5}u_3 & -\f{9}{4}u_1u_2 & \f{27}{40}u_1^3+\f{2592}{5}u_2^2.
\end{pmatrix}
-\lambda\begin{pmatrix}
0 & 0 & \f{12}{5} \\
0 & -\f{1}{160} & 0\\
\f{12}{5} & 0 & 0
\end{pmatrix},
$$
is an exact flat pencil of metrics. From the formula \eqref{Ffromg} one easily obtains the Frobenius potential
$$F= \f{1}{96768}u_1^7+\f{5}{3456}u_2^2u_1^3+\f{1}{290304}(840u_2^3+2240u_3^2)u_1+\f{5}{432}u_3u_2^2.$$

\subsubsection{Shephard groups of rank $2$ and $3$}
Applying Dubrovin's procedure one can easily reconstruct the Frobenius potential for each Shephard group. Below we present the list of the Frobenius potential for {\em all} Shephard groups of rank $2$ and $3$, not just for the exceptional ones. 

\LTcapwidth=\textwidth
\arraycolsep=4.5pt
\setlength{\tabcolsep}{3.5pt}
{\footnotesize
\begin{longtable}{| c | c |}
\caption{Frobenius potential for all Shephard groups of rank $2$ and $3$.}\label{tab1}\\
\hline
\begin{tabular}{c} Type \end{tabular} & \begin{tabular}{c} Potential \end{tabular} \\
\hline\\[-10pt]
$G_{4}$ 
&
$F= \f{1}{24}u_1u_2^2-\f{i}{4608}\sqrt{3}u_1^4$
\smallskip\\
\hline\\[-10pt]
$G_{5}$ 
&
$F= -\f{1}{4}u_1^5+\f{5}{288}u_1u_2^2$
\smallskip\\
\hline\\[-10pt]
$G_{6}$ 
&
$F= \f{1}{96}u_2^2u_1-\f{1}{1935360}u_1^7$
\smallskip\\
\hline\\[-10pt]
$G_{8}$ 
&
$F= \f{7}{288}u_2^2u_1+\f{7}{1536}u_1^4$
\smallskip\\
\hline\\[-10pt]
$G_{9}$ 
&
$F=\f{7}{1152}u_1u_2^2+\f{1}{7680}u_1^7$
\smallskip\\
\hline\\[-10pt]
$G_{10}$ 
&
$F=\f{11}{1152}u_1u_2^2+\f{11}{34560}u_1^5$
\smallskip\\
\hline\\[-10pt]
$G_{14}$ 
&
$F=\f{5}{1152}u_1u_2^2+\f{45}{56}u_1^9$
\smallskip\\
\hline\\[-10pt]
$G_{16}$ 
&
$F= -\f{19}{2880000}\sqrt{5}u_1^4+\f{19}{1800}u_2^2u_1$
\smallskip\\
\hline\\[-10pt]
$G_{17}$ 
&
$F= \f{19}{7200}u_2^2u_1+\f{19}{6048000000}u_1^7$
\smallskip\\
\hline\\[-10pt]
$G_{18}$ 
&
$F= \f{29}{7200}u_1u_2^2+\f{29}{12}u_1^5$
\smallskip\\
\hline\\[-10pt]
$G_{20}$ 
&
$F= \f{11}{1800}u_2^2u_1+\f{11}{5184000}\sqrt{5}u_1^6$
\smallskip\\
\hline\\[-10pt]
$G_{21}$ 
&
$F= \f{11}{7200}u_1u_2^2+\f{1}{933120000}u_1^{11}$
\smallskip\\
\hline\\[-10pt]
$G_{23}$ 
&
$F= \f{1}{400}u_1u_3^2-\f{187}{70560}u_3u_2^2
+\f{34969}{1764000}u_2^2u_1^5-\f{34969}{4445280}u_2^3u_1^2$
\smallskip\\
\hline\\[-10pt]
$G_{25}$ 
&
$F= \f{5}{288}u_1u_3^2-\f{20}{3}u_3u_2^2+\f{5}{2}u_1^2u_2^2+\f{1}{3072}u_1^5$
\smallskip\\
\hline\\[-10pt]
$G_{26}$ 
&
$F=\f{1}{96768}u_1^7+\f{5}{3456}u_2^2u_1^3+\f{1}{290304}(840u_2^3+2240u_3^2)u_1+\f{5}{432}u_3u_2^2 $
\smallskip\\
\hline\\[-10pt]
$G(m,1,2)$
&
$F= \f{1}{24}u_1u_2^2+\f{1}{2880}u_1^5$
\smallskip\\
\hline\\[-10pt]
$G(m,1,3)$
&
$F= \f{1}{54}u_1u_3^2+\f{1}{72}u_2^2u_3
+\f{1}{2592}u_1u_2^2(u_1^2+3u_2)+\f{1}{816480}u_1^7$
\smallskip\\
\hline
\end{longtable}
}

Recall that any Shephard group $G$ admits a presentation of the form \cite{Coxeter2}
\begin{eqnarray*}
s_i^{p_i}&=&1\\
s_is_js_i\cdots &=&s_j s_is_j\cdots
\end{eqnarray*}
where $s_i$ are the generating (pseudo)-reflections and $p_i$ are integers $\ge 2$. The Coxeter group obtained replacing $p_i$ with $2$
 in the above presentation is usually called the Coxeter group associated with $G$. Due to the Hertling's theorem \cite{CH}, since the Frobenius manifold structure on the space of orbits of a Shephard group is polynomial, it must be isomorphic to the Frobenius manifold structure on the space of orbits of a Coxeter group. For each Shephard group  $G$, this Coxeter group is exactly the Coxeter group associated with $G$ \cite{Dad}.
 This means that the above Frobenius  potentials coincide up to rescaling of the variables with the potentials of the  the associated Coxeter group. For instance, the Frobenius potentials in the last two examples ($G_{(m,1,2)}$ and $G_{(m,1,3)}$) do not depend on $m$ and thus coincide with the Frobenius potentials associated with the Coxeter groups $B_2$ and $B_3$, respectively. 

\section{$\vee$-systems}
$\vee$-systems were introduced by A. Veselov in \cite{Ve} to construct new solutions of WDVV associativity equations, starting from a special set of covectors. 
The conditions defining a $\vee$-system are precisely the conditions that guarantee that a particular function constructed from a special set of covectors satisfies the WDVV equations. 

Now we recall the notion of $\vee$-systems (see \cite{Ve}). Let $V$ be a finite dimensional real vector space (the notion has been generalized also to complex vector spaces in \cite{FV}), and denote with $\mathcal{V}$ a finite set of non-collinear covectors $\alpha\in V^*$.  We assume that the the symmetric bilinear form defined by $g:=\sum_{\alpha\in  \mathcal{V}} \alpha\otimes \alpha$ is non-degenerate. We will denote by $\check{\alpha}$ the vector uniquely defined by: $(\check{\alpha},\cdot)=\alpha(\cdot)$ where $(\cdot,\cdot)$
 is the bilinear form defined by $g$.
\begin{definition}\label{veecondition}
We say that  $\mathcal{V}$  is a $\vee$-system if for each two-dimensional plane $
\Pi\subset V^*$ we have
\begin{equation}\label{veeequation}
\sum_{\beta\in \Pi\cap \mathcal{V}}\beta(\check{\alpha})\check{\beta}=\mu \check{\alpha},
\end{equation}
for each $\alpha \in \Pi\cap \mathcal{V}$ and for some $\mu$, which may depend on $\Pi$ and $\alpha$. 
\end{definition}
Let $\mathcal{H}$ be the family of hyperplanes associated with the covectors contained in $\mathcal{V}$. By definition, $H\in\mathcal{H}$ if and only if ${\rm ker}(\alpha)=H$ for one of the
 covectors in $\mathcal{V}$. We denote by $\alpha_H$ a covector associated to $H$ and by $\pi_H: V\rightarrow H^{\perp}$ is the linear map having as kernel $H$ and range $H^{\perp}$, the orthogonal complement of $H$ in $V$, where the orthogonality is defined via $g$. Due to the results
 of \cite{ALjmp,FV} the definition of $\vee$-system is equivalent to the requirement that the one parameter family of
 connections
\beq\nabla-\lambda\sum_{H\in \mathcal{H}}\frac{d\alpha_H}{\alpha_H}\otimes\pi_H,\eeq
(where $\nabla$ is the trivial affine structure defined by the coordinates$\{p^1,...,p^n\}$) is flat for any value of the parameter $\lambda$.

Given  a set of covectors defining a $\vee$-system the corresponding solution of WDVV associativity equations is given by the formula:
\beq\label{Fpotential}
F(p):=\frac{1}{2}\sum_{H\in \mathcal{H}}(\alpha_H(p))^2\log{\alpha_H(p)}.\eeq
One of the main example of $\vee$-systems is given by Coxeter systems \cite{Ve}. It was proved in \cite{Dad}
 that Veselov's solutions of WDVV equations constructed from a Coxeter group define the almost dual structure of the Frobenius manifolds defined on the space of orbits of these groups. Notice that the dual product has the form
\beq\label{veselovdual} *=\sum_{H\in \mathcal{H}}\frac{d\alpha_H}{\alpha_H}\otimes\pi_H.\eeq
Similarly, we will see in Section \ref{Section6} that for all exceptional well-generated complex reflection groups of rank $2$ and $3$ the dual connection (and the dual product) of a bi-flat $F$-manifold constructed on the relevant orbit spaces can be expressed using a Dunkl-Kohno-type connection. This connection is built using a generalization of the notion of $\vee$-system. 
\subsection{An example: $A_3$}
The $\vee$-system data that give rise to $A_3$ are the following covectors: 
\begin{eqnarray*}
&&\alpha_1 = [1,-1, 0],\qquad\alpha_2 = [1,0, -1],\qquad\alpha_3 = [0,1, -1]\\
&&\alpha_4 = [2,1, 1],\qquad\alpha_5 = [1,2, 1],\qquad\alpha_6 = [1,1, 2],\\
\end{eqnarray*}
and the associated projections:
\begin{eqnarray*}
&&\pi_1=
\begin{pmatrix}
1 &  -1 &  0\\
 -1 & 1 & 0\\ 
 0 & 0 & 0
\end{pmatrix},\qquad
\pi_2=
\begin{pmatrix}
1 & 0 & -1\\
 0 &  0 &  0\\ 
 -1 & 0 & 1
\end{pmatrix},\qquad
\pi_3 =
\begin{pmatrix}
0 & 0 & 0\\ 
0 & 1 &-1\\
0 &-1 & 1
\end{pmatrix},
\\
&&\pi_4=
\begin{pmatrix}
 2 & 1 & 1\\
 0 &  0&  0\\
 0 &  0 & 0
\end{pmatrix},\qquad
\pi_5 =
\begin{pmatrix}
0 & 0 & 0\\ 
1 & 2 &1\\
0 & 0 &0
\end{pmatrix},
\qquad 
\pi_6 =
\begin{pmatrix}
0 & 0 & 0\\
0 & 0 & 0\\ 
1 & 1 & 2
\end{pmatrix}.
\end{eqnarray*}

\section{Flat and bi-flat $F$-manifolds}
$F$-manifolds with compatible flat structure (for short {\it flat $F$-manifolds}) have been introduced by Manin as a generalization 
 of Frobenius manifolds. They are a particular instance of a more general class of manifolds, called $F$-manifolds, introduced 
 by Hertling and Manin in \cite{HM}. Let us recall their definition.
\begin{defi}
A flat $F$-manifold $(M, \circ, \nabla, e)$ is a manifold $M$ equipped with the following data: 
\begin{enumerate}
\item a commutative associate product $\circ : TM \times TM \rightarrow TM$ with flat unit $e$.
\item a flat torsionless affine connection $\nabla$ compatible with the product:
$$\nabla_k c^i_{jl}=\nabla_j c^i_{kl},$$
where $c^i_{jk}$ are the structure constants of $\circ$.
\end{enumerate}
\end{defi}
In {\it flat coordinates} for $\nabla$, as a consequence of the axioms defining a flat $F$-manifold, we have that the structure constants can be expressed via the second partial derivatives of a vector potential:
$$c^i_{jk}=\d_j\d_k A^i.$$
Furthermore, the vector potential $A^i$ satisfies the following equations, where one can choose as unit vector field $e=\d_1$:
\begin{eqnarray}
\label{oriented1}
\d_j\d_l A^i\d_k\d_mA^l&=&\d_k\d_l A^i\d_k\d_mA^l\\
\label{oriented2}
\d_1\d_i A^j&=&\delta_{i}^{j}.
\end{eqnarray}
They are called \emph{oriented associativity equations} \cite{LM}. 
Flat $F$-manifolds share several properties with Frobenius manifolds. Dubrovin's deformed connections, Dubrovin's duality
 and Dubrovin's principal hierarchy are well defined also for these manifolds (for the last point see \cite{LPR}). 
The missing data are the invariant metric $\eta$ end the Euler vector field $E$. A natural generalization of Frobenius 
  manifolds with Euler vector field can be obtained replacing the flat invariant metric $\eta$ with a flat connection $\nabla$ satisfying
 the usual conditions. 
\begin{defi} 
\label{defi:fmancc}
A  \emph{Frobenius manifold without metric} $(M,\circ,\nabla,e,E)$ is a manifold equipped with an associative commutative product $\circ$ on sections of its tangent bundle, two distinguished vector fields $e$ (unit) and $E$ (Euler vector field) and 
 a flat connection $\nabla$ satisfying the following requirements:
\begin{itemize}
\item the connection $\nabla$ is  compatible with the product:
\begin{equation*}
\nabla_k c^i_{jl}=\nabla_j c^i_{kl}.
\end{equation*}
\item $e$ is the unit of the product and it is flat: $\nabla e=0$.
\item Furthermore, the following conditions must hold: $$\nabla\nabla E=0,\quad [e,E]=e, \quad {\rm Lie}_Ec^i_{jk}=c^i_{jk}.$$	
\end{itemize}
\end{defi}
\subsection{Almost duality and bi-flat $F$-manifolds}
In the semisimple case, Dubrovin's almost dual structure can be extended to Frobenius manifolds without metric using the following data:
\begin{itemize}
\item a dual product $*$ defined in the usual way as $X*Y=E^{-1}\circ X\circ Y$ for any pair of vector fields $X$ and $Y$. 
\item a dual connection  $\nabla^{(2)}$ satisfying the following properties
\begin{enumerate}
\item $\nabla^{(2)} E=0$.
\item $\nabla^{(2)}$ is compatible with  $*$.
\item $\nabla^{(1)}$ and $\nabla^{(2)}$ are almost hydrodynamically equivalent \cite{AL2012}:
$$(d_{\nabla^{(1)}}-d_{\nabla^{(2)}})(X\,\circ)=0,$$
for every vector field $X$,  where $d_{\nabla}$ denotes the exterior covariant derivative associated to connection $\nabla.$
\end{enumerate}
\end{itemize}
\begin{rmk}
In the case of Frobenius manifolds the dual connection, in general, does coincide with 
 the Levi-Civita connection of the intersection form.  The difference between these two connections is proportional to the dual product.
\end{rmk}
\begin{theorem}
The dual connection defined above  is uniquely defined in terms of $\nabla^{(1)}$. Moreover 
 the flatness of the dual connection is equivalent to the linearity of the Euler vector field: $\nabla^{(1)}\nabla^{(1)} E=0$.
\end{theorem}
\emph{Proof}. From the above definition it follows that, in canonical coordinates for $\circ$ (the product compatible with $\nabla^{(1)}$), we have (see \cite{AL2012, AL-Bi-flat}):
\begin{eqnarray*}
&&c^i_{jk}=\delta^i_j\delta^i_k,\qquad c^{*i}_{jk}=\f{1}{u^i}\delta^i_j\delta^i_k,\\ 
&&e=\sum_k\d_k,\qquad E=\sum_k u^k\d_k,\\
&&\Gamma^{(1)i}_{ij}=\Gamma^{(2)i}_{ij}=\Gamma^i_{ij},\qquad i\neq j
\end{eqnarray*}
Moreover (see \cite{AL-Bi-flat}):
\begin{equation}\label{naturalc}
\begin{split}
\Gamma^{(1)i}_{jk}&:=0,\qquad\Gamma^{(2)i}_{jk}:=0,\qquad\forall\; i\ne j\ne k \ne i,\\
\Gamma^{(1)i}_{jj}&:=-\Gamma^{(1)i}_{ij},\qquad\Gamma^{(2)i}_{jj}:=-\f{u^i}{u^j}\Gamma^{(2)i}_{ij},\qquad i\ne j,\\
\Gamma^{(1)i}_{ii}&:=-\sum_{l\ne i}\Gamma^{(1)i}_{li},\qquad 
\Gamma^{(2)i}_{ii}:=-\sum_{l\ne i}\f{u^l}{u^i}\Gamma^{(2)i}_{li}-\f{1}{u^i}.
\end{split}
\end{equation} 
This proves the first part of the theorem. Let us prove the second part. The conditions expressing zero curvature are given by:
\begin{eqnarray}
\label{curv1}
&&\d_j\Gamma^i_{ik}+\Gamma^i_{ij}\Gamma^i_{ik}-\Gamma^i_{ik}\Gamma^k_{kj}
-\Gamma^i_{ij}\Gamma^j_{jk}=0, \quad \mbox{if $i\ne k\ne j\ne i$,}\\
\label{curv2}
&&e^l\d_l\Gamma^{i}_{ik}=0,\\
\label{curv3}
&&E^l\d_l\Gamma^{i}_{ik}=-\Gamma^{i}_{ik}.
\end{eqnarray}
The conditions \eqref{curv1} and \eqref{curv2} provide the flatness of the natural connection and thus are equivalent to oriented
 associativity equations \eqref{oriented1} and \eqref{oriented2}. The flatness of the dual connection follows from the condition \eqref{curv3}.

On the other hand, using the vanishing of the curvature for $\nabla^{(1)}$, after a straightforward computation, we obtain
\begin{eqnarray}\label{nablanabla}
(\nabla^{(1)}\nabla^{(1)} E)^i_{kj}=\d_k\d_j E^i+\Gamma^i_{jl}\d_k E^l+\Gamma^i_{km}\d_j E^m-\Gamma^m_{kj}\d_m E^i+E^l\d_l\Gamma^i_{kj}=0.
\end{eqnarray}
In canonical coordinates, the condition \eqref{nablanabla} reads
\begin{eqnarray*}
\delta^i_k\delta^i_j+\Gamma^{i}_{jk}  +E^l\d_l\Gamma^{i}_{kj}=0,
\end{eqnarray*}
and one can easily check that it is equivalent to the condition \eqref{curv3}.
\endproof
Bi-flat $F$-manifolds are Frobenius manifolds without metric endowed with the almost dual structure introduced above. They 
 have been introduced in \cite{AL-Bi-flat} motivated by the theory of integrable systems of hydrodynamic type. 
The definition is as follows:

\begin{defi}\label{multiflatdefi}
A \emph{bi-flat}  $F$-manifold $(M,\nabla^{(1)},\nabla^{(2)},\circ,*,e,E)$
 is a manifold $M$ equipped with a pair
 of flat connections $\nabla^{(1)}$ and $\nabla^{(2)}$, a pair of products $\circ$ and $*$ on sections of the tangent bundle $TM$ and  a pair of vector fields $e$ and  $E$ satisfying the following axioms:
\begin{itemize}
\item $E$ behaves like a Euler vector field: $[e,E]=e,\,{\rm Lie}_E c^i_{jk}=c^i_{jk}$, where $c^i_{jk}$ are the structure constants of $\circ$.
\item the product $\circ$ is commutative, associative and with unity $e$. Moreover $\nabla^{(1)} e=0$. 
\item The product $*$ is commutative, associative and with unity $E$. It is defined as: $X*Y=E^{-1}\circ X\circ Y,\,\forall X,Y$ vector fields on $M$. 
  Moreover $\nabla^{(2)} E=0$.
\item $\nabla^{(1)}$ is compatible with  $\circ$ and $\nabla^{(2)}$ is compatible with  $*$ in the following sense:
\begin{equation*}
\nabla^{(l)}_X \circ_{(l)}\left(Y,Z\right)=\nabla^{(l)}_Y \circ_{(l)}\left(X,Z\right),\,l=1,2,
\end{equation*}
for all $X, Y, Z$ are vector fields on $M$, where $\circ_{(1)}=\circ$ and $\circ_{(2)}=\star.$
\item $\nabla^{(1)}$ and $\nabla^{(2)}$ are almost hydrodynamically equivalent.
\end{itemize}
\end{defi}
The first connection $\nabla^{(1)}$ is often called \emph{the natural connection}. Frobenius manifolds are bi-flat $F$-manifolds equipped with an invariant metric $\eta$.


\begin{remark}
The manifold $M$ in the above definition might be  a real or complex manifold. In
the latter case $TM$ is intended as the holomorphic tangent bundle and
all the geometric data are supposed to be holomorphic. In the present  paper we will deal with 
 complex manifolds.
\end{remark}

\section{Bi-flat $F$-structure on the space of orbits of complex reflection groups}\label{Section5}
In this Section we present a procedure that allows us to reconstruct (in principle) the bi-flat $F$-manifold structure and the vector potential starting from a well-generated complex reflection group (recall that any finite real reflection group is automatically well-generated).  Let us consider the pair $(V, G)$, where $V$ is a complex vector space of complex dimension $n$ and $G$ is a well-generated complex reflection group. We identify $V$ with $\mathbb{C}^n$ with standard coordinates $\{p^1, \dots, p^n\}$. We also consider the coordinates given by a set of basic invariant polynomials $\{u^1, \dots, u^n\}$ in the ring of invariant polynomials $\mathbb{C}[V]^G$, which can be viewed as coordinates on the space of orbits $\mathbb{C}^n/G$.  Since the algorithm is based on the interplay between the standard coordinates $\{p^1 \dots, p^n\}$ on $\mathbb{C}^n$ and the coordinates $\{u^1, \dots, u^n\}$, we denote with a superscript tilde the components written in the coordinates $\{u^1, \dots, u^n\}$.

Although $\{p^1, \dots, p^n\}$ and $\{u^1, \dots, u^n\}$ are coordinates on different spaces, since the map $\phi: \mathbb{C}^n \rightarrow \mathbb{C}^n/G$ is a diffeomorphism on any fundamental domain of the action of $G$ on $\mathbb{C}^n$ and since we are working locally, we will treat $\{p^1, \dots, p^n\}$ and $\{u^1, \dots, u^n\}$ just as different coordinates systems. Notice also that the choice of basic invariant polynomials  $\{u^1, \dots, u^n\}$ is not unique in general. 

Our construction relies on the hypotheses that the dual connection $\nabla^{(2)}$ and the dual product $*$ are related  by the formula 
\beq\label{main-hp}
\nabla^{(2)}=\nabla^{(0)}+\lambda *, 
\eeq
where $\nabla^{(0)}$ is the standard affine flat connection on $\mathbb{C}^n$  and $\lambda$ is a suitable constant. Taking into account the condition $\nabla^{(2)}E=0$ and
 the form of the Euler vector field $E$, it is immediate to prove that $\lambda=-1$ (see remark \ref{normalization} below).

The hypotheses above were chosen because this is exactly what happens in many examples of $\vee$-systems related to Coxeter groups, which constitute our inspiring model. The fact that with these assumptions one indeed can equip the orbit space with a bi-flat $F$-manifold structure is then verified a posteriori. 

In this Section we explain how to define the dual connection and the dual product under the assumption \eqref{main-hp}. We assume that the flat coordinates for the natural connection we are looking for are basic invariants for the complex reflection groups and we call them \emph{generalized Saito flat coordinates}.  Combining this assumption with the assumption \eqref{main-hp} we obtain an explicit formula for the Christoffel symbols of the dual connection and an explicit expression for the structure constants of the dual product, in terms of flat basic invariants. In practice, to select the generalized Saito flat coordinates among basic invariants, we have to impose that the dual connection defined by this formula is flat and that the natural connection associated to generalized Saito flat coordinates is compatible with the natural product. We then show that, in all the examples, these two conditions fix uniquely the natural connection. We will see later in Section \ref{Section7} that  dropping  the hypothesis \eqref{main-hp} the natural connection and the associated generalized Saito flat coordinates in general are no longer uniquely determined.



Before explaining the procedure we present the following fundamental Lemma which provides an important consequence  of the fact that $\nabla^{(1)}$ and $\nabla^{(2)}$ are forced to be almost hydrodynamically equivalent:
\begin{lemma} Let $\nabla^{(1)}$ and $\nabla^{(2)}$ be two almost hydrodynamically equivalent flat connections associated to the products $\circ$ and $*$ respectively. Then in flat coordinates for $\nabla^{(1)}$, we have:
\beq\label{eqalhydeq}
 \tilde{\Gamma}^{(2)k}_{il} \tilde{c}^{*l}_{mj}=\tilde\Gamma^{(2)k}_{jl} \tilde{c}^{*l}_{mi}.
\eeq
\end{lemma}
\proof
By definition of almost hydrodynamical equivalence, we have
$(d_{\nabla^{(1)}}-d_{\nabla^{(2)}})(X*)=0$ for every vector fields $X$ (and equivalently with $X\circ$). 
By straightforward computation we get
$$\Gamma^{(1)k}_{lj}(X*)^l_i-\Gamma^{(1)k}_{li}(X*)^l_j=\Gamma^{(2)k}_{lj}(X*)^l_i-\Gamma^{(2)k}_{li}(X*)^l_j.$$
In flat coordinates for $\nabla^{(1)}$ the above condition reduces to \eqref{eqalhydeq}.

\endproof

We are now able to find a general formula for the dual product and the dual connection under the assumption \eqref{main-hp}.

\begin{theorem} In the generalized Saito flat coordinates $\{u^1, 
\dots, u^n\}$, the Christoffel symbols of the dual connection and the Christoffel symbols of the trivial connection on $\mathbb{C}^n$ are related by the formula
\beq\label{fundamental1}
\tilde \Gamma^{(2)i}_{jk}=-\frac{d_i}{d_i-1}\f{\d^2 u^i}{\d p^m\d p^n}(J^{-1})^m_j(J^{-1})^n_k,
\eeq
where $J^i_j=\f{\d u^i}{\d p^j}$ and $d_i={\rm degree}(u^i),$ the degree of the invariant polynomial $u^i$. 
\end{theorem}
\proof
Since $\nabla^{(1)}$ and $\nabla^{(2)}$ are almost hydrodynamically equivalent, relation \eqref{eqalhydeq} holds. 
 Moreover, it is immediate to prove that the Euler vector field 
$E=\sum_k p^k \frac{\d}{\d p^k}$ expressed in the Saito flat coordinates 
 is given by $\tilde{E}=\sum_k d_k u^k \frac{\d}{\d u^k}$, where $d_k$ are the degrees of the invariant polynomials.
 
 \noindent From the condition $\nabla^{(2)}E=0$, we obtain:
\beq\label{nablaE}
\tilde \Gamma^{(2)k}_{jl}\tilde E^l=-d_k\delta^k_j.
\eeq
If we multiply equation \eqref{eqalhydeq} by $\tilde E^i$ on both sides we get
$$\tilde \Gamma^{(2)k}_{il}\tilde E^i \tilde c^{*l}_{mj}=\tilde \Gamma^{(2)k}_{jl}\tilde c^{*l}_{mi}\tilde E^i.$$
Taking into account that $E$ is the unity of the dual product, the right hand side reduces to $\tilde \Gamma^{(2)k}_{jm},$ 
while, due to \eqref{nablaE}, the left hand side reduces to $-d_k \tilde c^{*k}_{mj}$.

\noindent In this way we get: 
\beq\label{intermiediateeq1}\tilde \Gamma^{(2)k}_{jm}=-d_k\tilde c^{*k}_{mj}.\eeq
The Christoffel symbols of the dual connection are also given by the formula
$$\tilde \Gamma^{(2)i}_{jk}=-\f{\d^2 u^i}{\d p^m\d p^n}(J^{-1})^m_j(J^{-1})^n_k-\tilde c^{*i}_{jk},$$
which is simply obtained by \eqref{main-hp} applying the transformation formula for Christoffel symbols passing from the coordinates $\{p^1, \dots, p^n\}$ to the coordinates $\{u^1, \dots, u^n\}$.
Combining these two relations we get 
$$\tilde c^{*i}_{jk}=-\frac{1}{1-d_i}\f{\d^2 u^i}{\d p^m\d p^n}(J^{-1})^m_j(J^{-1})^n_k,$$
and from this using \eqref{intermiediateeq1} and the standing hypothesis \eqref{main-hp}, we find the fundamental formula \eqref{fundamental1}.

\endproof

It is important to observe that the fundamental relation \eqref{fundamental1} holds  only in the generalized Saito flat coordinates: indeed in deriving it we used relation \eqref{eqalhydeq}, which is true only in a coordinate system given by a distinguished set of basic of invariant polynomials in which the Christoffel symbols of $\nabla^{(1)}$ vanish identically, i.e. the generalized Saito flat coordinates. This means that \eqref{fundamental1} can be used to define the dual connection only in this special set of coordinates. On the other hand, we can use this fact to select the  generalized Saito coordinates among other coordinate systems. Indeed, among the basic invariant polynomials we choose those for which the formula  \eqref{fundamental1} \emph{defines a flat} connection and for which the structure constants  $\tilde c^i_{jk}$ of $\circ$, defined via 
$$\tilde c^i_{jk}=\tilde c^{*i}_{jl}\tilde c^{*l}_{km}(e^{-1})^m,$$
are compatible with the natural connection $\nabla^{(1)}$. In this way, in the Saito flat coordinates, we have that the Christoffel symbols of the natural connection, the Christoffel symbols of the dual connection and the structure constants of the dual product can all be reconstructed directly and unambiguously from the Christoffel symbols of the trivial connection expressed in such coordinates:
$$\tilde\Gamma^{(1)k}_{ij}=0,\quad\tilde\Gamma^{(2)k}_{ij}=\frac{d_i}{1-d_i}\f{\d^2 u^i}{\d p^m\d p^n}(J^{-1})^m_j(J^{-1})^n_k,\quad\tilde c^{*i}_{jk}=\frac{1}{d_i-1}\f{\d^2 u^i}{\d p^m\d p^n}(J^{-1})^m_j(J^{-1})^n_k.$$
Coming back to the original coordinates on $\mathbb{C}^n$ given by $\{p^1, \dots, p^n\}$ we get the following theorem, whose proof is a straightforward application of tensorial transformations. 
\begin{theorem} In the coordinates $\{p^1,...,p^n\}$ the structure constants of the dual product and the Christoffel symbols of the dual connection are given by by the formulas
\beq\label{fundamental2}
c^{*i}_{jk}=-\Gamma^{(2)i}_{jk}=\frac{1}{d_l-1}\f{\d^2 u^l}{\d p^j\d p^k}(J^{-1})^i_l,
\eeq
where $J^i_j=\f{\d u^i}{\d p^j}$.
\end{theorem}
In practice, to select the Saito flat coordinates it is convenient to impose the flatness of the dual connection and the compatibility between $\nabla^{(1)}$ and $\circ$ in the coordinates $\{p^1,...,p^n\}$, where the Christoffel symbols of $\nabla^{(2)}$ are given by the simple formula \eqref{fundamental2}. 
Once the generalized Saito coordinates have been selected, one can easily reconstruct the full bi-flat $F$-manifold structure. It turns out that in all the cases we have investigated the unit of the product $\circ$ is the vector field $e=\f{\d}{\d u^n}$ where $u^n$ is the highest degree basic invariant polynomial.

\begin{rmk}\label{normalization}
With this choice, one has  $ [e,E]=d_n e$. In order to obtain $ [e,E]=e$ one should normalize the Euler vector field and the dual product as $ E\to\f{E}{d_n}$ and $c^{*i}_{jk}\to d_n c^{*i}_{jk}$.  Using this normalization the r.h.s of \eqref{main-hp} must be replaced by $-\frac{1}{d_n}X*$.
\end{rmk}

Below we present the results of the procedure just described for {\em all} well-generated exceptional complex reflection groups of rank $2$ and $3$ and for the families $G(m,1,2)$ and $G(m,1,3)$. In particular, for each case we provide the Saito flat coordinates (the distinguished basis of the invariant polynomials) and the vector potential for the product $\circ$ in the Saito flat coordinates. The other data for the bi-flat $F$-structure can be easily reconstructed from these data using the formulas above.

\subsection{The case of $G_{4}$}
In this case the ring of invariants is generated by (see \cite{LT}):
$$U_1=p_1^4+2i\sqrt{3}p_1^2p_2^2+p_2^4,\quad U_2=p_1^5p_2-p_1p_2^5.$$
Here $U_1$ and $U_2$ coincide with the generalized Saito flat coordinates.

Up to inessential linear terms the vector potential for $\tilde c^i_{jk}$  is given by 
$$A^1=u_1u_2, \quad A^2=\frac{1}{2}u_2^2-\frac{1}{96}i\sqrt{3}u_1^3.$$
Lowering the indices with  the metric
$$\eta=
\begin{pmatrix}
0 & \f{1}{12} \\
\f{1}{12} & 0
\end{pmatrix}
$$
we get an exact 1-form whose potential coincides with the Frobenius potential for $G_4$.

\subsection{The case of $G_{5}$}
In this case the ring of invariants is generated by (see \cite{LT}):
$$U_1= p_1^5p_2-p_1p_2^5 ,\qquad U_2=(p_1^4+2i\sqrt{3}p_1^2p_2^2+p_2^4)^3.$$
We look for generalized Saito flat coordinates of the form $u_1=U_1,\, u_2=U_2+c U_1^2$. 
Under the assumption of the general algorithm we presented, we have that $c$ is fixed by the compatibility of the product $\circ$  with the natural connection $\nabla^{(1)}$ (in the coordinates $p$), and it is given by $c_1=-6i\sqrt{3}.$ Up to inessential linear terms the vector potential for $\circ$ in the generalized Saito flat coordinates  is given by $$A^1=u_1^2, \quad A^2=\frac{1}{2}u_2^2-\frac{141}{4}u_1^4.$$
It is easy to check that the above vector potential does not come from a Frobenius potential.

\subsection{The case of $G_{6}$}
$G_ 6$ is a well-generated complex reflection group of rank two. Its ring of invariants is generated by (see \cite{LT}):
$$U_1=p_1^4+2i\sqrt{3}p_1^2p_2^2+p_2^4,\qquad U_2=p_1^{10}p_2^2-2p_1^6p_2^6+p_1^2p_2^{10}.$$
We look for generalized Saito flat coordinates of the form $u_1=U_1,\, u_2=U_2+c U_1^3$.  
 The compatibility between the natural connection $\nabla^{(1)}$ and $\circ$ fixes a unique value of $c$, namely $c=\frac{5}{288}i\sqrt{3}$ and consequently a distinguished set of basic invariants. From these we get immediately the vector potential $A^k$, $k=1,2$ for $\circ$ in the flat coordinates is given by (up to inessential linear terms):
$$A^1=u_1u_2-\frac{1}{576} i \sqrt{3} u_1^4,\quad A^2=\frac{1}{2}u_2^2-\frac{1}{6144}u_1^6.$$
It is easy to check that the above vector potential does not come from a Frobenius potential.

\subsection{The case of $G_{8}$}
$G_8$ is a well-generated complex reflection group of rank two. Basic invariants are given by (see \cite{LT}):
$$u_1=p_1^8+14p_1^4p_2^4+p_2^8,\qquad u_2=p_1^{12}-33p_1^8p_2^4-33p_1^4p_2^8+p_2^{12}.$$
These are automatically generalized Saito coordinates and there are no arbitrary constants to be fixed. 
The compatibility between the natural connection $\nabla^{(1)}$ and $\circ$ is automatically satisfied, as well all the other properties characterizing a bi-flat $F$ structure. 
The vector potential $A^k$, $k=1,2$ for $\circ$ in the flat coordinates is given by (up to inessential linear terms)
$$A^1=u_1u_2,\qquad A^2=\frac{1}{2}u_2^2+\frac{3}{8}u_1^3.$$
Lowering the indices with  the metric:
$$\eta=
\begin{pmatrix}
0 & \f{1}{12} \\
\f{1}{12} & 0
\end{pmatrix}
$$
we get an exact 1-form, whose potential coincides with the Frobenius potential for $G_8$.

\subsection{The case of $G_{9}$}
$G_9$ is a well-generated complex reflection group of rank two. Basic polynomial invariants are (see \cite{LT}):
$$U_1=p_1^8+14p_1^4p_2^4+p_2^8,\qquad U_2=(p_1^{12}-33p_1^8p_2^4-33p_1^4p_2^8+p_2^{12})^2.$$
We look for Saito flat coordinates of the form $u_1=U_1,\, u_2=U_2+c U_1^3$.  
 Imposing the compatibility of the natural connection $\nabla^{(1)}$ with $\circ$ we obtain a unique value of $c$, namely $c=-11/16$ and consequently a distinguished basis of invariant polynomials (generalized Saito coordinates). In this way we obtain 
the vector potential for $\circ$ (up to inessential linear terms) given by:
$$A^1=u_1u_2+\frac{3}{32}u_1^4,\qquad A^2=\frac{1}{2}u_2^2+\frac{33}{512}u_1^6.$$
It is easy to check that the above vector potential does not come from a Frobenius potential.

\subsection{The case of $G_{10}$}
$G_{10}$ is a well-generated complex reflection group of rank two. Basic polynomial invariants are given by (see \cite{LT}):
$$U_1=p_1^{12}-33p_1^8p_2^4-33p_1^4p_2^8+p_2^{12},\qquad U_2=(p_1^8+14p_1^4p_2^4+p_2^8)^3.$$
We look for generalized Saito flat coordinates of the form $u_1=U_1,\, u_2=U_2+c U_1^2$.  
 The compatibility between the natural connection $\nabla^{(1)}$ and $\circ$ determines a unique value of $c$, namely $c=-\frac{7}{12}$ and consequently a distinguished basis of invariant polynomials (Saito coordinates). In this way we obtain the vector potential (up to inessential linear terms)
$$A^1=u_1u_2+\frac{1}{18}u_1^3,\qquad A^2=\frac{1}{2}u_2^2+\frac{35}{432}u_1^4.$$
It is easy to check that the above vector potential does not come from a Frobenius potential.

\subsection{The case of $G_{14}$}
$G_{14}$ is a well-generated complex reflection group of rank two. Basic polynomial invariants are provided by (see \cite{LT}):
$$U_1=p_1^5p_2-p_1p_2^5,\qquad U_2=(p_1^{12}-33p_1^8p_2^4-33p_1^4p_2^8+p_2^{12})^2.$$
 We look for generalized Saito flat coordinates of the form $u_1=U_1,\, u_2=U_2+c U_1^4$.   
The compatibility between  the natural connection $\nabla^{(1)}$ and $\circ$ determines a unique value of $c$, namely $c=66$, and consequently a distinguished set of basic invariant polynomials. In this way we obtain 
  the vector potential (up to inessential linear terms) 
$$A^1=-\frac{24}{5}u_1^5+u_1u_2,\qquad A^2=\frac{1}{2}u_2^2+792u_1^8.$$
It is easy to check that the above vector potential does not come from a Frobenius potential.

\subsection{The case of $G_{16}$}
$G_{16}$ is a well-generated complex reflection group of rank two. Basic polynomial invariants are given by (see \cite{LT}):
\begin{eqnarray*}
u_1&=&p_1^{20}-\frac{38}{3}\sqrt{5}p_1^{18}p_2^2-19p_1^{16}p_2^4-152\sqrt{5}p_1^{14}p_2^6-494p_1^{12}p_2^8+\frac{988}{3}\sqrt{5}p_1^{10}p_2^{10}+
+p_2^{20}+\\
&&-\frac{38}{3}\sqrt{5}p_2^{18}p_1^2
-19p_2^{16}p_1^4-152\sqrt{5}p_2^{14}p_1^6-494p_2^{12}p_1^8,\\
u_2&=&p_1^{29}p_2-\frac{116}{45}\sqrt{5}p_1^{27}p_2^3+\frac{1769}{25}p_1^{25}p_2^5+\frac{464}{5}\sqrt{5}p_1^{23}p_2^7+
+\frac{2001}{5}p_1^{21}p_2^9-\frac{2668}{15}\sqrt{5}p_1^{19}p_2^{11}\\
&&+\frac{12673}{5}p_1^{17}p_2^{13}-p_2^{29}p_1
+\frac{116}{45}\sqrt{5}p_2^{27}p_1^3-\frac{1769}{25}p_2^{25}p_1^5-\frac{464}{5}\sqrt{5}p_2^{23}p_1^7-\frac{2001}{5}p_2^{21}p_1^9\\
&&+\frac{2668}{15}\sqrt{5}p_2^{19}p_1^{11}-\frac{12673}{5}p_2^{17}p_1^{13}.
\end{eqnarray*}
There are no arbitrary constants in this case, and these invariant polynomials are automatically generalized Saito coordinates.  All the conditions defining a bi-flat $F$ structure are automatically satisfied. 
 For the vector potential of $\circ$ we obtain (up to inessential linear terms)
$$A^1=u_1u_2,\qquad A^2=\frac{1}{2}u_2^2-\frac{1}{800}\sqrt{5}u_1^3.$$
Lowering the indices with  the metric:
$$\eta=
\begin{pmatrix}
0 & \f{19}{900} \\
\f{19}{900} & 0
\end{pmatrix}
$$
we get an exact 1-form whose potential coincides with the Frobenius potential for $G_{16}$.

\subsection{The case of $G_{17}$}
$G_{17}$ is a well-generated complex reflection group of rank two. Basic polynomial invariants are given by (see \cite{LT}):
\begin{eqnarray*}
U_1&=&p_1^{20}-\frac{38}{3}\sqrt{5}p_1^{18}p_2^2-19p_1^{16}p_2^4-152\sqrt{5}p_1^{14}p_2^6-494p_1^{12}p_2^8+\frac{988}{3}\sqrt{5}p_1^{10}p_2^{10}+p_2^{20}\\
&&-\frac{38}{3}\sqrt{5}p_2^{18}p_1^2
-19p_2^{16}p_1^4-152\sqrt{5}p_2^{14}p_1^6-494p_2^{12}p_1^8,\\
U_2&=&(p_1^{29}p_2-\frac{116}{45}\sqrt{5}p_1^{27}p_2^3+\frac{1769}{25}p_1^{25}p_2^5+\frac{464}{5}\sqrt{5}p_1^{23}p_2^7+\frac{2001}{5}p_1^{21}p_2^9-\frac{2668}{15}\sqrt{5}p_1^{19}p_2^{11}\\
&&+\frac{12673}{5}p_1^{17}p_2^{13}-p_2^{29}p_1+\frac{116}{45}\sqrt{5}p_2^{27}p_1^3-\frac{1769}{25}p_2^{25}p_1^5-\frac{464}{5}\sqrt{5}p_2^{23}p_1^7-\frac{2001}{5}p_2^{21}p_1^9\\
&&+\frac{2668}{15}\sqrt{5}p_2^{19}p_1^{11}-\frac{12673}{5}p_2^{17}p_1^{13})^2.
\end{eqnarray*}
We look for generalized Saito flat coordinates of the form $u_1=U_1,\, u_2=U_2+c U_1^3$. 
 Imposing the compatibility between the natural connection $\nabla^{(1)}$ and $\circ$ we obtain a unique value of $c$, namely $c=\frac{29}{12000}\sqrt{5}$ and consequently a distinguished basis of invariant polynomials (Saito coordinates).
All the other conditions defining a bi-flat $F$ structure are satisfied.  
 In this way we obtain the vector potential for $\circ$ (up to inessential linear terms)
$$A^1=u_1u_2-\frac{3}{8000}\sqrt{5}u_1^4,\qquad A^2=\frac{1}{2}u_2^2+\frac{319}{96000000}u_1^6.$$
It is easy to check that the above vector potential does not come from a Frobenius potential.

\subsection{The case of $G_{18}$}
$G_{18}$ is a well-generated complex reflection group of rank two. Basic polynomial invariants are given by (see \cite{LT}):
\begin{eqnarray*}
U_1&=&p_1^{29}p_2-\frac{116}{45}\sqrt{5}p_1^{27}p_2^3+\frac{1769}{25}p_1^{25}p_2^5+\frac{464}{5}\sqrt{5}p_1^{23}p_2^7+
+\frac{2001}{5}p_1^{21}p_2^9-\frac{2668}{15}\sqrt{5}p_1^{19}p_2^{11}\\
&&+\frac{12673}{5}p_1^{17}p_2^{13}-p_2^{29}p_1
+\frac{116}{45}\sqrt{5}p_2^{27}p_1^3-\frac{1769}{25}p_2^{25}p_1^5-\frac{464}{5}\sqrt{5}p_2^{23}p_1^7-\frac{2001}{5}p_2^{21}p_1^9\\
&&+\frac{2668}{15}\sqrt{5}p_2^{19}p_1^{11}-\frac{12673}{5}p_2^{17}p_1^{13},\\
U_2&=&(p_1^{20}-\frac{38}{3}\sqrt{5}p_1^{18}p_2^2-19p_1^{16}p_2^4-152\sqrt{5}p_1^{14}p_2^6-494p_1^{12}p_2^8+\frac{988}{3}\sqrt{5}p_1^{10}p_2^{10}+
p_2^{20}+\\
&&-\frac{38}{3}\sqrt{5}p_2^{18}p_1^2
-19p_2^{16}p_1^4-152\sqrt{5}p_2^{14}p_1^6-494p_2^{12}p_1^8)^3.
\end{eqnarray*}
We look for generalized Saito flat coordinates of the form $u_1=U_1,\, u_2=U_2+c U_1^2$. 
Imposing the compatibility of the natural connection $\nabla^{(1)}$ with $\circ$, we obtain a unique value of $c$, namely $c=38\sqrt{5}$ and consequently a distinguished basis of invariant polynomials (Saito coordinates).
In this way we obtain the vector potential for $\circ$ (up to inessential linear terms)
$$A^1=-\frac{16}{3}\sqrt{5}u_1^3+u_1u_2,\qquad A^2=\frac{1}{2}u_2^2+\frac{4180}{3}u_1^4.$$
It is easy to check that the above vector potential does not come from a Frobenius potential.

\subsection{The case of $G_{20}$}
$G_{20}$ is a well-generated complex reflection group of rank two. Basic polynomial invariants are given by (see \cite{LT}):
\begin{eqnarray*}
u_1&=&p_1^{12}+\frac{22}{5}\sqrt{5}p_1^{10}p_2^2-33p_1^8p_2^4-\frac{44}{5}\sqrt{5}p_1^6p_2^6-33p_1^4p_2^8+\frac{22}{5}\sqrt{5}p_1^2p_2^{10}+p_2^{12},\\
u_2&=&p_1^{29}p_2-\frac{116}{45}\sqrt{5}p_1^{27}p_2^3+\frac{1769}{25}p_1^{25}p_2^5+\frac{464}{5}\sqrt{5}p_1^{23}p_2^7+\frac{2001}{5}p_1^{21}p_2^9-\frac{2668}{15}\sqrt{5}p_1^{19}p_2^{11}\\
&&+\frac{12673}{5}p_1^{17}p_2^{13}-p_2^{29}p_1+\frac{116}{45}\sqrt{5}p_2^{27}p_1^3-\frac{1769}{25}p_2^{25}p_1^5-\frac{464}{5}\sqrt{5}p_2^{23}p_1^7-\frac{2001}{5}p_2^{21}p_1^9\\
&&+\frac{2668}{15}\sqrt{5}p_2^{19}p_1^{11}-\frac{12673}{5}p_2^{17}p_1^{13}.
\end{eqnarray*}
In this case, the general base does not depend on arbitrary constants, and all the conditions defining a bi-flat $F$-structure are automatically satisfied. 
The computation of the vector potential for $\circ$ gives (up to inessential linear terms):
$$A^1=u_1u_2,\qquad A^2=\frac{1}{2}u_2^2+\frac{1}{960}\sqrt{5}u_1^5.$$
Lowering the indices with  the metric:
$$\eta=
\begin{pmatrix}
0 & \f{11}{900} \\
\f{11}{900} & 0
\end{pmatrix}
$$
we get an exact 1-form whose potential coincides with the Frobenius potential for $G_{20}$.

\subsection{The case of $G_{21}$}
$G_{21}$ is a well-generated complex reflection group of rank two. Basic polynomial invariants are given by (see \cite{LT}):
\begin{eqnarray*}
U_1&=&p_1^{12}+\frac{22}{5}\sqrt{5}p_1^{10}p_2^2-33p_1^8p_2^4-\frac{44}{5}\sqrt{5}p_1^6p_2^6-33p_1^4p_2^8+\frac{22}{5}\sqrt{5}p_1^2p_2^{10}+p_2^{12},\\
U_2&=&(p_1^{29}p_2-\frac{116}{45}\sqrt{5}p_1^{27}p_2^3+\frac{1769}{25}p_1^{25}p_2^5+\frac{464}{5}\sqrt{5}p_1^{23}p_2^7+\frac{2001}{5}p_1^{21}p_2^9-\frac{2668}{15}\sqrt{5}p_1^{19}p_2^{11}\\
&&+\frac{12673}{5}p_1^{17}p_2^{13}-p_2^{29}p_1+\frac{116}{45}\sqrt{5}p_2^{27}p_1^3-\frac{1769}{25}p_2^{25}p_1^5-\frac{464}{5}\sqrt{5}p_2^{23}p_1^7-\frac{2001}{5}p_2^{21}p_1^9\\
&&+\frac{2668}{15}\sqrt{5}p_2^{19}p_1^{11}-\frac{12673}{5}p_2^{17}p_1^{13})^2.
\end{eqnarray*}
We look for generalized Saito flat coordinates of the form $u_1=U_1,\, u_2=U_2+c U_1^5$.  
The compatibility of the natural connection $\nabla^{(1)}$ with the product $\circ$ determines a unique value of $c$, namely $c=-\frac{29}{14400}\sqrt{5}$ and consequently a distinguished basis of invariant polynomials (Saito coordinates). All the other conditions defining a bi-flat $F$ structure are automatically satisfied. 

Computing the vector potential for $\circ$, we obtain (up to inessential linear terms)
$$A^1=u_1u_2+\frac{1}{8640}\sqrt{5}u_1^6,\qquad A^2=\frac{1}{2}u_2^2+\frac{551}{149299200}u_1^{10}.$$
It is easy to check that the above vector potential does not come from a Frobenius potential.

\subsection{The case of $G_{24}$}
The complex reflection group $G_{24}$  is a rank three well-generated complex reflection group. Basic invariants are given by the following homogeneous polynomials of degree $4,6,14$ (see for instance Section 6 of \cite{Klein} or \cite{E}):
\begin{eqnarray*}
U_1&=&p_1^3p_2+p_1p_3^3+p_2^3p_3,\\
U_2&=&p_1^5p_3+p_1p_2^5+p_2p_3^5-5p_1^2p_2^2p_3^2,\\
U_3&=&p_1^{14}+p_2^{14}+p_3^{14}+18p_1^7p_2^7+18p_1^7p_3^7+18p_2^7p_3^7\\
&&-126p_1^6p_2^3p_3^5-126p_1^5p_2^6p_3^3-126p_1^3p_2^5p_3^6+375p_1^8p_2^4p_3^2
+375p_1^4p_2^2p_3^8+375p_1^2p_2^8p_3^4\\
&&-250p_1^9p_2p_3^4-250p_1^4p_2^9p_3-250p_1p_2^4p_3^9-34p_1^{11}p_2^2p_3-34p_1^2p_2p_3^{11}-34p_1p_2^{11}p_3^2.
\end{eqnarray*}
Notice that $U_2$ is equal (up to a constant factor) to the determinant of the Hessian of $U_1$, while $U_3$ is constructed using a differential determinant involving $U_1$ and $U_2$ (see formula (1.14) in \cite{E}). 
We look for generalized Saito flat coordinates of the form 
$$u_1=U_1,\quad u_2=U_2,\quad u_3=U_3+c U_1^2U_2.$$
Imposing the flatness of the dual connection (in the original coordinates of $\mathbb{C}^n$) 
$$\Gamma^{(2)i}_{jk}=\frac{1}{1-d_l}\f{\d^2 u^l}{\d p^j\d p^k}(J^{-1})^i_l,$$
we get $c=-34$. In the flat coordinates we obtain that the vector potential for $\circ$  is given by (up to inessential linear terms)
\begin{eqnarray*}
A^1&=&-\frac{14}{3}u_1^3u_2-u_1u_2-\frac{14}{3}u_2^3,\\
A^2&=&-\frac{14}{5}u_1^5+14u_1^2u_2^2-u_2u_3,\\
A^3&=&-\frac{1}{2}u_3^2-56u_1^7-294u_1^4u_2^2+196u_1u_2^4.\\
\end{eqnarray*}
Rescaling the variables $v_1=-2^{-\frac{1}{3}}u_1, $ $v_2=u_2$, $v_3=-2^{\frac{1}{3}}\frac{u_3}{56}$,  we obtain
the vector potential found in \cite{KMS2} (Section 7.2). It is easy to check that the above vector potential does not come from a Frobenius potential.

\subsection{The case of $G_{25}$}
Basic invariants are given by the following homogeneous polynomials of degree $6,9,12$ (see for instance \cite{Ma}):
\begin{eqnarray*}
U_1&=&p_1^6+p_2^6+p_3^6-10p_1^3p_2^3-10p_1^3p_3^3-10p_2^3p_3^3\\
U_2&=&(p_1^3-p_2^3)(p_1^3-p_3^3)(p_2^3-p_3^3)\\
U_3&=&(p_1^3+p_2^3+p_3^3)((p_1^3+p_2^3+p_3^3)^3+216p_1^3p_2^3p_3^3).
\end{eqnarray*}
We look for generalized Saito flat coordinates of the form:
$$u_1=U_1,\quad u_2=U_2,\quad u_3=U_3+c_1 U_1^2.$$
Imposing the flatness of the dual connection in the original coordinates of $\mathbb{C}^n$ 
$$\Gamma^{(2)i}_{jk}=\frac{1}{1-d_l}\f{\d^2 u^l}{\d p^j\d p^k}(J^{-1})^i_l,$$
we obtain  $c=-\f{5}{8}$. In the Saito flat coordinates, the vector potential for $\circ$ is given by (up to inessential linear terms) 
$$A^1=\frac{1}{8}(3u_1^2-8u_3)u_2,\,A^2=-\frac{1}{2}u_3^2-144u_1u_2^2-\frac{3}{64}u_1^4,\,A^3=-u_1u_3+192u_2^2.$$
Lowering the indices with  the metric:
$$\eta=
\begin{pmatrix}
0 & 0 &-\f{5}{144} \\
0 & \f{40}{3} & 0\\
-\f{5}{144} & 0 & 0
\end{pmatrix}
$$
we get an exact 1-form whose potential coincides with the Frobenius potential for $G_{20}$.

\subsection{The case of $G_{26}$} 
The basic invariants are:
\begin{eqnarray*}
U_1 &=& p_1^6-10p_1^3p_2^3-10p_1^3p_3^3+p_2^6-10p_2^3p_3^3+p_3^6\\
U_2 &=& (p_1^3+p_2^3+p_3^3)(216p_1^3p_2^3p_3^3+(p_1^3+p_2^3+p_3^3)^3)\\
U_3 &=& (p_1^3+p_2^3+p_3^3)^6-540p_1^3p_2^3p_3^3(p_1^3+p_2^3+p_3^3)^3-5832p_1^6p_2^6p_3^6
\end{eqnarray*}
We look for generalized Saito flat coordinates of the form:
$$u_1=U_1,\quad u_2=U_2+c_1U_1^2,\quad u_3=U_3+c_2U_1U_2+c_3U_1^3.$$
Imposing the flatness of the dual connection in the original coordinates of $\mathbb{C}^n$
$$\Gamma^{(2)i}_{jk}=\frac{1}{1-d_l}\f{\d^2 u^l}{\d p^j\d p^k}(J^{-1})^i_l,$$
we obtain
$$c_1=-\f{1}{6},\qquad c_2=-\f{1}{2},\qquad c_3=\f{1}{9}.$$
In the Saito flat coordinates the vector potential for $\circ$ is given by
\begin{eqnarray*}
A^1&=& u_1u_3+\f{3}{4}u_2^2-\f{1}{4}u_1^2u_2+\f{1}{48}u_1^4,\\
A^2&=& \f{1}{4}u_1u_2^2+\f{1}{12}u_1^3u_2+u_2u_3-\f{1}{144}u_1^5,\\ 
A^3&=& \f{1}{2}u_3^2+\f{1}{4}u_2^3+\f{1}{4}u_2^2u_1^2-\f{1}{48}u_1^4u_2+\f{1}{216}u_1^6
\end{eqnarray*}
It is easy to check that this vector potential does not come from a Frobenius potential. In other words we do not recover the Frobenius manifold associated with the Shephard group $G_{26}$. We will see later how to obtain this structure 
 modifying slightly the above procedure. 

\subsection{The case of $G_{27}$}
$G_{27}$ is a well-generated complex reflection group. As a basis for its invariants, we use the invariants of Wiman \cite{Wiman}.
\begin{eqnarray*}
U_1&=& 10p_1^3p_2^3+9p_3(p_1^5+p_2^5)-45p_1^2p_2^2p_3^2-135p_1p_2p_3^4+27p_3^6\\
U_2&=&\f{1}{20250}{\rm det}(\rm{hessian}(U_1))\\
U_3&=&\f{1}{24300}{\rm Bord}(U_1,U_2),
\end{eqnarray*}
where $\rm{Bord}$ is a differential determinantal expression involving $U_1$ and $U_2$. 
Using the same procedure adopted in the previous cases involving rank three well-generatred complex reflection groups, we obtain generalized Saito flat coordinates 
$$u^1=U^1,\qquad u^2=U_2+c_1u_1^2,\qquad u_3=U_3+c_2U_2^2u_1+c_3U_2u_1^3+c_4u_1^5,$$
with
$$c_1 = -\f{1}{9},\qquad c_2=-\f{17}{18},\qquad c_3=\f{29}{27},\qquad c_4 = -\f{43}{486}.$$
The vector potential for $\circ$ is provided by:
\begin{eqnarray*}
A^1&=& u_1u_3-\f{5}{6}u_2^3-\f{5}{18}u_2^2u_1^2+\f{5}{81}u_2u_1^4+\f{2}{2187}u_1^6,\\
A^2&=&u_2u_3+\f{25}{54}u_2^3 u_1-\f{25}{81} u_2^2 u_1^3-\f{10}{729} u_2u_1^5-\f{100}{19683} u_1^7,\\
A^3&=& \f{1}{2}u_3^2-\f{35}{72}u_2^5+\f{875}{648}u_1^2u_2^4+\f{875}{6561}u_2^2u_1^6+\f{24000}{1417176}u_1^8 u_2-\f{350}{177147}u_1^{10}.
\end{eqnarray*}
Rescaling the variables  we obtain the vector potential found in \cite{KMS2} (Section 7.3). It is easy to check that the above vector potential does not come from a Frobenius potential.

\subsection{The case of $G(m,1,2)$}
We can choose as basic invariants
$$U_1=p_1^m+p_2^m,\qquad U_2=p_1^{2m}+p_2^{2m}.$$
Generalized Saito flat coordinates are $u_1=U_1$ and $u_2=U_2+cU_1^2$ with $c=-\f{1}{2}\f{2m-1}{m}$ and the vector potential is
\begin{eqnarray*}
A^1&=&u_1u_2+\f{1}{6}\f{m-2}{m}u_1^3,\\
A^2&=&\f{1}{2}u_2^2+\f{1}{12}\f{m-1}{m^2}u_1^4.
\end{eqnarray*}
It is easy to check that the above vector potential does not come from a Frobenius potential.

\subsection{The case of $G(m,1,3)$}

We can choose as basic invariants
$$U_1=p_1^m+p_2^m+p_3^m,\qquad U_2=p_1^{2m}+p_2^{2m}+p_3^{2m},\qquad U_3=p_1^{3m}+p_2^{3m}+p_3^{3m}.$$
Generalized Saito flat coordinates are 
$$u_1=U_1,\quad u_2=U_2+c_1U_1^2\quad u_3=U_3+c_2U_1U_2+c_3U_1^3$$ 
with 
$$c_1=-\f{1}{3}\f{2m-1}{m},\qquad c_2=-\f{1}{2}\f{3m-1}{m},\qquad c_3=\f{2}{9}c_2^2$$ 
and the vector potential is
\begin{eqnarray*}
A^1&=& u_1u_3+\left(\f{1}{2}\f{3m-1}{m}-\f{5}{4}\right)u_1^2u_2+\left(\f{1}{12}\f{(3m-1)^2}{m^2}-\f{31}{72}\f{3m-1}{m}+\f{5}{9}\right)u_1^4+\f{3}{8}u_2^2,\\
A^2&=&u_2u_3+\f{1}{180}\f{m^3-2m^2-m+2}{m^3}u_1^5+\f{1}{180}\f{10m^3-10m^2}{m^3}u_1^3u_2+\\
&&+\f{1}{180}\f{45m^3-45m^2}{m^3}u_1u_2^2\\
A^3&=& \f{1}{2160}\f{4m^3-12m^2+16m-8}{m^4}u_1^6+\f{1}{2160}\f{45m^3-135m^2
+90m}{m^4}u_1^4u_2+\\
&&+\f{1}{2160}\f{270m^3-270m^2}{m^4}u_1^2u_2^2+\f{1}{16}\f{u_2^3}{m}+\f{1}{2}u_3^2.
\end{eqnarray*}

\section{A universal formula for the dual product}\label{Section6}

It is well-know that in the case of the Frobenius structure associated to any Coxeter group, the dual product is immediately recovered via Veselov's $\vee$-system structure associated with the roots of the Coxeter group itself. In this Section we show that, similarly to the Coxeter case, in the case of irreducible well-generated complex reflection groups of rank $2$ and $3$ the dual product and the dual connection are given by a universal formula of Dunkl-Kohno type. This is a generalization of the $\vee$-system construction, since the formula \eqref{Dunkl-Kohno-eq1} below is obtained considering  unitary projections onto the unitary complement of the hyperplanes defined by the covectors $\alpha_s$,  computed with respect to a suitable Hermitian metric (in all cases analyzed it is standard Hermitian metric, except for the case of $G_{27}$). 

In order to obtain this formula, it is necessary to identify the reflecting hyperplanes in all cases under consideration. This is achieved using the fact that
Shephard and Todd proved that the determinant of the Jacobian matrix ${\rm det}\f{\d u^i}{\d p^j}$ factorizes into the product of linear forms defining the reflecting
 hyperplanes. The form corresponding to a (pseudo)-reflection  of order $p$ appears exactly with multiplicity $p-1$ in ${\rm det}\f{\d u^i}{\d p^j}$. Remarkably, for all rank 2 and rank 3 well-generated complex reflection groups of the Shephard Todd list  the "weights" $\kappa_s$ assigned at each hyperplane coincide with the order of the (pseudo)-reflection having as mirror exactly that hyperplane.   
\begin{theorem}\label{Dunkl-Kohno-th1}
In the cases $G_{4}$ $G_{5}$, $G_{6}$, $G_{8}$, $G_{9}$, $G_{10}$, $G_{14}$, $G_{16}$, $G_{17}$, $G_{18}$, $G_{20}$, $G_{21}$, $G_{23}$, $G_{24}$, $G_{25}$, $G_{26}$, $G_{27}$   the dual product and the dual connection at a point $p$ are expressed  by the following formula (in the coordinates $\{p^1,..,p^n\}$):
\begin{equation}
\label{Dunkl-Kohno-eq1} c^{*i}_{jk}(p)=-\Gamma^{*i}_{jk}(p)=h^{im}\left\{\f{1}{N}\sum_{s=1}^{M}\frac{\kappa_s}{||\alpha_s||^2}\frac{(\alpha_s)_j \,(\alpha_k)_p \, (\bar\alpha_s)_m}{\alpha_s(p)}\right\},\end{equation}
where 
\begin{itemize}
\item $\alpha_s$ are constant covectors in $\mathbb{C}^n$;
\item $h^{im}$ are the components of the inverse of a suitable Hermitian metric. In all the cases a part from one ($G_{27}$) $h$  is the standard Hermitian metric;
\item $N$ is a normalizing factor chosen in such a way that $c^{*i}_{jk}E^k=\delta^i_j$;
\item $M$ is the number of distinct factors of  ${\rm det}\f{\d u^i}{\d p^j}$ (the number of reflecting hyperplanes); 
\item $\kappa_s$ is the order of the (pseudo)-reflection defined by the hyperplane $\ker(\alpha_s)$.
\end{itemize}
 The list of values of $M,N,\kappa_s$ and the choice of the Hermitian metric are given below. The complete list of 1-forms $\alpha$ obtained factorizing ${\rm det}\f{\d u^i}{\d p^j}$  is given in the Appendix 1.

\LTcapwidth=\textwidth
\arraycolsep=4.5pt
\setlength{\tabcolsep}{3.5pt}
\begin{longtable}{| c | c|  c|  c| c|}
\caption{Rank two complex reflection groups.}\label{tab2}\\
\hline
\begin{tabular}{c} Type \end{tabular} & \begin{tabular}{c} Metric \end{tabular} &  \begin{tabular}{c} M \end{tabular} & \begin{tabular}{c} N
\end{tabular}  &  \begin{tabular}{c} $\kappa_s$ \end{tabular}\\
\hline\\[-10pt]
$G_{4}$ 
&
$h=
\begin{pmatrix}
1&0\\
0&1
\end{pmatrix}
$
&
4
&
4
&
2  (s=1..4)
\smallskip\\
\hline\\[-10pt]
$G_{5}$ 
&
$h=
\begin{pmatrix}
1&0\\ 
0&1
\end{pmatrix}
$
&
8
&
8
&
2  (s=1..8),
\smallskip\\
\hline\\[-10pt]
$G_{6}$ 
&
$h=
\begin{pmatrix}
1&0\\ 
0&1
\end{pmatrix}
$
&
10
&
12
&
2 (s=1..6), 3 (s=7..10)
\smallskip\\
\hline\\[-10pt]
$G_{8}$ 
&
$h=
\begin{pmatrix}
1&0\\ 
0&1
\end{pmatrix}
$
&
6
&
6
&
2 (s=1..6).
\smallskip\\
\hline\\[-10pt]
$G_{9}$ 
&
$h=
\begin{pmatrix}
1&0\\ 
0&1
\end{pmatrix}
$
&
18
&
24
&
2 (s=1..12), 4 (s=13..18)
\smallskip\\
\hline\\[-10pt]
$G_{10}$ 
&
$h=
\begin{pmatrix}
1&0\\ 
0&1
\end{pmatrix}
$
&
14
&
24
&
3 (s=1..8), 4 (s=9..14)
\smallskip\\
\hline\\[-10pt]
$G_{14}$ 
&
$h=
\begin{pmatrix}
1&0\\ 
0&1
\end{pmatrix}
$
&
20
&
24
&
2 (s=1..12), 3 (s=13..20)
\smallskip\\
\hline\\[-10pt]
$G_{16}$ 
&
$h=
\begin{pmatrix}
1&0\\ 
0&1
\end{pmatrix}
$
&
12
&
12
&
2 (s=1..12)
\smallskip\\
\hline\\[-10pt]
$G_{17}$ 
&
$h=
\begin{pmatrix}
1&0\\ 
0&1
\end{pmatrix}
$
&
42
&
60
&
2 (s=1..30), 5 (s=31..42)
\smallskip\\
\hline\\[-10pt]
$G_{18}$ 
&
$h=
\begin{pmatrix}
1&0\\ 
0&1
\end{pmatrix}
$
&
32
&
60
&
3 (s=1..20), 5 (s=21..32)
\smallskip\\
\hline\\[-10pt]
$G_{20}$ 
&
$h=
\begin{pmatrix}
1&0\\ 
0&1
\end{pmatrix}
$
&
20
&
20
&
2 (s=1..20)
\smallskip\\
\hline\\[-10pt]
$G_{21}$ 
&
$h=
\begin{pmatrix}
1&0\\ 
0&1
\end{pmatrix}
$
&
50
&
60
&
2 (s=1..30) 3 (s=31..50)
\smallskip\\
\hline
\end{longtable}


\LTcapwidth=\textwidth
\arraycolsep=4.5pt
\setlength{\tabcolsep}{3.5pt}
{\footnotesize
\begin{longtable}{| c|  c|  c|  c| c|}
\caption{Rank three complex reflection groups.}\label{tab3}\\
\hline
\begin{tabular}{c} Type \end{tabular} & \begin{tabular}{c} Metric \end{tabular} &  \begin{tabular}{c} n \end{tabular} & \begin{tabular}{c} N
\end{tabular}  &  \begin{tabular}{c} $\kappa_s$ \end{tabular}\\
\hline\\[-10pt]
$G_{24}$ 
&
$h=
\begin{pmatrix}
1&0&0\\ 
0&1&0\\
0&0&1
\end{pmatrix}
$
&
21
&
42
&
2  (s=1..21)
\smallskip\\
\hline\\[-10pt]
$G_{25}$ 
&
$h=
\begin{pmatrix}
1&0&0\\ 
0&1&0\\
0&0&1
\end{pmatrix}
$
&
12
&
24
&
2  (s=1..12),
\smallskip\\
\hline\\[-10pt]
$G_{26}$ 
&
$h=
\begin{pmatrix}
1&0&0\\ 
0&1&0\\
0&0&1
\end{pmatrix}
$
&
21
&
18
&
2 (s=1..9), 3 (s=10..21)
\smallskip\\
\hline\\[-10pt]
$G_{27}$ 
&
$h=
\begin{pmatrix}
1&0&0\\ 
0&1&0\\
0&0&3
\end{pmatrix}
$
&
45
&
15
&
2 (s=1..45)
\smallskip\\
\hline
\end{longtable}
}

\end{theorem} 

\proof
The proof boils down to check with a direct computation that the following identity holds true in all the cases examined 
\begin{equation}
\frac{1}{d_l-1}\f{\d^2 u^l}{\d p^j\d p^k}(J^{-1})^i_l=\f{1}{N}\sum_{H\in \mathcal{H}}\frac{d\alpha_H}{\alpha_H}\otimes\kappa_H\pi_H.
\end{equation}
where $\mathcal{H}$ is the collection of reflecting hyperplanes, 
the weights $\kappa_H$ coincide with the order of the corresponding (pesudo)-reflection fixing hyperplane $H$ and $\pi_{H}$ is a unitary projection onto a suitable unitary complement of $H$ in $\mathbb{C}^n$. The long symbolic computations have been performed using the symbolic capabilities of the program Maple. 

\endproof
\subsection{Remarks}
\begin{enumerate}
\item In all the cases examined except for $G_{27}$ the Hermitian metric is the standard one. 
\item The proof is based on long symbolic computations. We believe that it is a general fact that the dual product and the dual connection associated to well-generated complex groups can be expressed in this way, a fact not limited to the exceptional cases of rank $2$ and $3$. 
\end{enumerate}

\subsection{An analogue of Veselov's potential}
In the case of $\vee$-systems, there is a natural potential $F$ that allows  one to recover the dual product (see formula \eqref{Fpotential}). In complete analogy, in our case, we have a similar formula for a {\em vector} potential $F^i$. 

\begin{proposition}
The  structure constants of the dual product $c^{*i}_{jk}$ in the cases appearing in Theorem \ref{Dunkl-Kohno-th1}, can be expressed as 
$$c^{*i}_{jk}(p)=\frac{\d F^i(p)}{\d p^j \d p^k},$$
where the dual potential is given by the formula
\beq\label{Veselovvectorpotential}
F^i(p)=\f{1}{N}\sum_{H\in \mathcal{H}}\frac{\kappa_H}{||\alpha_H||^2}\,\alpha_H(p)\ln(\alpha_H(p))\check{\bar{ \alpha}}^i_H.
\eeq
The weights $\kappa_H$ coincide with the order of the (pesudo)-reflection fixing hyperplane $H$  and $\check{\bar{ \alpha}}^i_H$ is the $i$-component of the vector obtained from $\alpha_H$ via the Hermitian metric.
\end{proposition} 
\proof 
This is a straightforward computation.
\endproof

\subsection{Generalized Saito flat coordinates: the standard case}
We have seen that, in all the examples, the dual product is given by the formula \eqref{Dunkl-Kohno-eq1}. Given the dual product, the fundamental formula \eqref{fundamental2}  allows one to uniquely reconstruct the flat basic invariants. Indeed, the freedom in the choice of these invariants disappears if one requires that
\beq\label{fundamental3}
\f{\d^2 u^i}{\d p^j\d p^k}=(d_i-1)c^{*s}_{jk}\f{\d u^i}{\d p^s}.
\eeq
This means that $u^i$ must be a flat coordinate of the flat connection $\nabla^{(0)}+(d_i-1) *$. We will call basic invariants satisfying condition \eqref{fundamental3} the \emph{standard generalized Saito flat coordinates}. Non standard bi-flat structures  and related flat coordinates will be constructed in the next Section. 

\begin{rmk}
It was observed in \cite{FS} that the condition
 \eqref{fundamental3} is satisfied by standard Saito polynomials.
\end{rmk}

\section{Bi-flat $F$-manifold structure on the space of orbits of  groups revisited}\label{Section7}

If we look back at the results of Section \ref{Section5}, we observe that in the case of Shephard groups,   the standard bi-flat $F$-structure does not coincide, in general, with the Dubrovin's structure. In this Section we show that the bi-flat structure associated with a complex reflection group in general is not unique. In particular, 
we prove that in all the examples of Shephard groups where the standard structure does coincide with the Frobenius manifold structure, there is an underlying one-parameter family including both of them for special values of the parameter. As a consequence, we have a {\em one parameter family of generalized Saito flat coordinates}. 

\subsection{The case of $G_{26}$} 
 We treat in details the case of $G_{26}$ and we provide the main steps for the remaining groups.  All the results of this part have been obtained with the help of the symbolic package Maple.

In this case our starting point is the dual  product obtained in the Section \ref{Section6}
$$c^{*i}_{lp}(u)=\f{2}{18}\sum_{s=1}^{9}\frac{1}{||\alpha_s||^2}\frac{(\alpha_s)_l \,(\alpha_s)_p \, (\check{\bar\alpha}_s)^i}{\alpha_s(u)}+\f{3}{18}\sum_{s=10}^{21}\frac{1}{||\alpha_s||^2}\frac{(\alpha_s)_l \,(\alpha_s)_p \, (\check{\bar\alpha}_s)^i}{\alpha_s(u)}.$$
The Christoffel symbols of the natural connection are given by: 
$$\tilde \Gamma^{(1)i}_{jk}=\f{\d^2 u^m}{\d p^j\d p^k}(J^{-1})^i_m,$$
where $(u_1,u_2,u_3)$ are easily obtained from the basic invariants $(U_1,U_2,U_3)$ of Section \ref{Section5}:
\beq\label{SaitoFlatparameter} u_1=U_1,\quad u_2=U_2+c_1U_1^2,\quad u_3=U_3+c_2U_1U_2+c_3U_1^3.\eeq
By construction $(u_1,u_2,u_3)$ are flat coordinates of the flat connection  $\nabla^{(1)}$.

The product $\circ$  can be also be defined from the dual product $\star$, specifically its structure constants are given by:
$$\tilde c^i_{jk}=\tilde c^{*i}_{jl}\tilde c^{*l}_{km}(e^{-1})^m,$$
where $e^{-1}$ is the inverse of the unit of $\circ$, where the inverse is computed with respect to $\star$.

The compatibility of $\circ$ with the natural connection $\nabla^{(1)}$ imposes the following constraints on the constants $c_i$ appearing in the expressions of $(u_1, u_2, u_3)$ as a function of the basic invariants $(U_1, U_2, U_3)$:
$$c_1=c_1,\qquad c_2=3 c_1,\qquad c_3=\f{1}{2}c_1 (2 c_1-1).$$
Once these constraints are imposed to \eqref{SaitoFlatparameter}, we effectively obtain a one-parameter family (depending on $c_1$) of Saito flat coordinates. 
In particular,  it is straightforward to check that the product $\circ$ in the generalized Saito flat coordinates is given by $\tilde{c}^i_{jk}=\f{\d^2 A^i}{\d u^j\d u^k}$
 where
\begin{eqnarray*}
A^1 &=& \f{1}{8}(12c_1^2+7c_1+1)u_1^4+\f{1}{8}(-24c_1-6)u_2u_1^2+u_3u_1+\f{3}{4}u_2^2,\\
A^2 &=& \f{1}{20}(12c_1^3+15c_1^2+3c_1)u_1^5-\f{1}{2}u_2u_1^3c_1-\f{3}{2}c_1u_2^2u_1+u_3u_2,\\
A^3 &=&\f{1}{2}u_3^2
+\f{1}{80}(-96c_1^4-96c_1^3-36c_1^2-6c_1)u_1^6+\f{9}{2}(c_1+\f{1}{2})c_1(c_1+\f{1}{4})u_2u_1^4+\\
&&-\f{9}{2}(c_1+\f{1}{2})c_1u_2^2u_1^2+\f{3}{4}u_2^3c_1+\f{3}{8}u_2^3.
\end{eqnarray*}
In other words, we have equipped the orbit space of $G_{26}$ with a one parameter family of bi-flat $F$-structures. Among them there is only one that admits an invariant metric given by (up to multiplicative constant):
$$
\eta=
\begin{pmatrix}
0 &  0 & \f{5}{18}\\
0 & \f{5}{12} & 0\\
\f{5}{18} & 0 & 0
\end{pmatrix}.
$$
This corresponds to special choice $c_1=-\f{1}{4}$. In this case we have a Frobenius potential
$$F=\f{5}{36}u_1u_3^2+\f{5}{24}u_3u_2^2+\f{5}{96}u_1u_2^3+\f{5}{192}u_2^2u_1^3
+\f{1}{5376}u_1^7,$$
that coincides with the Frobenius potential obtained in Section 2.

In the coordinates $\{p^1,...,p^n\}$, for generic values of $c_1$, the Christoffel symbols of the dual connection can be obtained imposing the almost-hydrodynamical equivalence between $\nabla^{(1)}$ and $\nabla^{(2)}$. This gives rise to the following equation  
$$(\Gamma^{(2)k}_{il}-\Gamma^{(1)k}_{il})c_{mj}^l-(\Gamma^{(2)k}_{jl}-\Gamma^{(1)k}_{jl})c_{mi}^l,$$
which we can solve for $\Gamma^{(2)i}_{jk}$. 
In this case, the Christoffel symbols of $\nabla^{(2)}$ are not related anymore to the structure constants of the dual product $\star$ by the simple formula
$$\Gamma^{(2)i}_{jk}=-c^{*i}_{jk}.$$
However, the Christoffel symbols of $\nabla^{(2)}$ can still be expressed via a Dunkl-Kohno-type connection associated with a different choice of the weights. We have
 $$\Gamma^{(2)i}_{jk}=-c^{*i}_{jk}+\left(\f{2}{3}c_1+\f{1}{9}\right)C^{i}_{jk},$$
where
$$C^i_{lp}(u)=4\sum_{s=1}^{9}\frac{1}{||\alpha_s||^2}\frac{(\alpha_s)_l \,(\alpha_s)_p \, (\check{\bar\alpha}_s)^i}{\alpha_s(u)}-3\sum_{s=10}^{21}\frac{1}{||\alpha_s||^2}\frac{(\alpha_s)_l \,(\alpha_s)_p \, (\check{\bar\alpha}_s)^i}{\alpha_s(u)}.$$
 
Let us briefly summarize the remaining cases.

\subsection{The case of $G_{5}$}

The Christoffel symbols of the dual connection are
$$\Gamma^{(2)i}_{jk}=-c^{*i}_{jk}+\left(-\f{1}{4}+\f{i}{72}\sqrt{3}c\right)C^{i}_{jk},$$
where
$$C^i_{lp}(u)=3\sum_{s=1}^{4}\frac{1}{||\alpha_s||^2}\frac{(\alpha_s)_l \,(\alpha_s)_p \, (\check{\bar\alpha}_s)^i}{\alpha_s(u)}-3\sum_{s=5}^{8}\frac{1}{||\alpha_s||^2}\frac{(\alpha_s)_l \,(\alpha_s)_p \, (\check{\bar\alpha}_s)^i}{\alpha_s(u)}.$$
Up to inessential linear terms the vector potential is given by:
\begin{eqnarray*}
A^1&=& -(4i)\sqrt{3}u_1^3-\f{2}{3}u_1^3 c+u_1u_2,\\ 
A^2&=&-(4i)\sqrt{3}u_1^4 c-\f{1}{3}u_1^4 c^2+\f{1}{2}u_2^2.
\end{eqnarray*}
Finally, for $c=-6i\sqrt{3}$ one recovers the Frobenius structure listed in Table 1.
\subsection{The case of $G_{6}$}
The Christoffel symbols of the dual connection are
$$\Gamma^{(2)i}_{jk}=-c^{*i}_{jk}+\left(-8ic\sqrt{3}-\frac{5}{12}\right)C^{i}_{jk},$$
where
$$C^i_{lp}(u)=2\sum_{s=1}^{6}\frac{1}{||\alpha_s||^2}\frac{(\alpha_s)_l \,(\alpha_s)_p \, (\check{\bar\alpha}_s)^i}{\alpha_s(u)}-3\sum_{s=7}^{10}\frac{1}{||\alpha_s||^2}\frac{(\alpha_s)_l \,(\alpha_s)_p \, (\check{\bar\alpha}_s)^i}{\alpha_s(u)}.$$
Up to inessential linear terms the vector potential is given by:
\begin{eqnarray*}
A^1&=& u_2u_1+\frac{1}{144}u_1^4(i\sqrt{3}-72c),\\ 
A^2&=&\frac{1}{2}u_2^2+\frac{1}{120}u_1^6c(i\sqrt{3}-36c).
\end{eqnarray*}
Finally, for $c=\frac{1}{72}i\sqrt{3}$ one recovers the Frobenius structure listed in Table 1.

\subsection{The case of $G_{9}$}
The Christoffel symbols of the dual connection are
$$\Gamma^{(2)i}_{jk}=-c^{*i}_{jk}+\left(-\frac{4}{3}c-\frac{11}{12}\right)C^{i}_{jk},$$
where
$$C^i_{lp}(u)=\sum_{s=1}^{12}\frac{1}{||\alpha_s||^2}\frac{(\alpha_s)_l \,(\alpha_s)_p \, (\check{\bar\alpha}_s)^i}{\alpha_s(u)}-2\sum_{s=13}^{18}\frac{1}{||\alpha_s||^2}\frac{(\alpha_s)_l \,(\alpha_s)_p \, (\check{\bar\alpha}_s)^i}{\alpha_s(u)}.$$
Up to inessential linear terms the vector potential is given by:
\begin{eqnarray*}
A^1&=& \frac{1}{4}(-2c-1)u_1^4+u_2u_1,\\ 
A^2&=& \frac{1}{10}(-3c-3)u_1^6+\frac{1}{2}u_2^2.
\end{eqnarray*}
Finally, for $c=-\frac{1}{2}$ one recovers the Frobenius structure listed in Table 1.

\subsection{The case of $G_{10}$}
The Christoffel symbols of the dual connection are
$$\Gamma^{(2)i}_{jk}=-c^{*i}_{jk}+\left(-\frac{1}{2}c-\frac{7}{24}\right)C^{i}_{jk},$$
where
$$C^i_{lp}(u)=3\sum_{s=1}^{8}\frac{1}{||\alpha_s||^2}\frac{(\alpha_s)_l \,(\alpha_s)_p \, (\check{\bar\alpha}_s)^i}{\alpha_s(u)}-4\sum_{s=9}^{14}\frac{1}{||\alpha_s||^2}\frac{(\alpha_s)_l \,(\alpha_s)_p \, (\check{\bar\alpha}_s)^i}{\alpha_s(u)}.$$
Up to inessential linear terms the vector potential is given by:
\begin{eqnarray*}
A^1&=& \frac{1}{3}(-2c-1)u_1^3+u_2u_1,\\ 
A^2&=& \frac{1}{6}(-2c^2-2c)u_1^4+\frac{1}{2}u_2^2.
\end{eqnarray*}
Finally, for $c=-\frac{1}{2}$ one recovers the Frobenius structure listed in Table 1.

\subsection{The case of $G_{14}$}
The Christoffel symbols of the dual connection are
$$\Gamma^{(2)i}_{jk}=-c^{*i}_{jk}+\left(\frac{1}{144}c-\frac{11}{24}\right)C^{i}_{jk},$$
where
$$C^i_{lp}(u)=2\sum_{s=1}^{12}\frac{1}{||\alpha_s||^2}\frac{(\alpha_s)_l \,(\alpha_s)_p \, (\check{\bar\alpha}_s)^i}{\alpha_s(u)}-3\sum_{s=13}^{20}\frac{1}{||\alpha_s||^2}\frac{(\alpha_s)_l \,(\alpha_s)_p \, (\check{\bar\alpha}_s)^i}{\alpha_s(u)}.$$
Up to inessential linear terms the vector potential is given by:
\begin{eqnarray*}
A^1&=& \frac{1}{5}(-2c+108)u_1^5+u_2u_1,\\ 
A^2&=& \frac{1}{14}(-4c^2+432c)u_1^8+\frac{1}{2}u_2^2.
\end{eqnarray*}
Finally, for $c=54$ one recovers the Frobenius structure listed in Table 1.

\subsection{The case of $G_{17}$}
The Christoffel symbols of the dual connection are
$$\Gamma^{(2)i}_{jk}=-c^{*i}_{jk}+\left(\f{29}{60}-40\sqrt{5}c\right)C^{i}_{jk},$$
where
$$C^i_{lp}(u)=2\sum_{s=1}^{30}\frac{1}{||\alpha_s||^2}\frac{(\alpha_s)_l \,(\alpha_s)_p \, (\check{\bar\alpha}_s)^i}{\alpha_s(u)}-5\sum_{s=31}^{42}\frac{1}{||\alpha_s||^2}\frac{(\alpha_s)_l \,(\alpha_s)_p \, (\check{\bar\alpha}_s)^i}{\alpha_s(u)}.$$
Up to inessential linear terms the vector potential is given by:
\begin{eqnarray*}
A^1&=& u_1u_2+\f{1}{1200}(\sqrt{5}-600 c)u_1^4,\\ 
A^2&=& \f{1}{2}u_2^2+\f{1}{1000}c u_1^6(\sqrt{5}-300 c).
\end{eqnarray*}
Finally, for $c=\f{1}{600}\sqrt{5}$ one recovers the Frobenius structure listed in Table 1.

\subsection{The case of $G_{18}$}
The Christoffel symbols of the dual connection are
$$\Gamma^{(2)i}_{jk}=-c^{*i}_{jk}+\left(-\f{19}{60}+\f{1}{600}\sqrt{5}c\right)C^{i}_{jk},$$
where
$$C^i_{lp}(u)=3\sum_{s=1}^{20}\frac{1}{||\alpha_s||^2}\frac{(\alpha_s)_l \,(\alpha_s)_p \, (\check{\bar\alpha}_s)^i}{\alpha_s(u)}-5\sum_{s=21}^{32}\frac{1}{||\alpha_s||^2}\frac{(\alpha_s)_l \,(\alpha_s)_p \, (\check{\bar\alpha}_s)^i}{\alpha_s(u)}.$$
Up to inessential linear terms the vector potential is given by:
\begin{eqnarray*}
A^1&=& 20u_1^3\sqrt{5}-\f{2}{3}u_1^3 c+u_1u_2,\\
A^2&=& 20 c u_1^4\sqrt{5}-\f{1}{3}c^2u_1^4+\f{1}{2}u_2^2.
\end{eqnarray*}
Finally, for $c=-6i\sqrt{3}$ one recovers the Frobenius structure listed in Table 1.

\subsection{The case of $G_{21}$}
The Christoffel symbols of the dual connection are
$$\Gamma^{(2)i}_{jk}=-c^{*i}_{jk}+\left(-\frac{1}{300}(29\sqrt{5}+14400c)\sqrt{5}\right)C^{i}_{jk},$$
where
$$C^i_{lp}(u)=2\sum_{s=1}^{30}\frac{1}{||\alpha_s||^2}\frac{(\alpha_s)_l \,(\alpha_s)_p \, (\check{\bar\alpha}_s)^i}{\alpha_s(u)}-3\sum_{s=31}^{50}\frac{1}{||\alpha_s||^2}\frac{(\alpha_s)_l \,(\alpha_s)_p \, (\check{\bar\alpha}_s)^i}{\alpha_s(u)}.$$
Up to inessential linear terms the vector potential is given by:
\begin{eqnarray*}
A^1&=& -\frac{1}{1800}u_1^6(\sqrt{5}+600c)+u_1u_2,\\
A^2&=& -\frac{1}{1080}c(\sqrt{5}+300c)u_1^{10}+\f{1}{2}u_2^2.
\end{eqnarray*}
Finally, for $c=-\frac{1}{600}\sqrt{5}$ one recovers the Frobenius structure listed in Table 1.

\subsection{The case of $G(3,1,2)$}
The dual connection is
$$\Gamma^{(2)i}_{jk}=-c^{*i}_{jk}+\left(c+\f{5}{6}\right)C^{i}_{jk},$$
where
$$c^{*i}_{lp}(u)=\f{1}{6}\left(3\sum_{s=1}^{2}\frac{1}{||\alpha_s||^2}\frac{(\alpha_s)_l \,(\alpha_s)_p \, (\check{\bar\alpha}_s)^i}{\alpha_s(u)}+2\sum_{s=3}^{5}\frac{1}{||\alpha_s||^2}\frac{(\alpha_s)_l \,(\alpha_s)_p \, (\check{\bar\alpha}_s)^i}{\alpha_s(u)}\right)$$
and
$$C^i_{lp}(u)=3\sum_{s=1}^{2}\frac{1}{||\alpha_s||^2}\frac{(\alpha_s)_l \,(\alpha_s)_p \, (\check{\bar\alpha}_s)^i}{\alpha_s(u)}-2\sum_{s=3}^{5}\frac{1}{||\alpha_s||^2}\frac{(\alpha_s)_l \,(\alpha_s)_p \, (\check{\bar\alpha}_s)^i}{\alpha_s(u)}.$$
Up to inessential linear terms the vector potential is
\begin{eqnarray*}
A^1&=& \f{1}{6}(-3-4c)u_1^3+u_1u_2,\\
A^2&=&\f{1}{6}(-2c^2-3c-1)u_1^4+\f{1}{2}u_2^2.
\end{eqnarray*}
Finally, for $c=-\frac{3}{4}$ one recovers the Frobenius structure listed in Table 1 with $m=3$.

\subsection{The case of $G(3,1,3)$}

The flat coordinates  are
\begin{eqnarray*}
u_1 &=& p_1^3+p_2^3+p_3^3\\
u_2 &=&p_1^6+p_2^6+p_3^6+c_1(p_1^3+p_2^3+p_3^3)^2\\
u_3 &=& p_1^9+p_2^9+p_3^9+c_2(p_1^3+p_2^3+p_3^3)(p_1^6+p_2^6+p_3^6)+c_3(p_1^3+p_2^3+p_3^3)^3
\end{eqnarray*}
with $c_2 = \f{3}{2}c_1-\f{1}{2}$ and $c_3 := \f{1}{2}c_1^2-\f{1}{3}c_1+\f{1}{18}$. The dual connection is
$$\Gamma^{(2)i}_{jk}=-c^{*i}_{jk}+\left(2c_1+\f{10}{9}\right)C^{i}_{jk},$$
where
$$c^{*i}_{lp}(u)=\f{1}{9}\left(3\sum_{s=1}^{3}\frac{1}{||\alpha_s||^2}\frac{(\alpha_s)_l \,(\alpha_s)_p \, (\check{\bar\alpha}_s)^i}{\alpha_s(u)}+2\sum_{s=4}^{12}\frac{1}{||\alpha_s||^2}\frac{(\alpha_s)_l \,(\alpha_s)_p \, (\check{\bar\alpha}_s)^i}{\alpha_s(u)}\right)$$
and
$$C^i_{lp}(u)=3\sum_{s=1}^{3}\frac{1}{||\alpha_s||^2}\frac{(\alpha_s)_l \,(\alpha_s)_p \, (\check{\bar\alpha}_s)^i}{\alpha_s(u)}-\sum_{s=4}^{12}\frac{1}{||\alpha_s||^2}\frac{(\alpha_s)_l \,(\alpha_s)_p \, (\check{\bar\alpha}_s)^i}{\alpha_s(u)}.$$
Up to inessential linear terms the vector potential is
\begin{eqnarray*}
A^1&=& \f{1}{24}(18c_1^2+19c_1+5)u_1^4+\f{1}{24}(-36c_1-18)u_2u_1^2+
u_1u_3+\f{3}{8}u_2^2,\\
A^2&=&\f{1}{180}(54c_1^3+99c_1^2+54c_1+9)u_1^5+\f{1}{180}(-30c_1-10)u_2u_1^3
+\\
&&\f{1}{180}(-135c_1-45)u_2^2u_1+u_2u_3,\\
A^3&=&\f{1}{2}u_3^2+\f{1}{2160}(-648c_1^4-1296c_1^3-972c_1^2-324c_1-40)u_1^6+\\
&&+\f{9}{8}\left(c_1+\f{1}{3}\right)\left(c_1+\f{2}{3}\right)u_2\left(c_1+\f{1}{2}\right)u_1^4-\f{9}{8}\left(c_1+\f{1}{3}\right)\left(c_1+\f{2}{3}\right)u_2^2u_1^2+\\
&&\f{1}{2160}(405c_1+270)u_2^3
\end{eqnarray*}
Finally, for $c_1=-\frac{1}{2}$ one recovers the Frobenius structure listed in Table 1.

\begin{remark}
In all the cases we dealt with in this Section, the dual connection has the general form
$$\Gamma^{(2)i}_{jk}=-c^{*i}_{jk}+\lambda C^{i}_{jk},$$
where $ C^{i}_{jk}$ are the Christoffel symbol of a Dunkl-Kohno-type connection
\begin{equation}
\nabla+\f{1}{N}\sum_{H\in \mathcal{H}}\frac{d\alpha_H}{\alpha_H}\otimes\kappa_H\pi_H.
\end{equation}
where the projectors $\pi_H$ and the weights $k_{H}$ satisfy the condition
\beq\label{normc2}
\sum_{H\in \mathcal{H}}\kappa_H\pi_H=0.
\eeq
This follows from the fact that the dual connection  has to satisfy the condition $\nabla^{(2)}E=0$.  
 Moreover the collection $\{k_H\}_{H\in \mathcal{H}}$ of complex weights define a $G$-invariant function. 
 It would be interesting to prove that in general the dual structure is always defined by the standard dual product for a suitable choice of the hermtian metric and by a dual connection of the form above. Indeed this would have as an immediate consequence the existence of an upper bound for the number of parameters:  taking into account the condition \eqref{normc2}, the number of parameters should be strictly less than the number of $G$-orbits in $\mathcal{H}$.
\end{remark}

\subsection{The bi-flat $F$-structures of associated Coxeter groups}

 The Frobenius manifold structure on the space of orbits of Shephard groups coincides with the Frobenius structure on the space of  orbits of the associated Coxeter group. We prove a similar theorem for the bi-flat $F$-structures of the exceptional Shephard groups $G_5$, $G_6$, $G_9$, $G_{10}$, $G_{14}$, $G_{17}$, $G_{18}$, $G_{21}$, $G_{26}$. In other words, the Frobenius structure of  each orbit space of these  groups sits inside a one-parameter family of bi-flat $F$-manifold structures, which depends only on the associated Coxeter groups and can be viewed as a natural deformation of the Frobenius structure itself. 
\begin{theorem}
The bi-flat $F$-structure on the space of orbits of each exceptional Shephard groups of rank $2$ and $3$ coincides with the bi-flat $F$-structure on the space of orbits of the associated Coxeter group.
\end{theorem}
\n
\proof The proof is not trivial only for the groups $G_5$, $G_6$, $G_9$, $G_{10}$, $G_{14}$, $G_{17}$, $G_{18}$, $G_{21}$, $G_{26}$. 

For the groups $G_5$, $G_{10}$, $G_{18}$  we have to prove that the vector potential coincides with the
 vector potential of $B_2$. The generalized Saito flat coordinates are $u_1=p_1^2+p_2^2$ and $u_2=p_1^4+p_4^2+c_1u_1^2$ and in these coordinates we obtain:
\begin{eqnarray*}
A_{B_2}^1&=& -\left(\f{2}{3}c_1+\f{1}{2}\right)u_1^3+u_1u_2,\\
A_{B_2}^2&=& -\left(\f{1}{3}c_1^2+\f{1}{2}c_1+1\right)u_1^4+\f{1}{2}u_2^2.
\end{eqnarray*}
These coincides with the vector potential of $G_5$, $G_{10}$, $G_{18}$ up to rescaling of the variables. 

For the groups $G_6$, $G_{9}$, $G_{17}$ we have to prove that the vector potential coincides with the
 vector potential of  $I_2(6)$.  In the generalized Saito flat coordinates 
\begin{eqnarray*}
u_1&=&p_1^2+p_2^2,\\
u_2&=&2p_1^6-30p_1^4p_2^2+30p_1^2p_2^4-2p_2^6+c_1u_1^3
\end{eqnarray*}
we obtain
\begin{eqnarray*}
A_{I_2(6)}^1&=& -\left(\f{2}{3}c_1+\f{1}{2}\right)u_1^3+u_1u_2,\\
A_{I_2(6)}^2&=& -\left(\f{1}{3}c_1^2+\f{1}{2}c_1+1\right)u_1^4+\f{1}{2}u_2^2,
\end{eqnarray*}
that coincides with the vector potential of $G_6$, $G_{9}$, $G_{17}$ up to rescaling of the variables. 

For the group $G_{14}$ we have to prove that the vector potential coincides with the
 vector potential of  $I_2(8)$.  In the generalized Saito flat coordinates 
\begin{eqnarray*}
u_1&=&p_1^2+p_2^2,\\
u_2&=&2p_1^8-56p_1^6p_2^2+140p_1^4p_2^4-56p_1^2p_2^6+2p_2^8+c_1u_1^4,
\end{eqnarray*}
 we get
\begin{eqnarray*}
A_{I_2(8)}^1&=& -\f{2}{5}c_1u_1^5+u_1u_2,\\
A_{I_2(8)}^2&=& -\f{2}{7}u_1^8c_1^2+\f{8}{7}u_1^8+\f{1}{2}u_2^2,
\end{eqnarray*}
that coincides with the vector potential of $G_{14}$ up to rescaling of the variables. 

For the group $G_{21}$ we have to prove that the vector potential coincides with the
 vector potential of  $I_2(10)$.  In the generalized Saito flat coordinates 
\begin{eqnarray*}
u_1&=&p_1^2+p_2^2,\\
u_2&=&2p_1^{10}-90p_1^8p_2^2+420p_1^6p_2^4-420p_1^4p_2^6+90p_1^2p_2^8-2p_2^{10}+c_1u_1^5,
\end{eqnarray*}
we obtain
\begin{eqnarray*}
A_{I_2(10)}^1&=& -\left(\f{2}{3}c_1+\f{1}{2}\right)u_1^3+u_1u_2,\\
A_{I_2(10)}^2&=& -\left(\f{1}{3}c_1^2+\f{1}{2}c_1+1\right)u_1^4+\f{1}{2}u_2^2,
\end{eqnarray*}
that coincides with the vector potential of $G_{21}$ up to rescaling of the variables. 

For the group $G_{26}$  we have to prove that the vector potential coincides with the vector potential of $B_3$. The generalized Saito flat 
 coordinates are
\begin{eqnarray*}
u_1&=&p_1^6+p_2^6+p_3^6,\\
u_2&=&p_1^4+p_2^4+p_3^4+c_1(p_1^2+p_2^2+p_3^2)^2,\\
u_3&=&p_1^6+p_2^6+p_3^6+c_2(p_1^2+p_2^2+p_3^2)(p_1^4+p_2^4+p_3^4)+c_3(p_1^2+p_2^2+p_3^2)^3,
\end{eqnarray*}
with $c_1= \f{1}{3}+\f{2}{3}c_2$ and $c_3 = \f{2}{9}c_2^2$. After some computations we get
 \begin{eqnarray*}
A^1_{B_3}&=&\f{1}{72}(24c_2^2+62c_2+40)u_1^4+\f{1}{72}(-72c_2-90)u_2u_1^2
+u_1u_3+\f{3}{8}u_2^2,\\
A^2_{B_3}&=&\f{1}{90}(8c_2^3+34c_2^2+46c_2+20)u_1^5-\f{1}{9}(c_2+1)u_2u_1^3+\f{1}{90}(-45c_2-45)u_2^2u_1+u_2u_3,\\
A^3_{B_3}&=&\f{1}{2}u_3^2+\f{1}{2160}(-128c_2^4-640c_2^3-1200c_2^2-1000c_2-312)u_1^6+\f{1}{3}\left(c_2+\f{5}{4}\right)(c_2+1)+\\
&&\left(c_2+\f{3}{2}\right)u_2u_1^4-\f{1}{2}(c_2+1)\left(c_2+\f{3}{2}\right)u_2^2u_1^2+\f{1}{2160}(270c_2+405)u_2^3,
\end{eqnarray*}
that coincides with the vector potential of $G_6$, $G_{9}$, $G_{17}$ up to rescaling of the variables.
\endproof

\begin{rmk}
The dual connection has the form
$$\Gamma^{(2)i}_{jk}=-c^{*i}_{jk}+\lambda C^{i}_{jk},$$
where
$$c^{*i}_{lp}(u)=\f{2}{N}\left(\sum_{s=1}^{n}\frac{1}{||\alpha_s||^2}\frac{(\alpha_s)_l \,(\alpha_s)_p \, (\check{\bar\alpha}_s)^i}{\alpha_s(u)}\right)$$
and
$$C^i_{lp}(u)=\sum_{s=1}^{k}\frac{\kappa_s}{||\alpha_s||^2}\frac{(\alpha_s)_l \,(\alpha_s)_p \, (\check{\bar\alpha}_s)^i}{\alpha_s(u)}.$$
\newline
In the case of $B_2$ we have $N=4$, $n=4$, $\kappa_s=1$ if $s=1,2$, $\kappa_s=-1$ if $s=3,4$ and $\lambda=2c_1+\f{3}{2}$.
\newline
In the case of $I_2(6)$ we have $N=6$, $n=6$, $\kappa_s=1$ if $s=1,4,6$, $\kappa_s=-1$ if $s=2,3,5$ and $\lambda=-\f{1}{3}c_1$.
\newline
In the case of $I_2(8)$ we have $N=8$, $n=8$, $\kappa_s=1$ if $s=1,2,3,4$, $\kappa_s=-1$ if $s=5,6,7,8$ and $\lambda=-\f{3}{8}c_1$.
\newline
In the case of $I_2(10)$ we have $N=10$, $n=10$, $\kappa_s=1$ if $s=1,7,8,9,10$, $\kappa_s=-1$ if $s=2,3,4,5,6$  and $\lambda=\f{2}{5}c_1$.
\newline
In the case of $B_3$ we have $N=6$, $n=9$,  $\kappa_s=2$ if $s=1,2,3$, $\kappa_s=-1$ if $s=4,5,6,7,8,9$ and $\lambda=2c_1+\f{3}{2}$.
\newline
In all cases the metric is the standard hermitean metric. The list of $\alpha$ is given in the Appendix 2.
\end{rmk}

\section{Bi-flat $F$-manifolds and generalized WDVV equations}
In this Section we prove that under a mild assumption semisimple bi-flat $F$-manifold are described in flat coordinates for the natural  connection by solutions of generalized WDVV equations
\begin{eqnarray}
\label{genWDVV1}
\d_j\d_l A^i\d_k\d_mA^l&=&\d_k\d_l A^i\d_k\d_mA^l,\\
\label{genWDVV2}
\d_1\d_i A^j&=&\delta_{i}^{j},\\
\label{genWDVV3}
E(A^i)&=&(1+w_i)A^i,
\end{eqnarray}
where $w_1=1$. This system of equations was obtained recently by Kato, Mano and Sekiguchi in the study of Saito structures without metrics \cite{KMS2}.

Comparing equations \eqref{genWDVV1}, \eqref{genWDVV2}, \eqref{genWDVV3} with \eqref{oriented1} and \eqref{oriented2}, we see that we have only the extra condition \eqref{genWDVV3}. We want to show  that this extra condition
 is related to the presence of a second compatible flat structure. 
 
We have seen that bi-flat $F$-manifolds, in the semisimple case,
 are equivalent to Frobenius manifolds without metric. This means that
 one can replace the conditions involving the dual connection with the condition 
$$\nabla^{(1)}\nabla^{(1)}E=0.$$
 Assuuming $e=\f{\d}{\d t^1}$ (this is not restrictive) and  
  $E$ of the form 
\begin{equation}\label{Eulervf}
E=\sum_iw^it^i\f{\d}{\d t^i},
\end{equation}
where $(t^1,...,t^n)$ are flat coordinates (this is not too restrictive if $\nabla^{(1)} E$ is semisimple) and imposing the conditions on the Euler vector field  (see Definition \ref{multiflatdefi}): 
$$[e,E]=e,\qquad ({\rm Lie}_E c)^i_{jk}=c^i_{jk},$$
we obtain $w_1=1$ and
$$({\rm Lie}_E c)^i_{jk}=E^l\d_l c^i_{jk}-(\d_l E^i)c^l_{jk}+(\d_j E^l)c^i_{lk}+(\d_k E^l)c^i_{jl}=$$
$$=E^l\d_l c^i_{jk}-w_ic^i_{jk}+w_jc^i_{jk}+w_kc^i_{jk}=c^i_{jk}.$$
From this relation we have
$$E(c^i_{jk})=(1+w_i-w_j-w_k)c^i_{jk}.$$
Writing $c^i_{jk}$ in terms of derivative of the vector potential, we obtain
$$E^l\d_l \d_j \d_k A^i=(1+w_i-w_j-w_k)\d_j \d_k A^i.$$
It is immediate to check that 
$$\d_j \d_k (E(A^i))=w_k c^i_{jk}+w_jc^i_{jk}+E(c^i_{jk}),$$
and substituting in the previous expression we find 
$$\d_j \d_k E(A^i))=(1+w_i)\d_j \d_k A^i,$$
or 
$$\d_j \d_k (E(A^i)-(1+w_i)A^i)=0,$$
which is exactly equation \eqref{genWDVV3} up to affine terms in $t^i$. But since the vector potential itself is defined up to affine terms in $t^i$ we are done. 
\newline
\newline
All the vector potentials obtained in this paper satisfy by construction the generalized WDVV equations \eqref{genWDVV1}, \eqref{genWDVV2} and \eqref{genWDVV3} where  in  \eqref{genWDVV2} one has to replace $1$ with $n$ (since the unit vector field is $\f{\d}{\d u_n}$) and 
 in  \eqref{genWDVV3} one has to replace $E$ with the normalized Euler vector field $\f{E}{d_n}$ and, as a consequence, $w_i=\f{d_i}{d_n}$. Vector potentials associated with ${\rm rank}>2$ well-generated exceptional complex reflection groups have been found in \cite{KMS1} (with the exception of Shephard groups). The results of the present Section, combined with the results of \cite{KMS2} imply the existence of a bi-flat structure for any well-generated complex reflection group. In particular, given the flat structure $(\nabla,\circ,e)$ in flat coordinates $(u^1,...,u^n)$ one can immediately renconstruct the dual structure using the standard formula for the dual product and the formula \eqref{intermiediateeq1}. In the present paper we 
 have adopted  a ``dual" approach starting from the dual structure that has a universal beautiful form in the coordinates $(p^1,...,p^n)$.  This approach leads naturally to discover the presence of parameters in several examples. 
As far as we know, the presence of these parameters has never been observed before. The results of the present Section also prove that the relation between generalized WDVV equations and the full family of Painlev\'e VI equation (in standard form)  obtained in \cite{KMS2} is the counterpart in flat coordinates of the relation between three dimensional semisimple bi-flat $F$-manifolds and the full family of Painlev\'e VI equation (in sigma form) obtained in canonical coordinates in \cite{Limrn,ALmulti} (see also \cite{AL-Bi-flat}).

\section{Appendix 1. Reflecting mirrors}
We give the list, case by case, of the covectors $\alpha_s$ obtained factorizing ${\rm det}\f{\d u^i}{\d p^j}$.

\subsection{The $G_{4}$ case}
We have 4  mirrors defining reflections of order 3:
\begin{eqnarray*}
&&\alpha_1= \left[ -1-i, \sqrt{3}-1\right],\qquad \alpha_2= \left[ 1-i, \sqrt{3}+1\right],\\
&&\alpha_3=  \left[ -1+i, \sqrt{3}+1\right],\qquad \alpha_4= \left[ 1+i, \sqrt{3}-1\right].\
\end{eqnarray*}

\subsection{The $G_{5}$ case}
We have 8  mirrors defining reflections of order 3:
\begin{eqnarray*}
&\alpha_1= \left[1,-\frac{1}{2}\sqrt{3}+\frac{1}{2}+\frac{1}{2}i-\frac{1}{2}i\sqrt{3}\right],\,
&\alpha_2=\left[ 1,\frac{1}{2}\sqrt{3}+\frac{1}{2}-\frac{1}{2}i-\frac{1}{2} i \sqrt{3}\right],\\
&\alpha_3=  \left[ 1,\frac{1}{2}\sqrt{3}-\frac{1}{2}-\frac{1}{2} i+\frac{1}{2} i \sqrt{3}\right],\,
&\alpha_4= \left[ 1,-\frac{1}{2}\sqrt{3}-\frac{1}{2}+\frac{1}{2}i+\frac{1}{2}i \sqrt{3}\right],\\
&\alpha_5= \left[1,\frac{1}{2}\sqrt{3}+\frac{1}{2}+\frac{1}{2}i+\frac{1}{2}i \sqrt{3}\right],\,
&\alpha_6= \left[ 1,-\frac{1}{2}\sqrt{3}+\frac{1}{2}-\frac{1}{2}i+\frac{1}{2}i \sqrt{3}\right],\\
&\alpha_7=  \left[1,\frac{1}{2}\sqrt{3}-\frac{1}{2}+\frac{1}{2}i-\frac{1}{2}i \sqrt{3}\right],\,
&\alpha_8= \left[1,-\frac{1}{2}\sqrt{3}-\frac{1}{2}-\frac{1}{2}i-\frac{1}{2}i\sqrt{3}\right].\
\end{eqnarray*}

\subsection{The $G_{6}$ case}
We have 6  mirrors defining reflections of order 2:
\begin{eqnarray*}
\alpha_1= \left[1,-1\right],\,
\alpha_2= \left[ 1,-i\right],\,
\alpha_3=  \left[ 1,i\right],\,
\alpha_4= \left[ 1,1\right],\,
\alpha_5= \left[1,0\right],\,
\alpha_6= \left[ 0, 1\right],
\end{eqnarray*}
and  4  mirrors defining reflections of order 3:
\begin{eqnarray*}
&\alpha_7=  \left[-1-i , \sqrt{3}-1\right],\qquad
&\alpha_8= \left[-1+i,\sqrt{3}+1\right],\\
&\alpha_9= \left[1+i,\sqrt{3}-1\right],\qquad
&\alpha_{10}= \left[1-i,\sqrt{3}+1\right].
\end{eqnarray*}

\subsection{The $G_{8}$ case}
We have 6  mirrors defining reflections of order 4:
\begin{eqnarray*}
\alpha_1= \left[1,1\right],\,
\alpha_2= \left[ 1,-1\right],\,
\alpha_3=  \left[ 0,1\right],\,
\alpha_4= \left[ 1,-i\right],\,
\alpha_5= \left[1,i\right],\,
\alpha_6= \left[ 1, 0\right].\\
\end{eqnarray*}

\subsection{The $G_{9}$ case}
We have 12  mirrors defining reflections of order 2:
\begin{eqnarray*}
&\alpha_1= \left[1,\frac{1}{2}\sqrt{2}-\frac{1}{2}i\sqrt{2}\right],\qquad
&\alpha_2= \left[1,\frac{1}{2}\sqrt{2}+\frac{1}{2} i \sqrt{2}\right],\\
&\alpha_3= \left[1,-\frac{1}{2}\sqrt{2}-\frac{1}{2}i\sqrt{2}\right],\qquad
&\alpha_4=\left[ 1,-\frac{1}{2}\sqrt{2}+\frac{1}{2}i\sqrt{2}\right],\\
&\alpha_5=\left[1,-\sqrt{2}+1\right],\qquad
&\alpha_6=\left[1,\sqrt{2}+1\right],\\
&\alpha_7= \left[1,i \sqrt{2}+i\right],\qquad
&\alpha_8=\left[1,-i\sqrt{2}-i\right],\\
&\alpha_9=\left[1,i \sqrt{2}-i\right],\qquad
&\alpha_{10}= \left[1,-i \sqrt{2}+i\right],\\
&\alpha_{11}= \left[1,\sqrt{2}-1\right],\qquad
&\alpha_{12}= \left[1,-\sqrt{2}-1\right],\\
\end{eqnarray*}
and 6  mirrors defining reflections of order 4:
\begin{eqnarray*}
\alpha_{13}= \left[1,1\right],\,
\alpha_{14}= \left[0,1\right],\,
\alpha_{15}= \left[1, -1\right],\,
\alpha_{16}= \left[1 , i \right],\,
\alpha_{17}= \left[1, -i\right],\,
\alpha_{18}= \left[1, 0\right].
\end{eqnarray*}

\subsection{The $G_{10}$ case}
We have 8 mirrors defining reflections of order 3:
\begin{eqnarray*}
\alpha_1&=& \left[1,-\frac{1}{2}i \sqrt{3}-\frac{1}{2} \sqrt{2+2i \sqrt{3}}-\frac{1}{2}\right],\\
\alpha_2&=& \left[1, \frac{1}{2} i \sqrt{3}+\frac{1}{4} \sqrt{2+2 i \sqrt{3}}-\frac{1}{2}-\frac{1}{4}i\sqrt{3} \sqrt{2+2i\sqrt{3}}\right],\\
\alpha_3&=&  \left[1,\frac{1}{2} i \sqrt{3}-\frac{1}{4}\sqrt{2+2 i \sqrt{3}}-\frac{1}{2}+\frac{1}{4} i\sqrt{3} \sqrt{2+2i \sqrt{3}}\right],\\
\alpha_4&=& \left[ 1,-\frac{1}{2} i \sqrt{3}+\frac{1}{2} \sqrt{2+2 i \sqrt{3}}-\frac{1}{2}\right],\\
\alpha_5&=& \left[1,\frac{1}{2} i \sqrt{3}+\frac{1}{2} \sqrt{2+2i \sqrt{3}}+\frac{1}{2}\right],\\
\alpha_6&=& \left[1,-\frac{1}{2}i\sqrt{3}-\frac{1}{4}\sqrt{2+2i \sqrt{3}}+\frac{1}{2}+\frac{1}{4}i\sqrt{3}\sqrt{2+2i \sqrt{3}}\right],\\
\alpha_7&=&  \left[1,-\frac{1}{2}i\sqrt{3}+\frac{1}{4} \sqrt{2+2i \sqrt{3}}+\frac{1}{2}-\frac{1}{4}i\sqrt{3} \sqrt{2+2i \sqrt{3}}\right],\\
\alpha_8&=& \left[1,\frac{1}{2}i \sqrt{3}-\frac{1}{2}\sqrt{2+2i \sqrt{3}}+\frac{1}{2}\right],\\
\end{eqnarray*}
and 6  mirrors defining reflections of order 4:
\begin{eqnarray*}
&&\alpha_9= \left[1,1\right],\,\alpha_{10}= \left[1,0\right],\,
\alpha_{11}= \left[1,-1\right],\,\alpha_{12}=\left[0,1\right],\\
&&\alpha_{13}= \left[1,-\frac{1}{4} \sqrt{2+2i \sqrt{3}}-\frac{1}{4}i \sqrt{3} \sqrt{2+2i \sqrt{3}}\right],\\
&&\alpha_{14}= \left[1,\frac{1}{4}\sqrt{2+2i \sqrt{3}}+\frac{1}{4}i \sqrt{3} \sqrt{2+2i \sqrt{3}}\right].\\
\end{eqnarray*}

\subsection{The $G_{14}$ case}
We have 12  mirrors defining reflections of order 2:
\begin{eqnarray*}
&\alpha_1= \left[1,-\frac{1}{2}\sqrt{2}+\frac{1}{2}i\sqrt{2} \right],\qquad
&\alpha_2= \left[1,-\frac{1}{2}\sqrt{2}-\frac{1}{2}i\sqrt{2}\right],\\
&\alpha_3=  \left[ 1,\frac{1}{2}\sqrt{2}-\frac{1}{2}i\sqrt{2}\right],\qquad
&\alpha_4= \left[ 1,\frac{1}{2}\sqrt{2}+\frac{1}{2}i\sqrt{2} \right],\\
&\alpha_5= \left[1,\sqrt{2}-1\right],\qquad
&\alpha_6= \left[1,-\sqrt{2}-1 \right],\\
&\alpha_7=  \left[1,-i\sqrt{2}+i \right],\qquad
&\alpha_8= \left[ 1,i \sqrt{2}-i\right],\\
&\alpha_9=\left[1,-i\sqrt{2}-i \right],\qquad
&\alpha_{10}= \left[1,i \sqrt{2}+i \right],\\
&\alpha_{11}= \left[1,-\sqrt{2}+1 \right],\qquad
&\alpha_{12}= \left[1,\sqrt{2}+1 \right],\\
\end{eqnarray*}
and 8  mirrors defining reflections of order 3:
\begin{eqnarray*}
&\alpha_{13}= \left[1,-\frac{1}{2}+\frac{1}{2}i+\frac{1}{2}\sqrt{3}-\frac{1}{2}i \sqrt{3}\right],\qquad
&\alpha_{14}= \left[ 1,-\frac{1}{2}+\frac{1}{2}i-\frac{1}{2}\sqrt{3}+\frac{1}{2}i \sqrt{3}\right], \\
&\alpha_{15}= \left[1,-\frac{1}{2}\sqrt{3}-\frac{1}{2}i\sqrt{3}-\frac{1}{2}-\frac{1}{2} i\right], \qquad
&\alpha_{16}= \left[1,\frac{1}{2}\sqrt{3}+\frac{1}{2}i\sqrt{3}-\frac{1}{2}-\frac{1}{2} i \right], \\
&\alpha_{17}= \left[ 1,\frac{1}{2}\sqrt{3}+\frac{1}{2} i\sqrt{3}+\frac{1}{2}+\frac{1}{2} i\right], \qquad
&\alpha_{18}= \left[1,-\frac{1}{2}\sqrt{3}-\frac{1}{2}i\sqrt{3}+\frac{1}{2}+\frac{1}{2} i\right], \\
&\alpha_{19}= \left[1,\frac{1}{2}-\frac{1}{2} i+\frac{1}{2} \sqrt{3}-\frac{1}{2} i \sqrt{3}\right], \qquad
&\alpha_{20}= \left[ 1,\frac{1}{2}-\frac{1}{2} i-\frac{1}{2}\sqrt{3}+\frac{1}{2} i \sqrt{3}\right], \\
\end{eqnarray*}

\subsection{The $G_{16}$ case}
We have 12 mirrors defining reflections of order 2:
\begin{eqnarray*}
&&\alpha_1=\left[ 1,-\frac{1}{10}\mu -\frac{1}{10}\sigma \right],\quad
\alpha_2=\left[ 1,-(\frac{1}{10}\mu +\frac{1}{10}\sigma)^3-(\frac{1}{25}\mu + \frac{1}{25}\sigma) \sqrt{5} \right],\\
&&\alpha_3=\left[ 1,(\frac{1}{25}\mu +\frac{1}{25}\sigma) \sqrt{5}+(\frac{1}{10} \mu + \frac{1}{10}\sigma)^3 \right],\quad
\alpha_4=\left[1,\frac{1}{10} \mu +\frac{1}{10} \sigma \right],\\
&&\alpha_5=\left[ 1,-(-\frac{1}{2} i-\frac{1}{2} i\sqrt{5}+\frac{1}{2}\nu)^3-(-i-i \sqrt{5}+\nu) \sqrt{5}+4i+4i \sqrt{5}-4 \nu \right],\\
&&\alpha_6=\left[ 1,(-\frac{1}{2} i-\frac{1}{2} i \sqrt{5}+\frac{1}{2} \nu)^3+(-i-i \sqrt{5}+\nu)\sqrt{5}-4i-4i\sqrt{5}+4 \nu\right],\\
&&\alpha_7=\left[ 1,\frac{1}{2} i+\frac{1}{2} i \sqrt{5}-\frac{1}{2} \nu\right],\quad
\alpha_8=\left[1,-\frac{1}{2}i-\frac{1}{10}i\sqrt{5}+\frac{1}{2}\nu \right],\\
&&\alpha_9=\left[ 1,\frac{1}{2}\sqrt{5}-\frac{1}{2}-\frac{1}{2} \lambda\right],\quad
\alpha_{10}= \left[1,\frac{1}{2} \sqrt{5}-\frac{1}{2}+\frac{1}{2} \lambda\right],\\
&&\alpha_{11}= \left[1,-\frac{1}{2}\sqrt{5}+\frac{1}{2}-\frac{1}{2} \lambda  \right],\quad
\alpha_{12}= \left[ 1,-\frac{1}{2}\sqrt{5}+\frac{1}{2}+\frac{1}{2} \lambda \right],\\
\end{eqnarray*}
with 
$$\quad \nu=\sqrt{-10-2\sqrt{5}}, \quad \lambda=\sqrt{10-2\sqrt{5}},\quad \sigma=\sqrt{5}\nu,\mu=\sqrt{5}\lambda.$$

\subsection{The $G_{17}$ case}
We have 30  mirrors defining reflections of order 2:
\begin{eqnarray*}
&\alpha_1= \left[ 1,\frac{1}{2} \sqrt{5}-\frac{1}{2}+\frac{1}{2} i+\frac{1}{2} i \sqrt{5} \right],\quad
&\alpha_2= \left[ 1,\frac{1}{2} \sqrt{5}-\frac{1}{2}-\frac{1}{2} i-\frac{1}{2} i \sqrt{5} \right],\\
&\alpha_3= \left[  1,-\frac{1}{2} \sqrt{5}+\frac{1}{2}+\frac{3}{2} i-\frac{1}{2} i \sqrt{5} \right],\quad
&\alpha_4= \left[ 1,-\frac{1}{2} \sqrt{5}+\frac{1}{2}-\frac{3}{2} i+\frac{1}{2} i \sqrt{5} \right],\\
&\alpha_5= \left[ 1,\frac{1}{2} \sqrt{5}-\frac{1}{2}+\frac{3}{2} i-\frac{1}{2} i \sqrt{5}  \right],\quad
&\alpha_6= \left[1, \frac{1}{2} \sqrt{5}-\frac{1}{2}-\frac{3}{2} i+\frac{1}{2} i \sqrt{5} \right],\\
&\alpha_7= \left[ 1, -i   \right],\quad
&\alpha_8= \left[ 1, i  \right],\\
&\alpha_9= \left[ 1,\frac{1}{10} \sqrt{5}-\frac{1}{2}-\frac{1}{2} i+\frac{3}{10} i \sqrt{5} \right],\quad
&\alpha_{10}= \left[  1,\frac{1}{10} \sqrt{5}-\frac{1}{2}+\frac{1}{2} i-\frac{3}{10} i \sqrt{5} \right],\\
&\alpha_{11}= \left[ 1,-\frac{1}{6} \sqrt{5}+\frac{1}{6}+\frac{1}{6} i+\frac{1}{6} i \sqrt{5}  \right],\quad
&\alpha_{12}= \left[ 1,-\frac{1}{6} \sqrt{5}+\frac{1}{6}-\frac{1}{6} i-\frac{1}{6} i \sqrt{5} \right],\\
&\alpha_{13}= \left[  1,-\frac{1}{2} \sqrt{5}-\frac{3}{2}-\frac{1}{2} i-\frac{1}{2} i \sqrt{5} \right],\quad
&\alpha_{14}= \left[ 1,-\frac{1}{2} \sqrt{5}-\frac{3}{2}+\frac{1}{2} i+\frac{1}{2} i \sqrt{5} \right],\\
&\alpha_{15}= \left[  1,-\frac{3}{10} \sqrt{5}-\frac{1}{2}-\frac{1}{2} i-\frac{1}{10} i \sqrt{5} \right],\quad
&\alpha_{16}= \left[ 1,-\frac{3}{10} \sqrt{5}-\frac{1}{2}+\frac{1}{2} i+\frac{1}{10} i \sqrt{5}  \right],\\
&\alpha_{17}= \left[ 1,\frac{3}{10} \sqrt{5}+\frac{1}{2}-\frac{1}{2} i-\frac{1}{10} i \sqrt{5} \right],\quad
&\alpha_{18}= \left[ 1,\frac{3}{10}\sqrt{5}+\frac{1}{2}+\frac{1}{2} i+\frac{1}{10} i \sqrt{5} \right],\\
&\alpha_{19}= \left[ 0, 1 \right],\quad
&\alpha_{20}= \left[ 1, -1 \right],\\
&\alpha_{21}= \left[  1,\frac{1}{2} \sqrt{5}+\frac{3}{2}+\frac{1}{2} i+\frac{1}{2} i\sqrt{5} \right],\quad
&\alpha_{22}= \left[  1,\frac{1}{2} \sqrt{5}+\frac{3}{2}-\frac{1}{2} i-\frac{1}{2} i \sqrt{5}\right],\\
&\alpha_{23}= \left[  1,-\frac{1}{10} \sqrt{5}+\frac{1}{2}+\frac{1}{2} i-\frac{3}{10} i \sqrt{5} \right],\quad
&\alpha_{24}= \left[  1,-\frac{1}{10} \sqrt{5}+\frac{1}{2}-\frac{1}{2} i+\frac{3}{10} i \sqrt{5}  \right],\\
&\alpha_{25}= \left[ 1,1  \right],\quad
&\alpha_{26}= \left[ 1,-\frac{1}{2} \sqrt{5}+\frac{1}{2}-\frac{1}{2} i-\frac{1}{2} i \sqrt{5} \right],\\
&\alpha_{27}= \left[  1,-\frac{1}{2} \sqrt{5}+\frac{1}{2}+\frac{1}{2} i+\frac{1}{2} i \sqrt{5} \right],\quad 
&\alpha_{28}= \left[ 1,\frac{1}{6} \sqrt{5} -\frac{1}{6}-\frac{1}{6} i-\frac{1}{6} i \sqrt{5} \right],\\
&\alpha_{29}= \left[ 1,\frac{1}{6} \sqrt{5}-\frac{1}{6}+\frac{1}{6} i+\frac{1}{6} i \sqrt{5} \right],\quad
&\alpha_{30}= \left[  1,0 \right],\\
\end{eqnarray*}
and 12  mirrors defining reflections of order 5:
\begin{eqnarray*}
&&\alpha_{31}= \left[ 1,\frac{1}{2} \sqrt{5}-\frac{1}{2}-\frac{1}{2} \sigma   \right],\quad
\alpha_{32}= \left[  1,\frac{1}{2} \sqrt{5}-\frac{1}{2}+\frac{1}{2} \sigma \right],\\
&&\alpha_{33}= \left[ 1,-\frac{1}{2} \sqrt{5}+\frac{1}{2}+\frac{1}{2} \sigma \right],\quad
\alpha_{34}= \left[ 1,-\frac{1}{2} \sqrt{5}+\frac{1}{2}-\frac{1}{2} \sigma  \right],\\
&&\alpha_{35}= \left[ 1,\frac{1}{10} \mu +\frac{1}{10} \nu \right],\quad
\alpha_{36}= \left[ 1,-(\frac{1}{25} \mu +\frac{1}{25} \nu) \sqrt{5}-(\frac{1}{10} \mu +\frac{1}{10} \nu)^3 \right],\\
&&\alpha_{37}= \left[  1,-\frac{1}{10} \mu -\frac{1}{10} \nu \right],\quad
\alpha_{38}= \left[ 1,(\frac{1}{10} \mu +\frac{1}{10} \nu)^3+(\frac{1}{25} \mu +\frac{1}{25} \nu) \sqrt{5} \right],\\
&&\alpha_{39}= \left[  1,\frac{1}{2} i+\frac{1}{2} i \sqrt{5}-\frac{1}{2} \lambda \right],\quad
\alpha_{42}= \left[  1,-\frac{1}{2} i-\frac{1}{2} i \sqrt{5}+\frac{1}{2}\lambda\right],\\
&&\alpha_{40}= \left[  1,(-\frac{1}{2} i-\frac{1}{2} i \sqrt{5}+\frac{1}{2}\lambda)^3+(-i- i \sqrt{5}+\lambda) \sqrt{5}-4 i-4 i \sqrt{5}+4 \lambda \right],\\
&&\alpha_{41}= \left[ 1,-(-\frac{1}{2} i-\frac{1}{2} i \sqrt{5}+\frac{1}{2}\lambda)^3-(-i-i \sqrt{5}+\lambda) \sqrt{5}+4 i+4 i \sqrt{5}-4 \lambda \right],\\
\end{eqnarray*}
with 
$$\sigma=\sqrt{10-2\sqrt{5}}, \quad \lambda=\sqrt{-10-2\sqrt{5}},\quad\mu=\sqrt{5}\sigma,\quad \nu=\sqrt{5}\lambda.$$

\subsection{The $G_{18}$ case}
We have 20  mirrors defining reflections of order 3:
\begin{footnotesize}
\begin{eqnarray*}
&\alpha_1= \left[ 1,-\frac{1}{2} \sqrt{3}+\frac{1}{2}-\frac{1}{2} i+\frac{1}{2} i \sqrt{3} \right],\,
&\alpha_2= \left[ 1,\frac{1}{2} \sqrt{3}+\frac{1}{2}-\frac{1}{2} i-\frac{1}{2} i \sqrt{3} \right],\\
&\alpha_3=\left[  1,-\frac{1}{2}\sqrt{3}+\frac{1}{2}+\frac{1}{2} i-\frac{1}{2} i \sqrt{3}   \right],\,
&\alpha_4=\left[  1,\frac{1}{2} \sqrt{3}+\frac{1}{2}+\frac{1}{2} i+\frac{1}{2} i \sqrt{3} \right],\\
&\alpha_5=\left[ 1,\frac{1}{2} \sqrt{3}-\frac{1}{2}-\frac{1}{2} i+\frac{1}{2} i \sqrt{3} \right],\,
&\alpha_6=\left[ 1,-\frac{1}{2} \sqrt{3}-\frac{1}{2}-\frac{1}{2} i-\frac{1}{2} i \sqrt{3} \right],\\
&\alpha_7=\left[  1,-\frac{1}{2}\sqrt{3}-\frac{1}{2}+\frac{1}{2} i+\frac{1}{2} i \sqrt{3} \right],\,
&\alpha_8=\left[ 1,\frac{1}{2} \sqrt{3}-\frac{1}{2}+\frac{1}{2} i-\frac{1}{2} i \sqrt{3} \right],\\
&\alpha_9=\left[  1,\frac{1}{2} \sqrt{3}+\frac{1}{2} \sqrt{5} \sqrt{3}-\frac{1}{2} \sqrt{5}-\frac{3}{2}  \right],\,
&\alpha_{10}=\left[  1,-\frac{1}{2}\sqrt{3}-\frac{1}{2} \sqrt{5} \sqrt{3}-\frac{1}{2} \sqrt{5}-\frac{3}{2} \right],\\
&\alpha_{11}=\left[ 1,-\frac{1}{2} i (\sqrt{5} \sqrt{3}-\sqrt{3}+\sqrt{5}-3) \right],\,
&\alpha_{12}=\left[ 1,\frac{1}{2} i (\sqrt{5} \sqrt{3}-\sqrt{3}-\sqrt{5}+3) \right],\\
&\alpha_{13}=\left[ 1,\frac{1}{2} i (\sqrt{5} \sqrt{3}-\sqrt{3}+\sqrt{5}-3)\right],\,
&\alpha_{14}=\left[ 1,-\frac{1}{2} i (\sqrt{5} \sqrt{3}-\sqrt{3}-\sqrt{5}+3)   \right],\\
&\alpha_{15}=\left[ 1,\frac{3}{2}-\frac{1}{2} \sqrt{5} \sqrt{3}-\frac{1}{2} \sqrt{3}+\frac{1}{2} \sqrt{5}  \right],\,
&\alpha_{16}=\left[  1,\frac{1}{2} \sqrt{3}+\frac{1}{2} \sqrt{5} \sqrt{3}+\frac{1}{2} \sqrt{5}+\frac{3}{2} \right],\\
&\alpha_{17}=\left[  1,-\frac{1}{6} \sqrt{5} \sqrt{3}-\frac{1}{6} \sqrt{3}-\frac{1}{6} i \sqrt{3}+\frac{1}{6} i \sqrt{3} \sqrt{5} \right],\,
&\alpha_{18}=\left[  1,-\frac{1}{6}\sqrt{5} \sqrt{3}-\frac{1}{6} \sqrt{3}+\frac{1}{6} i \sqrt{3}-\frac{1}{6} i \sqrt{5} \sqrt{3} \right],\\
&\alpha_{19}= \left[ 1,\frac{1}{6} \sqrt{5} \sqrt{3}+\frac{1}{6} \sqrt{3}+\frac{1}{6} i \sqrt{3}-\frac{1}{6} i \sqrt{5} \sqrt{3}\right],\,
&\alpha_{20}= \left[  1,\frac{1}{6} \sqrt{5}\sqrt{3}+\frac{1}{6} \sqrt{3}-\frac{1}{6} i \sqrt{3}+\frac{1}{6} i \sqrt{3} \sqrt{5} \right],\\
\end{eqnarray*}
\end{footnotesize}
and  12  hyperplanes defining reflections of order 5:
\begin{footnotesize}
\begin{eqnarray*}
&\alpha_{21}= \left[ 1,\frac{1}{20} i \sqrt{5} \mu +\frac{1}{4} i \mu +\frac{1}{10} \sqrt{5} \mu \right],\,
&\alpha_{22}= \left[  1,-\frac{1}{20} i \sqrt{5} \mu -\frac{1}{4} i \mu -\frac{1}{10} \sqrt{5} \mu \right],\\
&\alpha_{23}= \left[ 1,-\frac{1}{20} i \sqrt{5} \mu -\frac{1}{4} i \mu +\frac{1}{10} \sqrt{5} \mu  \right],\,
&\alpha_{24}= \left[ 1,\frac{1}{20} i \sqrt{5} \mu +\frac{1}{4} i \mu -\frac{1}{10} \sqrt{5} \mu  \right],\\
&\alpha_{25}= \left[ 1,-\frac{1}{2}\sqrt{5}+\frac{1}{2} \mu +\frac{1}{2} \right],\,
&\alpha_{26}= \left[  1,-\frac{1}{2}\sqrt{5}-\frac{1}{2} \mu +\frac{1}{2} \right],\\
&\alpha_{27}= \left[  1,\frac{1}{2} \sqrt{5}-\frac{1}{2} \mu -\frac{1}{2} \right],\,
&\alpha_{28}= \left[ 1,\frac{1}{2} \sqrt{5}+\frac{1}{2} \mu -\frac{1}{2} \right],\\
&\alpha_{29}= \left[  1,-\frac{1}{16} \nu (i \sqrt{3}+1) (\mu +2) (1+\sqrt{5})  \right],\,
&\alpha_{30}= \left[  1,\frac{1}{16} \nu (i \sqrt{3}+1) (\mu -2) (1+\sqrt{5}) \right],\\
&\alpha_{31}= \left[ 1,\frac{1}{16} \nu (i \sqrt{3}+1) (\mu +2) (1+\sqrt{5})  \right],\,
&\alpha_{32}= \left[ 1,-\frac{1}{16} \nu (i \sqrt{5}+1) (\mu -2) (1+\sqrt{5})  \right],\\
\end{eqnarray*}
\end{footnotesize}
with $$\mu=\sqrt{10-2\sqrt{5}}, \quad \nu=\sqrt{2+2i\sqrt{3}}.$$

\subsection{The $G_{20}$ case}
We have 20  mirrors defining reflections of order 2:
\begin{footnotesize}
\begin{eqnarray*}
&&\alpha_1= \left[   1,-i\sqrt{-2 \sqrt{5}\sqrt{3}+4 \sqrt{3}-3\sqrt{5}+8} \right],\,
\alpha_2= \left[  1,i \sqrt{-2\sqrt{5} \sqrt{3}+4 \sqrt{3}-3\sqrt{5}+8} \right],\\
&&\alpha_3=  \left[ 1,-i \sqrt{2\sqrt{5}\sqrt{3}-4\sqrt{3}-3\sqrt{5}+8} \right],\,
\alpha_4= \left[ 1, i\sqrt{2\sqrt{5}\sqrt{3}-4\sqrt{3}-3\sqrt{5}+8} \right],\\
&&\alpha_5= \left[ 1,\frac{1}{2}i\sqrt{3}+\frac{1}{2}\sqrt{2-2 i \sqrt{3}}-\frac{1}{2}   \right],\,
\alpha_6= \left[  1,\frac{1}{2}i\sqrt{3}-\frac{1}{2}\sqrt{2-2i \sqrt{3}}-\frac{1}{2} \right],\\
&&\alpha_7=  \left[ 1,-\frac{1}{2}i\sqrt{3}+\frac{1}{4}\sqrt{2-2i \sqrt{3}}-\frac{1}{2}+\frac{1}{4}i\sqrt{3} \sqrt{2-2 i \sqrt{3}}  \right],\\
&&\alpha_8= \left[  1,-\frac{1}{2}i\sqrt{3}-\frac{1}{4}\sqrt{2-2i\sqrt{3}}-\frac{1}{2}-\frac{1}{4}i \sqrt{2-2i \sqrt{3}} \sqrt{3}  \right],\\
&&\alpha_9= \left[   1,-\frac{1}{2}\sqrt{5}-\frac{3}{2}+\frac{1}{2}\sqrt{5} \sqrt{3}+\frac{1}{2}\sqrt{3} \right],\,
\alpha_{10}= \left[  1,-\frac{1}{2}\sqrt{5}-\frac{3}{2}-\frac{1}{2}\sqrt{5} \sqrt{3}-\frac{1}{2}\sqrt{3}  \right],\\
&&\alpha_{11}= \left[   1,-\frac{1}{3}\sqrt{6 i+3 \sqrt{5}} \right],\,
\alpha_{12}= \left[  1,\frac{1}{3} \sqrt{6 i+3 \sqrt{5}}    \right],\\
&&\alpha_{13}= \left[   1,-\frac{1}{3}\sqrt{-6 i+3 \sqrt{5}}    \right],\,
\alpha_{14}= \left[   1,\frac{1}{3}\sqrt{-6 i+3 \sqrt{5}} \right], \\
&&\alpha_{15}= \left[   1,\frac{1}{2} i\sqrt{3}-\frac{1}{4}\sqrt{2-2i \sqrt{3}}+\frac{1}{2}-\frac{1}{4}i\sqrt{2-2i \sqrt{3}} \sqrt{3}   \right], \\
&&\alpha_{16}= \left[   1,-\frac{1}{2} i \sqrt{3}-\frac{1}{2}\sqrt{2-2i\sqrt{3}}+\frac{1}{2} \right],\,
\alpha_{17}= \left[   1,-\frac{1}{2} i \sqrt{3}+\frac{1}{2} \sqrt{2-2 i \sqrt{3}}+\frac{1}{2} \right], \\
&&\alpha_{18}= \left[ 1,\frac{1}{2} i \sqrt{3}+\frac{1}{4}\sqrt{2-2i \sqrt{3}}+\frac{1}{2}+\frac{1}{4}i\sqrt{3} \sqrt{2-2 i \sqrt{3}}  \right], \\
&&\alpha_{19}= \left[  1,\frac{1}{2}\sqrt{5}+\frac{3}{2}-\frac{1}{2}\sqrt{5} \sqrt{3}-\frac{1}{2}\sqrt{3} \right],\,
\alpha_{20}= \left[  1,\frac{1}{2}\sqrt{5}+\frac{3}{2}+\frac{1}{2}\sqrt{5} \sqrt{3}+\frac{1}{2}\sqrt{3}\right]. \\
\end{eqnarray*}
\end{footnotesize}

\subsection{The $G_{21}$ case}
We have 30  mirrors defining reflections of order 2:
\begin{eqnarray*}
&\alpha_1= \left[ 1,-\frac{1}{2}\sqrt{5}+\frac{1}{2}+\frac{3}{2} i-\frac{1}{2} i \sqrt{5}  \right],\quad
&\alpha_2= \left[  1,-\frac{1}{2} \sqrt{5}+\frac{1}{2}-\frac{3}{2} i+\frac{1}{2} i \sqrt{5} \right],\\
&\alpha_3= \left[1,1    \right],\quad
&\alpha_4= \left[  1,\frac{1}{6} \sqrt{5}-\frac{1}{6}-\frac{1}{6} i-\frac{1}{6} i \sqrt{5}  \right],\\
&\alpha_5= \left[ 1,\frac{1}{6} \sqrt{5}-\frac{1}{6}+\frac{1}{6} i+\frac{1}{6} i \sqrt{5}  \right],\quad
&\alpha_6= \left[  1,-\frac{3}{10} \sqrt{5}-\frac{1}{2}+\frac{1}{2} i+\frac{1}{10} i \sqrt{5} \right],\\
&\alpha_7= \left[ 1,-\frac{3}{10} \sqrt{5}-\frac{1}{2}-\frac{1}{2} i-\frac{1}{10} i \sqrt{5}   \right],\quad
&\alpha_8= \left[1, -1  \right],\\
&\alpha_9= \left[0,1    \right],\quad
&\alpha_{10}= \left[  1,-\frac{1}{10} \sqrt{5}+\frac{1}{2}+\frac{1}{2} i-\frac{3}{10} i \sqrt{5} \right],\\
&\alpha_{11}= \left[ 1,-\frac{1}{10} \sqrt{5}+\frac{1}{2}-\frac{1}{2} i+\frac{3}{10} i \sqrt{5}   \right],\quad
&\alpha_{12}= \left[  1,-\frac{1}{6} \sqrt{5}+\frac{1}{6}+\frac{1}{6} i+\frac{1}{6} i \sqrt{5} \right],\\
&\alpha_{13}= \left[ 1,-\frac{1}{6} \sqrt{5}+\frac{1}{6}-\frac{1}{6} i-\frac{1}{6} i \sqrt{5}   \right],\quad
&\alpha_{14}= \left[  1,\frac{3}{10}\sqrt{5}+\frac{1}{2}-\frac{1}{2} i-\frac{1}{10} i \sqrt{5}  \right],\\
&\alpha_{15}= \left[ 1,\frac{3}{10} \sqrt{5}+\frac{1}{2}+\frac{1}{2} i+\frac{1}{10} i \sqrt{5} \right],\quad
&\alpha_{16}= \left[  1,\frac{1}{2} \sqrt{5}-\frac{1}{2}-\frac{1}{2} i-\frac{1}{2} i \sqrt{5}  \right],\\
&\alpha_{17}= \left[ 1,\frac{1}{2} \sqrt{5}-\frac{1}{2}+\frac{1}{2} i+\frac{1}{2} i \sqrt{5}  \right],\quad
&\alpha_{18}= \left[  1,\frac{1}{2}\sqrt{5}+\frac{3}{2}+\frac{1}{2} i+\frac{1}{2} i \sqrt{5}  \right],\\
&\alpha_{19}= \left[ 1,\frac{1}{2} \sqrt{5}+\frac{3}{2}-\frac{1}{2} i-\frac{1}{2} i \sqrt{5}  \right],\quad
&\alpha_{20}= \left[ 1,-\frac{1}{2}\sqrt{5}-\frac{3}{2}-\frac{1}{2} i-\frac{1}{2} i \sqrt{5}  \right],\\
&\alpha_{21}= \left[ 1,-\frac{1}{2}\sqrt{5}-\frac{3}{2}+\frac{1}{2} i+\frac{1}{2} i \sqrt{5}  \right],\quad
&\alpha_{22}= \left[  1,-\frac{1}{2} \sqrt{5}+\frac{1}{2}+\frac{1}{2} i+\frac{1}{2} i \sqrt{5}\right],\\
&\alpha_{23}= \left[  1,-\frac{1}{2} \sqrt{5}+\frac{1}{2}-\frac{1}{2} i-\frac{1}{2} i \sqrt{5}  \right],\quad
&\alpha_{24}= \left[ 1,0   \right],\\
&\alpha_{25}= \left[  1, -i  \right],\quad
&\alpha_{26}= \left[ 1, i   \right],\\
&\alpha_{27}= \left[  1,\frac{1}{2} \sqrt{5}-\frac{1}{2}+\frac{3}{2} i-\frac{1}{2} i \sqrt{5}  \right],\quad
&\alpha_{28}= \left[ 1,\frac{1}{2} \sqrt{5}-\frac{1}{2}-\frac{3}{2} i+\frac{1}{2} i \sqrt{5}   \right],\\
&\alpha_{29}= \left[  1,-\frac{1}{2}+\frac{1}{2} i+\frac{1}{10} \sqrt{5}-\frac{3}{10} i \sqrt{5}  \right],\quad
&\alpha_{30}= \left[ 1,-\frac{1}{2}-\frac{1}{2} i+\frac{1}{10} \sqrt{5}+\frac{3}{10} i \sqrt{5}   \right],\\
\end{eqnarray*}
and 20  mirrors defining reflections of order 3:
\begin{footnotesize}
\begin{eqnarray*}
&\alpha_{31}= \left[  1,\frac{1}{2}+\frac{1}{2} i-\frac{1}{2} \sqrt{6i}  \right],\,
&\alpha_{32}= \left[ 1,\frac{1}{2}+\frac{1}{2} i+\frac{1}{2} \sqrt{6i}  \right],\\
&\alpha_{33}= \left[  1,-\frac{1}{2} i \sqrt{6i}+\frac{1}{2}-\frac{1}{2} i  \right],\,
&\alpha_{34}= \left[1,\frac{1}{2} i \sqrt{6i}+\frac{1}{2}-\frac{1}{2} i    \right],\\
&\alpha_{35}= \left[  1,-\frac{1}{2} \sqrt{5}-\frac{3}{2}-\frac{1}{2} \sqrt{5}\sqrt{3}-\frac{1}{2} \sqrt{3}  \right],\,
&\alpha_{36}= \left[ 1,-\frac{3}{2}-\frac{1}{2}\sqrt{5}+\frac{1}{2} \sqrt{5} \sqrt{3}+\frac{1}{2} \sqrt{3} \right],\\
&\alpha_{37}= \left[ 1, \sigma   \right],\,
&\alpha_{38}= \left[ 1,-\nu \sqrt{5}+\sigma^3  \right],\\
&\alpha_{39}= \left[ 1,-\sigma^3+\nu \sqrt{5} \right],\,
&\alpha_{40}= \left[ 1,-\frac{1}{6} \sqrt{5} \sqrt{3}-\frac{1}{6} \sqrt{3}-\frac{1}{6} i \sqrt{5} \sqrt{3}+\frac{1}{6} i \sqrt{3}  \right],\\
&\alpha_{41}= \left[  1,-\frac{1}{2} i \sqrt{6i}-\frac{1}{2}+\frac{1}{2} i \right],\,
&\alpha_{42}= \left[  1,\frac{1}{2} i \sqrt{6i}-\frac{1}{2}+\frac{1}{2} i  \right],\\
&\alpha_{43}= \left[  1,-\frac{1}{2}-\frac{1}{2} i+\frac{1}{2} \sqrt{6i}\right],\,
&\alpha_{44}= \left[ 1,-\frac{1}{2}-\frac{1}{2} i-\frac{1}{2} \sqrt{6 i}  \right],\\
&\alpha_{45}= \left[  1,-\mu^3 +\lambda \sqrt{5}+24 i-8 i \sqrt{5}-8 i \sqrt{5} \sqrt{3}+8 i \sqrt{3} \right],\,
&\alpha_{46}= \left[  1, \mu  \right],\\
&\alpha_{47}= \left[ 1,\mu^3 -\lambda \sqrt{5}-24 i+8 i \sqrt{5}+8 i \sqrt{5} \sqrt{3}-8 i \sqrt{3}   \right],\,
&\alpha_{48}= \left[  1,\frac{3}{2} i-\frac{1}{2} i \sqrt{5}-\frac{1}{2} i \sqrt{5} \sqrt{3}+\frac{1}{2} i \sqrt{3} \right],\\
&\alpha_{49}= \left[  1,\frac{1}{2}\sqrt{5}+\frac{3}{2}-\frac{1}{2} \sqrt{5} \sqrt{3}-\frac{1}{2}\sqrt{3}  \right],\,
&\alpha_{50}= \left[1,\frac{1}{2} \sqrt{5}+\frac{3}{2}+\frac{1}{2} \sqrt{5} \sqrt{3}+\frac{1}{2} \sqrt{3} \right],\\
\end{eqnarray*}
\end{footnotesize}
with 
\begin{footnotesize}
$$\mu =-\frac{3}{2}i+\frac{1}{2}i\sqrt{5}+\frac{1}{2}i\sqrt{5}\sqrt{3}-\frac{1}{2}i\sqrt{3},\,\nu=\frac{1}{9}\sqrt{5}\sqrt{3}+\frac{1}{9}\sqrt{3}+\frac{1}{9}i\sqrt{5}\sqrt{3}-\frac{1}{9}i\sqrt{3},\,\sigma=\frac{3}{2}\nu,\, \lambda=6\mu.$$
\end{footnotesize}

\subsection{The $G_{23}$ case}
We have 15  mirrors defining reflections of order 2:
\begin{footnotesize}
\begin{eqnarray*}
&&\alpha_1=\left[1,\, 0,\, 0\right]\quad\alpha_2=\left[0,\, 1,\, 0\right]\quad\alpha_3=\left[0,\, 0,\, 1\right]\\
&&\alpha_4=\left[-2,\,\sqrt {5}+3,\,-\sqrt {5}-1\right],\quad\alpha_5=\left[-2,\,\sqrt {5}-1,\,-\sqrt {5}+3\right],\quad
\alpha_6= \left[2,\,\sqrt {5}-1,\,\sqrt {5}-3\right]\\
&&\alpha_7=\left[-2,\,\sqrt {5}-1,\,\sqrt {5}-3\right]\quad,\alpha_8=\left[2,\,\sqrt {5}+3,\,-\sqrt {5}-1\right],\quad
\alpha_9=\left[-2,\,\sqrt {5}+3,\,\sqrt {5}+1\right]\\
&&\alpha_{10}=\left[2,\,\sqrt {5}-1,\,-\sqrt {5}+3\right],\quad\alpha_{11}=\left[2,\,\sqrt {5}+3,\,\sqrt {5}+1\right],\quad
\alpha_{12}=\left[2,\,\sqrt {5}-1,\,-\sqrt {5}-1\right]\\
&&\alpha_{13}=\left[-2,\,\sqrt {5}-1,\,-\sqrt {5}-1\right],\quad
\alpha_{14}=\left[2,\,\sqrt {5}-1,\,\sqrt {5}+1\right],\quad\alpha_{15}=\left[-2,\,\sqrt {5}-1,\,\sqrt {5}+1\right]
\end{eqnarray*}
\end{footnotesize}
\subsection{The $G_{24}$ case}
We have 21  mirrors defining reflections of order 2:
\begin{footnotesize}
\begin{eqnarray*}
\alpha_1&=& \left[1,-\f{1}{9}\mu^2-\f{1}{9}\mu+\f{11}{9},\f{1}{9}\mu^2+\f{1}{9}\mu-\f{20}{9}\right]\\
\alpha_2&=&\left[1,\f{1}{3}\mu-\f{1}{3},-\f{1}{3}\mu-\f{2}{3}\right]\\
\alpha_3&=& \left[1,\f{1}{9}\mu^2-\f{2}{9}\mu-\f{17}{9},-\f{1}{9}\mu^2+\f{2}{9}\mu+\f{8}{9}\right]\\
\alpha_4&=&\left[1,\f{1}{27}\nu\mu^2-\f{2}{27}\nu\mu+\f{1}{6}\mu-\f{8}{27}\nu+\f{5}{6},\f{1}{2}+\f{1}{27}\nu\mu^2-\f{17}{27}\nu-\f{2}{27}\nu\mu\right]\\
\alpha_5&=&\left[1,-\f{1}{27}\nu\mu^2-\f{2}{27}\nu\mu+\f{1}{6}\mu+\f{8}{27}\nu+\f{5}{6},\f{1}{2}-\f{1}{27}\nu\mu^2+\f{17}{27}\nu+\f{2}{27}\nu\mu\right]\\
\alpha_6&=&\left[1,\f{1}{27}\nu\mu^2-\f{1}{18}\mu^2-\f{5}{27}\nu\mu-\f{1}{18}\mu+\f{4}{27}\nu
+\f{29}{18},-\f{4}{9}\nu\mu+\f{1}{9}\nu\mu^2-\f{1}{3}\nu+\f{1}{2}\right]\\
\alpha_7&=&\left[1,-\f{1}{27}\nu\mu^2-\f{1}{18}\mu^2+\f{5}{27}\nu\mu-\f{1}{18}\mu-\f{4}{27}\nu
+\f{29}{18},\f{4}{9}\nu\mu-\f{1}{9}\nu\mu^2+\f{1}{3}\nu+\f{1}{2}\right]\\
\alpha_8&=&\left[1,\f{1}{27}\nu\mu^2+\f{1}{18}\mu^2-\f{5}{27}\nu\mu-\f{1}{9}\mu-\f{5}{27}\nu
+\f{1}{18},\f{2}{27}\nu\mu-\f{1}{27}\nu\mu^2-\f{1}{27}\nu+\f{1}{2}\right]\\
\alpha_9&=&\left[1,-\f{1}{27}\nu\mu^2+\f{1}{18}\mu^2+\f{5}{27}\nu\mu-\f{1}{9}\mu+\f{5}{27}\nu
+\f{1}{18},-\f{2}{27}\nu\mu+\f{1}{27}\nu\mu^2+\f{1}{27}\nu+\f{1}{2}\right]\\
\alpha_{10}&=&\left[1,\f{2}{27}\nu\mu^2+\f{1}{18}\mu^2-\f{7}{27}\nu\mu+\f{1}{18}\mu-\f{13}{27}\nu
-\f{10}{9},-\f{1}{27}\nu\mu^2+\f{1}{18}\mu^2+\f{5}{27}\mu\nu-\f{5}{18}\mu-\f{4}{27}\nu-\f{5}{18}\right]\\
\alpha_{11}&=&\left[1,-\f{2}{27}\nu\mu^2+\f{1}{18}\mu^2+\f{7}{27}\nu\mu+\f{1}{18}\mu+\f{13}{27}\nu
-\f{10}{9},\f{1}{27}\nu\mu^2+\f{1}{18}\mu^2-\f{5}{27}\mu\nu-\f{5}{18}\mu+\f{4}{27}\nu-\f{5}{18}\right]\\
\alpha_{12}&=& \left[1,-\f{1}{6}\mu-\f{1}{3}+\f{1}{3}\nu,\f{1}{27}\nu\mu^2-\f{1}{9}\mu^2-\f{2}{27}\nu\mu+\f{1}{18}\mu-\f{8}{27}\nu+\f{37}{18}\right]\\
\alpha_{13}&=&\left[1,-\f{1}{6}\mu-\f{1}{3}-\f{1}{3}\nu,-\f{1}{27}\nu\mu^2-\f{1}{9}\mu^2+\f{2}{27}\nu\mu+\f{1}{18}\mu+\f{8}{27}\nu+\f{37}{18}\right]\\
\alpha_{14}&=&\left[1,-\f{2}{27}\nu\mu^2-\f{1}{18}\mu^2+\f{7}{27}\nu\mu+\f{1}{9}\mu+\f{4}{27}\nu
+\f{4}{9},-\f{1}{27}\nu\mu^2+\f{1}{18}\mu^2+\f{5}{27}\mu\nu+\f{2}{9}\mu+\f{5}{27}\nu-\f{5}{18}\right]\\
\alpha_{15}&=&\left[1,\f{2}{27}\nu\mu^2-\f{1}{18}\mu^2-\f{7}{27}\nu\mu+\f{1}{9}\mu-\f{4}{27}\nu
+\f{4}{9},\f{1}{27}\nu\mu^2+\f{1}{18}\mu^2-\f{5}{27}\mu\nu+\f{2}{9}\mu-\f{5}{27}\nu-\f{5}{18}\right]\\
\alpha_{16}&=&\left[1,\f{1}{18}\mu^2+\f{1}{9}\nu\mu+\f{1}{18}\mu-\f{1}{9}\nu-\f{10}{9},\f{1}{18}\mu^2-\f{1}{9}\nu\mu-\f{1}{9}\mu+\f{4}{9}\nu-\f{13}{9}\right]\\
\alpha_{17}&=&\left[1,\f{1}{18}\mu^2-\f{1}{9}\nu\mu+\f{1}{18}\mu+\f{1}{9}\nu-\f{10}{9},\f{1}{18}\mu^2+\f{1}{9}\nu\mu-\f{1}{9}\mu-\f{4}{9}\nu-\f{13}{9}\right]\\
\alpha_{18}&=&\left[1,\f{2}{27}\nu\mu^2-\f{10}{27}\nu\mu-\f{1}{6}\mu-\f{1}{27}\nu-\f{1}{3},-\f{1}{18}\mu^2-\f{1}{18}\mu+\f{1}{9}\mu\nu-\f{1}{9}\nu+\f{1}{9}\right]\\
\alpha_{19}&=&\left[1,-\f{2}{27}\nu\mu^2+\f{10}{27}\nu\mu-\f{1}{6}\mu+\f{1}{27}\nu-\f{1}{3},-\f{1}{18}\mu^2-\f{1}{18}\mu-\f{1}{9}\mu\nu+\f{1}{9}\nu+\f{1}{9}\right]\\
\alpha_{20}&=&\left[1,-\f{1}{18}\mu^2-\f{1}{9}\nu\mu+\f{1}{9}\mu+\f{4}{9}\nu+\f{4}{9},-\f{2}{27}\nu\mu^2+\f{10}{27}\nu\mu+\f{1}{6}\mu+\f{1}{27}\nu-\f{2}{2}\right]\\
\alpha_{21}&=&\left[1,-\f{1}{18}\mu^2+\f{1}{9}\nu\mu+\f{1}{9}\mu+\f{4}{9}\nu-\f{4}{9},\f{2}{27}\nu\mu^2-\f{10}{27}\nu\mu+\f{1}{6}\mu-\f{1}{27}\nu-\f{2}{2}\right]\\
\end{eqnarray*}
\end{footnotesize}
where $\mu$ is a root of $z^3-21z-7=0$ and $\nu=\pm\sqrt{\f{\mu^2}{4}-2\mu-14}$.

\subsection{The $G_{25}$ case}
We have 12  mirrors defining reflections of order 2:
\begin{eqnarray*}
&&\alpha_1= \left[1,0,0\right],\,\alpha_2=\left[0,1,0\right],\,\alpha_3= \left[0,0,1\right],\\
&&\alpha_4=\left[1,1,1\right]\,\alpha_5=\left[-2,1+\mu,1-\mu\right],\,\alpha_6=\left[2,-1+\mu-1-\mu\right],\\
&&\alpha_7=\left[-2,-2,1+\mu\right],\,\alpha_8=\left[2,2,-1+\mu\right],\,
\alpha_9=\left[-2,1+\mu,-2\right],\\
&&\alpha_{10}=\left[2,-1+\mu,2\right],\,\alpha_{11}=\left[2,-1+\mu,-1+\mu\right],\,
\alpha_{12}= \left[-2,1+\mu,1+\mu\right]\\
\end{eqnarray*}
where $\mu=\pm i\sqrt{3}$. 

\subsection{The $G_{26}$ case}
We have 9  mirrors defining reflections of order 2:
\begin{eqnarray*}
&&\alpha_1=\left[0,1,-\f{1}{2}\mu+\f{1}{2}\right],\quad
\alpha_2=\left[0,1,\f{1}{2}\mu+\f{1}{2}\right],\quad
\alpha_3=\left[1,0,\f{1}{2}\mu+\f{1}{2}\right],\\
&&\alpha_4=\left[1,0,-\f{1}{2}\mu+\f{1}{2}\right],\quad
\alpha_5=\left[1,-\f{1}{2}\mu+\f{1}{2},0\right],\quad
\alpha_6=\left[1,\f{1}{2}\mu+\f{1}{2},0\right],\\
&&\alpha_7=\left[1,-1,0\right],\quad
\alpha_8=\left[1,0,-1\right],\quad
\alpha_9=\left[0,1,-1\right],\\
\end{eqnarray*}
and 12  hyperplanes defining reflections of order 3:
\begin{eqnarray*}
&&\alpha_{10}=\left[1,0,0\right],\,
\alpha_{11}=\left[0,1,0\right],\,
\alpha_{12}=\left[0,0,1\right],\\
&&\alpha_{13}=\left[1,1,1\right],\,
\alpha_{14}=\left[1,\f{1}{2}\mu-\f{1}{2},-\f{1}{2}\mu-\f{1}{2}\right],\,
\alpha_{15}=\left[1,-\f{1}{2}\mu-\f{1}{2},\f{1}{2}\mu-\f{1}{2}\right],\\
&&\alpha_{16}=\left[1,\f{1}{2}\mu-\f{1}{2},1\right],\,
\alpha_{17}=\left[1,-\f{1}{2}\mu-\f{1}{2},1\right],\,
\alpha_{18}=\left[1,1,\f{1}{2}\mu-\f{1}{2}\right],\\
&&\alpha_{19}=\left[1,1,-\f{1}{2}\mu-\f{1}{2}\right],\,
\alpha_{20}=\left[1,\f{1}{2}\mu-\f{1}{2},\f{1}{2}\mu-\f{1}{2}\right],\,
\alpha_{21}=\left[1,-\f{1}{2}\mu-\f{1}{2},-\f{1}{2}\mu-\f{1}{2}\right]
\end{eqnarray*}
where $\mu=\pm i\sqrt{3}$.

\subsection{The $G_{27}$ case}
We have 45 mirrors defining reflections of order 2:
\begin{eqnarray*}
\alpha_{{1}}&=& \left[1,\,-1,\,0\right]\\
\alpha_{{2}}&=& \left[-384,\,{\mu}^{3}-{\mu}^{2}-165\mu+2\,\sigma-291,\,0\right]\\ 
\alpha_{{3}}&=&\left[-384,\,{\mu}^{3}-{\mu}^{2}-165\mu-2\,\sigma-291,\,0\right]\\ 
\alpha_{{4}}&=&\left[-36864,\,{\mu}^{3}\sigma-96\,{
\mu}^{3}-{\mu}^{2}\sigma+96\,{\mu}^{2}-165\,\mu\,\sigma+15840\,\mu-291
\,\sigma+9504,\,0\right]\\ 
\alpha_{{5}}&=&\left[36864,\,{\mu}^{3}\sigma+96\,{
\mu}^{3}-{\mu}^{2}\sigma-96\,{\mu}^{2}-165\,\mu\,\sigma-15840\,\mu-291
\,\sigma-9504,\,0\right]\\ 
\alpha_{{6}}&=&\left[36864,\,36864,\,\right.\\
&&\left.{\mu}^{3}\sigma-
288\,{\mu}^{3}-9\,{\mu}^{2}\sigma+288\,{\mu}^{2}-117\,\mu\,\sigma+
47520\,\mu+405\,\sigma+28512\right]\\ 
\alpha_{{7}}&=&\left[-36864,\,-36864,\,\right.\\
&&\left.\mu^{3}\sigma
+288\,{\mu}^{3}-9\,{\mu}^{2}\sigma-288\,{\mu}^{2}-117\,\mu\,\sigma-
47520\,\mu+405\,\sigma-28512\right]\\ 
\alpha_{{8}}&=&\left[-36864,\,{\mu}^{3}\sigma+96\,{
\mu}^{3}-{\mu}^{2}\sigma-96\,{\mu}^{2}-165\,\mu\,\sigma-15840\,\mu-291
\,\sigma-9504,\,\right.\\
&&\left.-3\,{\mu}^{3}\sigma-288\,{\mu}^{3}+3\,{\mu}^{2}\sigma+
1440\,{\mu}^{2}+447\,\mu\,\sigma+36000\,\mu+441\,\sigma-2592\right]\\
\alpha_{{9}}&=&\left[36864,\,{\mu}^{3}\sigma-96\,{
\mu}^{3}-{\mu}^{2}\sigma+96\,{\mu}^{2}-165\,\mu\,\sigma+15840\,\mu-291
\,\sigma+9504,\,\right.\\
&&\left.-3\,{\mu}^{3}\sigma+288\,{\mu}^{3}+3\,{\mu}^{2}\sigma-
1440\,{\mu}^{2}+447\,\mu\,\sigma-36000\,\mu+441\,\sigma+2592\right]\\
\alpha_{{10}}&=&\left[36864,\,{\mu}^{3}\sigma-96\,{
\mu}^{3}-{\mu}^{2}\sigma+96\,{\mu}^{2}-165\,\mu\,\sigma+15840\,\mu-291
\,\sigma+9504,\,\right.\\
&&\left.-288\,{\mu}^{3}+1440\,{\mu}^{2}+48\,\mu\,\sigma+36000\,
\mu-144\,\sigma-57888\right]\\ 
\alpha_{{11}}&=&\left[-36864,\,{\mu}^{3}\sigma+96\,{
\mu}^{3}-{\mu}^{2}\sigma-96\,{\mu}^{2}-165\,\mu\,\sigma-15840\,\mu-291
\,\sigma-9504,\,\right.\\
&&\left.288\,{\mu}^{3}-1440\,{\mu}^{2}+48\,\mu\,\sigma-36000\,
\mu-144\,\sigma+57888\right]\\
\alpha_{{12}}&=&\left[36864,\,{\mu}^{3}\sigma-96\,{
\mu}^{3}-{\mu}^{2}\sigma+96\,{\mu}^{2}-165\,\mu\,\sigma+15840\,\mu-291
\,\sigma+9504,\,\right.\\
&&\left.4\,{\mu}^{3}\sigma-96\,{\mu}^{3}+864\,{\mu}^{2}-588\,\mu
\,\sigma+15840\,\mu-1224\,\sigma+2592\right]\\ 
\alpha_{{13}}&=&\left[-36864,\,{\mu}^{3}\sigma+96\,{
\mu}^{3}-{\mu}^{2}\sigma-96\,{\mu}^{2}-165\,\mu\,\sigma-15840\,\mu-291
\,\sigma-9504,\,\right.\\
&&\left.4\,{\mu}^{3}\sigma+96\,{\mu}^{3}-864\,{\mu}^{2}-588\,\mu
\,\sigma-15840\,\mu-1224\,\sigma-2592\right]\\ 
\alpha_{{14}}&=&\left[-36864,\,{\mu}^{3}\sigma+96\,{
\mu}^{3}-{\mu}^{2}\sigma-96\,{\mu}^{2}-165\,\mu\,\sigma-15840\,\mu-291
\,\sigma-9504,\,\right.\\
&&\left.-{\mu}^{3}\sigma+192\,{\mu}^{3}-3\,{\mu}^{2}\sigma+576\,
{\mu}^{2}+93\,\mu\,\sigma-31680\,\mu+351\,\sigma-136512\right]\\ 
\alpha_{{15}}&=&\left[36864,\,{\mu}^{3}\sigma-96\,{
\mu}^{3}-{\mu}^{2}\sigma+96\,{\mu}^{2}-165\,\mu\,\sigma+15840\,\mu-291
\,\sigma+9504,\,\right.\\
&&\left.-{\mu}^{3}\sigma-192\,{\mu}^{3}-3\,{\mu}^{2}\sigma-576\,
{\mu}^{2}+93\,\mu\,\sigma+31680\,\mu+351\,\sigma+136512\right]\\ 
\alpha_{{16}}&=&\left[768,\,2\,{\mu}^{3}-2\,{\mu}^{2
}-330\,\mu+4\,\sigma-582,\,8\,{\mu}^{3}-\mu\,\sigma-1080\,\mu-9\,\sigma-
2736\right]\\ 
\alpha_{{17}}&=&\left[768,\,2\,{\mu}^{3}-2\,{\mu}^{2
}-330\,\mu-4\,\sigma-582,\,8\,{\mu}^{3}+\mu\,\sigma-1080\,\mu+9\,\sigma-
2736\right]\\ 
\alpha_{{18}}&=&\left[-18432,\,-18432,\,\right.\\
&&\left.{\mu}^{3}
\sigma-144\,{\mu}^{3}+3\,{\mu}^{2}\sigma+144\,{\mu}^{2}-141\,\mu\,
\sigma+23760\,\mu-495\,\sigma+41904\right]\\ 
\end{eqnarray*}
\begin{eqnarray*} 
\alpha_{{19}}&=&\left[18432,\,18432,\,\right.\\
&&\left.{\mu}^{3}\sigma+
144\,{\mu}^{3}+3\,{\mu}^{2}\sigma-144\,{\mu}^{2}-141\,\mu\,\sigma-
23760\,\mu-495\,\sigma-41904\right]\\ 
\alpha_{{20}}&=&\left[-12288,\,-32\,{\mu}^{3}+32\,{
\mu}^{2}+5280\,\mu+64\,\sigma+9312,\,\right.\\
&&\left.{\mu}^{3}\sigma+128\,{\mu}^{3}-{\mu
}^{2}\sigma-149\,\mu\,\sigma-17280\,\mu-339\,\sigma-25344\right]\\ 
\alpha_{{21}}&=&\left[12288,\,32\,{\mu}^{3}-32\,{\mu
}^{2}-5280\,\mu+64\,\sigma-9312,\,\right.\\
&&\left.{\mu}^{3}\sigma-128\,{\mu}^{3}-{\mu}^{
2}\sigma-149\,\mu\,\sigma+17280\,\mu-339\,\sigma+25344\right]\\ 
\alpha_{{22}}&=&\left[-18432,\,-48\,{\mu}^{3}+48\,{
\mu}^{2}+7920\,\mu-96\,\sigma+13968,\,\right.\\
&&\left.{\mu}^{3}\sigma-240\,{\mu}^{3}-3\,
{\mu}^{2}\sigma+432\,{\mu}^{2}-129\,\mu\,\sigma+33840\,\mu-189\,\sigma
+11664\right]\\ 
\alpha_{{23}}&=&\left[18432,\,48\,{\mu}^{3}-48\,{\mu
}^{2}-7920\,\mu-96\,\sigma-13968,\,\right.\\
&&\left.{\mu}^{3}\sigma+240\,{\mu}^{3}-3\,{
\mu}^{2}\sigma-432\,{\mu}^{2}-129\,\mu\,\sigma-33840\,\mu-189\,\sigma-
11664\right]\\ 
\alpha_{{24}}&=&\left[-18432,\,-48\,{\mu}^{3}+48\,{
\mu}^{2}+7920\,\mu+96\,\sigma+13968,\,\right.\\
&&\left.{\mu}^{3}\sigma+96\,{\mu}^{3}-3\,{
\mu}^{2}\sigma-288\,{\mu}^{2}-129\,\mu\,\sigma-10080\,\mu+99\,\sigma-
25056\right]\\ 
\alpha_{{25}}&=&\left[18432,\,48\,{\mu}^{3}-48\,{\mu
}^{2}-7920\,\mu+96\,\sigma-13968,\,\right.\\
&&\left.{\mu}^{3}\sigma-96\,{\mu}^{3}-3\,{\mu
}^{2}\sigma+288\,{\mu}^{2}-129\,\mu\,\sigma+10080\,\mu+99\,\sigma+
25056\right]\\ 
\alpha_{{26}}&=&\left[-9216,\,{\mu}^{2}\sigma+96\,{
\mu}^{2}+6\,\mu\,\sigma-576\,\mu+21\,\sigma+864,\,\right.\\
&&\left.24\,{\mu}^{3}+72\,{\mu}^{2}+12\,\mu\,\sigma-3384\,\mu-36\,\sigma-4968\right]\\ 
\alpha_{{27}}&=&\left[-9216,\,-{\mu}^{2}\sigma+96\,{
\mu}^{2}-6\,\mu\,\sigma-576\,\mu-21\,\sigma+864,\,\right.\\
&&\left.24\,{\mu}^{3}+72\,{\mu
}^{2}-12\,\mu\,\sigma-3384\,\mu+36\,\sigma-4968\right]\\ 
\alpha_{{28}}&=&\left[-96,\,{\mu}^{3}-3\,{\mu}^{2}-
153\,\mu-21,\,-{\mu}^{3}+3\,{\mu}^{2}+153\,\mu+117\right]\\ 
\alpha_{{29}}&=&\left[-192,\,{\mu}^{3}+3\,{\mu}^{2}-
189\,\mu-447,\,-{\mu}^{3}-3\,{\mu}^{2}+189\,\mu+639\right]\\ 
\alpha_{{30}}&=&\left[36864,\,2\,{\mu}^{3}\sigma-96
\,{\mu}^{3}+2\,{\mu}^{2}\sigma+480\,{\mu}^{2}-306\,\mu\,\sigma+13536\,
\mu-1074\,\sigma-42336,\,\right.\\
&&\left.3\,{\mu}^{3}\sigma+96\,{\mu}^{3}-3\,{\mu}^{2}
\sigma+288\,{\mu}^{2}-447\,\mu\,\sigma-13536\,\mu-441\,\sigma-75168\right]\\ 
\alpha_{{31}}&=& \left[-36864,\,2\,{\mu}^{3}\sigma+96
\,{\mu}^{3}+2\,{\mu}^{2}\sigma-480\,{\mu}^{2}-306\,\mu\,\sigma-13536\,
\mu-1074\,\sigma+42336,\,\right.\\
&&\left.3\,{\mu}^{3}\sigma-96\,{\mu}^{3}-3\,{\mu}^{2}
\sigma-288\,{\mu}^{2}-447\,\mu\,\sigma+13536\,\mu-441\,\sigma+75168\right]\\ 
\alpha_{{32}}&=& \left[36864,\,3\,{\mu}^{3}\sigma-192
\,{\mu}^{3}-3\,{\mu}^{2}\sigma-192\,{\mu}^{2}-447\,\mu\,\sigma+29376\,
\mu-633\,\sigma+103104,\,\right.\\
&&\left.-{\mu}^{3}\sigma+192\,{\mu}^{3}-3\,{\mu}^{2}
\sigma-576\,{\mu}^{2}+141\,\mu\,\sigma-29376\,\mu+783\,\sigma+32832\right]\\ 
\alpha_{{33}}&=&\left[-36864,\,3\,{\mu}^{3}\sigma+
192\,{\mu}^{3}-3\,{\mu}^{2}\sigma+192\,{\mu}^{2}-447\,\mu\,\sigma-
29376\,\mu-633\,\sigma-103104,\,\right.\\
&&\left.-{\mu}^{3}\sigma-192\,{\mu}^{3}-3\,{\mu}
^{2}\sigma+576\,{\mu}^{2}+141\,\mu\,\sigma+29376\,\mu+783\,\sigma-
32832\right]\\ 
\alpha_{{34}}&=&\left[36864,\,384\,{\mu}^{2}-48\,\mu
\,\sigma+2304\,\mu-48\,\sigma+8064,\,\right.\\
&&\left.{\mu}^{3}\sigma+96\,{\mu}^{3}+3\,{
\mu}^{2}\sigma+288\,{\mu}^{2}-141\,\mu\,\sigma-18144\,\mu-207\,\sigma-
6048\right]\\ 
\alpha_{{35}}&=&\left[-36864,\,-384\,{\mu}^{2}-48\,
\mu\,\sigma-2304\,\mu-48\,\sigma-8064,\,\right.\\
&&\left.{\mu}^{3}\sigma-96\,{\mu}^{3}+3
\,{\mu}^{2}\sigma-288\,{\mu}^{2}-141\,\mu\,\sigma+18144\,\mu-207\,
\sigma+6048\right]\\ 
\alpha_{{36}}&=&\left[-4,\,\mu+1,\,-\mu+3\right]\\ 
\end{eqnarray*}
\begin{eqnarray*} 
\alpha_{{37}}&=&\left[64,{\mu}^{3}-{\mu}^{2}-149\mu-211,
-{\mu}^{3}+{\mu}^{2}+149\,\mu+147\right]\\ 
\alpha_{{38}}&=&\left[-18432,\,{\mu}^{3}\sigma+48\,{
\mu}^{3}-3\,{\mu}^{2}\sigma-48\,{\mu}^{2}-153\,\mu\,\sigma-5616\,\mu-
21\,\sigma+6768,\,\right.\\
&&\left.96\,{\mu}^{3}-288\,{\mu}^{2}+24\,\mu\,\sigma-12384\,
\mu+216\,\sigma+9504\right]\\ 
\alpha_{{39}}&=&\left[18432,\,{\mu}^{3}\sigma-48\,{
\mu}^{3}-3\,{\mu}^{2}\sigma+48\,{\mu}^{2}-153\,\mu\,\sigma+5616\,\mu-
21\,\sigma-6768,\,\right.\\
&&\left.-96\,{\mu}^{3}+288\,{\mu}^{2}+24\,\mu\,\sigma+12384\,
\mu+216\,\sigma-9504\right]\\ 
\alpha_{{40}}&=&\left[-36864,\,{\mu}^{3}\sigma-192\,
{\mu}^{3}+3\,{\mu}^{2}\sigma+192\,{\mu}^{2}-189\,\mu\,\sigma+27072\,
\mu-447\,\sigma+69696,\,\right.\\
&&\left.-3\,{\mu}^{3}\sigma-192\,{\mu}^{3}+3\,{\mu}^{2}
\sigma+576\,{\mu}^{2}+447\,\mu\,\sigma+24768\,\mu+1017\,\sigma+36288\right]\\ 
\alpha_{{41}}&=&\left[36864,\,{\mu}^{3}\sigma+192\,{
\mu}^{3}+3\,{\mu}^{2}\sigma-192\,{\mu}^{2}-189\,\mu\,\sigma-27072\,\mu
-447\,\sigma-69696,\,\right.\\
&&\left.-3\,{\mu}^{3}\sigma+192\,{\mu}^{3}+3\,{\mu}^{2}
\sigma-576\,{\mu}^{2}+447\,\mu\,\sigma-24768\,\mu+1017\,\sigma-36288\right]\\ 
\alpha_{{42}}&=& \left[-36864,\,{\mu}^{3}\sigma-288\,
{\mu}^{3}-{\mu}^{2}\sigma+672\,{\mu}^{2}-117\,\mu\,\sigma+40608\,\mu+
141\,\sigma-9504,\,\right.\\
&&\left.-2\,{\mu}^{3}\sigma+6\,{\mu}^{2}\sigma+306\,\mu\,
\sigma-4608\,\mu+234\,\sigma-41472\right]\\ 
\alpha_{{43}}&=&\left[36864,\,{\mu}^{3}\sigma+288\,{
\mu}^{3}-{\mu}^{2}\sigma-672\,{\mu}^{2}-117\,\mu\,\sigma-40608\,\mu+
141\,\sigma+9504,\,\right.\\
&&\left.-2\,{\mu}^{3}\sigma+6\,{\mu}^{2}\sigma+306\,\mu\,
\sigma+4608\,\mu+234\,\sigma+41472\right]\\ 
\alpha_{{44}}&=&\left[36864,\,2\,{\mu}^{3}\sigma-96
\,{\mu}^{3}-2\,{\mu}^{2}\sigma+480\,{\mu}^{2}-282\,\mu\,\sigma+8928\,
\mu-726\,\sigma+26784,\,\right.\\
&&\left.{\mu}^{3}\sigma-288\,{\mu}^{3}+3\,{\mu}^{2}
\sigma+288\,{\mu}^{2}-189\,\mu\,\sigma+42912\,\mu-639\,\sigma+97632\right]\\ 
\alpha_{{45}}&=&\left[-36864,\,2\,{\mu}^{3}\sigma+96
\,{\mu}^{3}-2\,{\mu}^{2}\sigma-480\,{\mu}^{2}-282\,\mu\,\sigma-8928\,
\mu-726\,\sigma-26784,\,\right.\\
&&\left.{\mu}^{3}\sigma+288\,{\mu}^{3}+3\,{\mu}^{2}
\sigma-288\,{\mu}^{2}-189\,\mu\,\sigma-42912\,\mu-639\,\sigma-97632\right]
\end{eqnarray*}
where $\mu$ is a root of $\mu^4-150\mu^2-360\mu+45=0,$ and $\sigma
=\sqrt{-48\mu^3+48\mu^2+7920\mu-13680}$.

\subsection{The $G(3,1,2)$ case}

We have 2 mirrors defining reflections of order 3:
$$\alpha_1=\left[1,0\right],\qquad \alpha_2 =\left[0,1\right],$$
and 3 mirrors defining reflections of order 2:
$$\alpha_3=\left[1,-1\right],\qquad \alpha_4 =\left[1,\f{1}{2}i\sqrt{3}+\f{1}{2}\right],\qquad \alpha_5 =\left[1,-\f{1}{2}i\sqrt{3}+\f{1}{2}\right].$$

\subsection{The $G(3,1,3)$ case}

We have 3 mirrors defining reflections of order 3:
$$\alpha_1=\left[1,0,0\right],\qquad \alpha_2 =\left[0,1,0\right],\qquad \alpha_3 =\left[0,0,1\right].$$
and 9 mirrors defining reflections of order 2:
\begin{eqnarray*}
&&\alpha_4=\left[1,-1,0\right],\qquad
\alpha_5=\left[1,0,-1\right], \qquad\quad
\alpha_6 =\left[0,1,-1\right]\\
&&\alpha_7 =\left[1,\f{1}{2}i\sqrt{3}+\f{1}{2},0\right],\qquad
\alpha_8 =\left[1,0,\f{1}{2}i\sqrt{3}+\f{1}{2}\right],\qquad
\alpha_9 =\left[0,1,\f{1}{2}i\sqrt{3}+\f{1}{2}\right]\\
&&\alpha_{10} =\left[1,-\f{1}{2}i\sqrt{3}+\f{1}{2},0\right], \qquad
\alpha_{11} =\left[1,0,-\f{1}{2}i\sqrt{3}+\f{1}{2}\right], \qquad
\alpha_{12} =\left[0,1,-\f{1}{2}i\sqrt{3}+\f{1}{2}\right]
\end{eqnarray*}

\subsection{The case of $I_2(6)$}

We have 6 mirrors defining reflections of order 2:
\begin{eqnarray*}
&&\alpha_1=\left[1,0\right],\qquad \alpha_2=\left[0,1\right],\qquad \alpha_3=\left[1,\f{1}{\sqrt{3}}\right]\\
&&\alpha_4=\left[1,-\sqrt{3}\right],\qquad \alpha_5=\left[1,-\f{1}{\sqrt{3}}\right],\qquad \alpha_6=\left[1,\sqrt{3}\right]
\end{eqnarray*}

\subsection{The case of $I_2(8)$}

We have 8 mirrors defining reflections of order 2:
\begin{eqnarray*}
&&\alpha_1=\left[1,0\right],\qquad \alpha_2=\left[0,1\right],\qquad \alpha_3=\left[1,-1\right]\qquad\alpha_4=\left[1,1\right]\\
&&\alpha_5=\left[1,-\sqrt{2}-1\right],\quad \alpha_6=\left[1,\sqrt{2}-1\right],\quad \alpha_7=\left[1,-\sqrt{2}+1\right],\quad
 \alpha_8=\left[1,\sqrt{2}+1\right]
\end{eqnarray*}

\subsection{The case of $I_2(10)$}
We have 10 mirrors defining reflections of order 2:
\begin{eqnarray*}
&&\alpha_1=\left[1,0\right],\qquad \alpha_2=\left[0,1\right],\\
&&\alpha_3=\left[1,\f{1}{2}\sqrt{5-2\sqrt{5}}(-4-2\sqrt{5})\right],\qquad\alpha_4=\left[1,-\sqrt{5-2\sqrt{5}}\right],\\
&&\alpha_5=\left[1,\sqrt{5-2\sqrt{5}}\right],\qquad \alpha_6=-\left[1,\f{1}{2}\sqrt{5-2\sqrt{5}}(-4-2\sqrt{5})\right],\\ &&\alpha_7=\left[1,-\f{1}{5}\sqrt{5-2\sqrt{5}}(-5-2\sqrt{5})\right],\qquad\alpha_8=\left[1,\f{1}{5}\sqrt{5-2\sqrt{5}}\sqrt{5}\right],\\
&&\alpha_9=\left[1,\f{1}{5}\sqrt{5-2\sqrt{5}}(-5-2\sqrt{5})\right],\qquad \alpha_{10}=\left[1,-\f{1}{5}\sqrt{5-2\sqrt{5}}\sqrt{5}\right]\\
\end{eqnarray*}

\subsection{The case of $B_2$}
We have 4 mirrors defining reflections of order 2:
\begin{eqnarray*}
\alpha_1=\left[1,0\right],\qquad \alpha_2=\left[0,1\right],\qquad \alpha_3=\left[1,-1\right]\qquad\alpha_4=\left[1,1\right]\\
\end{eqnarray*}

\subsection{The case of $B_3$}
We have 9 mirrors defining reflections of order 2:
\begin{eqnarray*}
&&\alpha_1=\left[1,0,0\right],\qquad \alpha_2=\left[0,1,0\right],\qquad \alpha_3=\left[0,0,1\right]\\
&&\alpha_4=\left[1,0,-1\right],\qquad \alpha_6=\left[0,1,-1\right],\qquad \alpha_7=\left[1,-1,0\right]\\
&&\alpha_7=\left[1,1,0\right],\qquad \alpha_8=\left[1,0,1\right],\qquad \alpha_9=\left[0,1,1\right].
\end{eqnarray*}

\section{Appendix 2. Saito coordinates in the cases of the groups $G_{29},G_{32},G_{33}$}
For higher rank complex reflection groups the computations become cumbersome.  
In the case of $G_{29},G_{32},G_{33}$ not all conditions have been checked.
However, imposing the flatness conditions at some special points, it is sufficient to fix uniquely the parameters 
 in the choice of the basic invariants. For this reason we conjecture that the basic invariants obtained in this way coincide with 
 the generalized Saito flat coordinates of a bi-flat structure. We expect also that  the vector potentials of the natural product of this structure coincide with the vector potentials obtained in \cite{KMS2} with a different approach. 

\subsection{The case of $G_{32}$}
Basic invariants are (see \cite{Ma2})
\begin{eqnarray*}
u_1&=&F_{12}\\
u_2&=&-54p_4^{18}+(17)(54)C_6p_4^{12}+(54)(1870)C_9p_4^9+\f{1}{2}(17)(27)
(19C_6^2-15C_{12})p_4^6+\\
&&(54)(170)C_6C_9p_4^3+C_6^3-30C_6C_{12}-25C_{18}=F_{18}\\
u_3&=&1728C_6p_4^{18}-(36)(1728)C_9p_4^{15}+(15)(144)(7C_{12}+C_6^2)p_4^{12}-(10)(1728)C_6C_9p_4^9+\\
&&(36)(178C_{18}-135C_6C_{12}+5C_6^3)p_4^6+432(41C_{12}-C_6^2)C_9p_4^3+C_6^4+6C_6^2C_{12}\\
&&-16C_6C_{18}+9C_{12}^2=F_{24}\\
u_4&=&F_{30}=-2(6^4)C_6p_4^{24}+312(6^4)C_9p_4^{21}
+216(715C_{12}-127C_6^2)p_4^{18}+272(6^4)C_6C_9p_4^{15}\\
&&+18(1306C_{18}+6045C_6C_{12}-295C_6^3)p_4^{12}
+216(73C_6^2-5473C_{12})C_9p_4^9+\\
&&\f{3}{2}(16648C_6C_{18}
+2334C_6^2C_{12}-20709C_{12}^2-C_6^4)p_4^6-36(1370C_{18}-657C_6C_{12}+\\
&&7C_6^3)C_9p_4^3
+C_6^5-19C_6^3C_{12}+29C_6^2C_{18}-6C_6C_{12}^2-5C_{12}C_{18},
\end{eqnarray*}
with
\begin{eqnarray*}
C_6 &=& p_1^6-10p_1^3p_2^3-10p_1^3p_3^3+p_2^6-10p_2^3p_3^3+p_3^6\\
C_9 &=& (p_1^3-p_2^3)(p_2^3-p_3^3)(-p_1^3+p_3^3)\\
C_{12} &=& (p_1^3+p_2^3+p_3^3)((p_1^3+p_2^3+p_3^3)^3+216p_1^3p_2^3p_3^3)\\
C_{18} &=& (p_1^3+p_2^3+p_3^3)^6-540p_1^3
p_2^3p_3^3(p_1^3+p_2^3+p_3^3)^3-5832p_1^6p_2^6p_3^6
\end{eqnarray*}
We conjecture that Saito flat coordinates are
$$u_1=F_{12},\quad
u_2=F_{18},\quad u_3=F_{24}+c_1F_{12}^2,\quad 
u_4=F_{30}+c_2F_{12}F_{18}.
$$
with
$$c_1 = -\frac{21}{25},\quad c_2 = -\frac{11}{25}.$$
Up to an inessential constant factor they coincide with the choice of basic invariants of Orlik and Terao.

\subsection{The case of $G_{33}$}
The basic invariants are (see \cite{Bu} with corrections of \cite{O})
\begin{eqnarray*}
U_1&=&p_1^4-8a_1p_1+48p_2p_3p_4p_5\\
U_2&=& p_1^6+(-20p_2^3-20p_3^3-20p_4^3-20p_5^3)p_1^3+360p_1^2p_2p_3p_4p_5-8p_2^6+80p_2^3p_3^3+80p_2^3p_4^3\\
&&+80p_2^3p_5^3-8p_3^6+80p_3^3p_4^3+80p_3^3p_5^3-8p_4^6+80p_4^3p_5^3-8p_5^6\\
U_3&=&\f{1}{63700992}{\rm det}(H(u_1))\\
U_4&=&5a_2p_1^6+(99a_3+a_1a_2)p_1^3+216a_4-36a_1a_3+24a_2^2-4a_1^2a_2+\\
&&p_2p_3p_4p_5(3p_1^8+33a_1p_1^5+(18a_2+30a_1^2)p_1^2)+(p_2p_3p_4p_5)^2
(243p_1^4+108a_1p_1)\\
U_5&=&4a_3p_1^9+(54a_4+12a_1a_3-a_2^2)p_1^6+(162a_1a_4-18a_2a_3+12a_1^2a_3-2a_1a_2^2)p_1^3+\\
&&27a_3^2-18a_1a_2a_3+4a_1^3a_3+4a_2^3-a_1^2a_2^2+\\
&&p_2p_3p_4p_5(6a_2p_1^8+p_1^5(54a_3+12a_1a_2)+p_1^2(243a_4+54a_1a_3-36a_2^2+6a_1^2a_2))\\
&&+(p_2p_3p_4p_5)^2(3p_1^{10}+18a_1p_1^7+p_1^4(54a_2+27a_1^2)+p_1(162a_3-54a_1a_2+12a_1^3)).
\end{eqnarray*}
with
\begin{eqnarray*}
a_1 &=& -p_2^3-p_3^3-p_4^3-p_5^3\\
a_2 &=& p_2^3p_3^3+p_2^3p_4^3+p_2^3p_5^3+p_3^3p_4^3+p_3^3p_5^3+p_4^3p_5^3\\
a_3 &=& -p_2^3p_3^3p_4^3-p_2^3p_3^3p_5^3-p_2^3p_4^3p_5^3-p_3^3p_4^3p_5^3\\
a_4 &=& p_2^3p_3^3p_4^3p_5^3
\end{eqnarray*}
We conjecture that Saito flat coordinates are
\begin{eqnarray*}
&&u_1=U_1,\quad u_2=U_2,\quad u_3=U_3+c_1U_1U_2,\quad u_4=U_4+c_2U_1^3+c_3U_2^2,\\
&&u_5=U_5+c_4U_1^3U_2+c_5U_1^2U_3+c_6U_2^3+c_7U_2U_4.
\end{eqnarray*}
with
\begin{eqnarray*}
&&c_1 = -\f{1}{768},\quad c_2 = -\f{5}{3072},\quad c_3 = -\f{5}{2304},\quad c_4 = \f{11}{884736},
\quad c_5 = -\f{1}{128},\\
&&\quad c_6 = \f{11}{1990656},\quad c_7 = -\f{1}{288}.
\end{eqnarray*}

\section{Appendix 3. The non-well generated cases} 
The procedure we have introduced works only partially for non-well generated complex reflection groups in the sense that it allows us to reconstruct flat connections, but it fails to provide a compatible product $\circ$. Let us present what happens if one tries to implement the algorithm in the case of $G_7$, a non-well generated complex reflection group of rank $2$, whose ring of invariants is generated by the following polynomials (see \cite{LT}): 
$$u_1=(p_1^4+(2i)\sqrt{3}p_1^2p_2^2+p_2^4)^3, \quad u_2=(p_1^5p_2-p_1p_2^5)^2.$$
Since $u_1$ and $u_2$ have the same degree, the connection $\nabla^{(1)}$ is uniquely determined. It turns out that it is almost hydrodynamically equivalent to the dual connection. Unfortunately, defining the unit of the product as a flat vector field and defining the product $\circ$ in  the standard way, we obtain that the compatibility is no longer satisfied. Similar results hold true in the case of the groups $G_{11}$, $G_{12}$, $G_{13}$, $G_{19}$
  and $G_{22}$. In the case of $G_{15}$,  there is one parameter in the basic invariants. It turns out that  the connection $\nabla^{(1)}$ is almost hydrodynamically equivalent to the dual connection for each value of this parameter and that the compatibility with the  product $\circ$  is never satisfied. To conclude let us consider the case of $G_{31}$. The basic invariants
 are
$$u_1=U_1,\quad u_2=U_2,\quad u_3=U_3+c_1U_1U_2,\quad u_4=U_4+c_2U_1^3+c_3U_2^2,$$
 where $U_2$ and $U_3$ coincide respectively with $U_3=F_{12}$ and $U_4=F_{20}$ of $G_{29}$ and
\begin{eqnarray*}
U_1&=& p_1^8+14p_1^4p_2^4+14p_1^4p_3^4+14p_1^4p_4^4
+168p_1^2p_2^2p_3^2p_4^2+p_2^8+
14p_2^4p_3^4.\\
&&+14p_2^4p_4^4+p_3^8+14p_3^4p_4^4+p_4^8\\
U_4&=&\f{1}{265531392}{\rm det}(H(U_1)).
\end{eqnarray*}
The values of these parameters are fixed using the same procedure applied to $G_{29}$, $G_{32}$ and $G_{33}$. They are
$$c_1 = -\f{3}{5},\qquad c_2 = -\f{1}{6480},\qquad c_3 = -\f{1}{4860}.$$

\end{document}